\documentclass[12pt]{JHEP3}

\newcommand{\ignore}[1]{}

\usepackage{graphicx}
\usepackage{amssymb}

\newcommand{\apm}{\alpha'}

\newcommand{\reffig}[1]{Fig.~\ref{fig:#1}}

\newcommand{\beq}{\begin{eqnarray}}% can be used as {equation} or {eqnarray}
\newcommand{\eeq}{\end{eqnarray}}
\newcommand{\nn}{\nonumber}

\def\be{\begin{equation}}
\newcommand{\bel}[1]{\be\label{#1}}
\def\ee{\end{equation}}
\newcommand{\eref}[1]{(\ref{#1})}
\newcommand{\Eref}[1]{Eq.~(\ref{#1})}
\newcommand{\rem}[1]{}

\def\half{\frac{1}{2}}
\def\Half{\frac{1}{2}}
\def\dd{\partial}
\newcommand{\<}{\langle}
\renewcommand{\>}{\rangle}
\def\NN{{\cal N}}
\def\kperp{k_\perp}
\def\pperp{p_\perp}
\def\xperp{x_\perp}

\def\nonestar{$\NN=1^*$}

\def\nfour{$\NN=4$}

\title{The Pomeron and  Gauge/String Duality}

\author{ Richard  C. Brower\footnote{Physics Department,
Boston University, Boston MA 02215},
Joseph Polchinski\footnote{Kavli
Institute for Theoretical Physics, University of
  California, Santa Barbara, CA 93106-4030}, %\\ 
Matthew J. Strassler\footnote{Department of Physics, University of Washington, Seattle,
  WA 98195} 
and  Chung-I Tan\footnote{Physics Department, Brown University,
Providence, RI 02912}
}
{}

\preprint{BROWN-HET-1462}

\abstract{
The traditional description of high-energy small-angle scattering in
QCD has two components --- a soft Pomeron Regge pole for the tensor
glueball, and a hard BFKL Pomeron in leading order at weak coupling.
On the basis of gauge/string duality, we present a coherent treatment
of the Pomeron.  In large-$N$ QCD-like theories, we use curved-space
string-theory to describe simultaneously both the BFKL regime and the
classic Regge regime.  The problem reduces to finding the spectrum of
a single $j$-plane Schr\"odinger operator.  For ultraviolet-conformal
theories,  the spectrum exhibits a set of Regge trajectories at
positive $t$, and a leading $j$-plane cut for negative $t$, the
cross-over point being model-dependent.  For theories with
logarithmically-running couplings, one instead finds a discrete
spectrum of poles at all $t$, where the Regge trajectories at positive
$t$ continuously become a set of slowly-varying and closely-spaced
poles at negative $t$.  Our results agree with expectations for the
BFKL Pomeron at negative $t$, and with the expected glueball spectrum
at positive $t$, but provide a framework in which they are unified.
Effects beyond the single Pomeron exchange are briefly discussed.
}
\begin{document}

%\maketitle

\baselineskip =16pt

\section{Introduction}

As a phenomenological model for hadrons in QCD, string theory
  in flat space has not been widely successful.
In what we may call the ``classic Regge regime''($s$ much greater than
$\Lambda_{QCD}^2$, with $|t|$ of order or smaller than $\Lambda_{QCD}^2$) 
\cite{ChewFrautschi,VNGribov:1961}, hadronic
scattering data suggests that QCD amplitudes exhibit ``Regge
behavior'' similar to that of flat-space classical string theory.  By
Regge behavior is meant that, {\it e.g.}  for $2\to2$ scattering
amplitudes, ${\cal A}(s,t)\sim s^{\alpha(t)}$, where the functions
$\alpha(t)$ are called ``Regge trajectories''.  For most kinematics,
however, strings in flat space disagree qualitatively with QCD. In
elastic scattering at large angles $(s\sim -t\gg \Lambda_{QCD}^2$),
QCD amplitudes are suppressed by powers of $s$, while amplitudes in
string theory are exponentially suppressed.  For scattering with $s\gg
-t \gg \Lambda_{QCD}$ (small fixed angles and ultra-high energies),
string amplitudes continue to show Regge behavior with a linear
trajectory, but QCD amplitudes behave differently.  The asymptotic
Regge regime is physically important, as it dominates total
cross-sections 
and differential
cross-sections $d\sigma/dt$ at small angle.  Unfortunately neither
direct perturbative computation nor lattice gauge theory methods can
be used to compute QCD amplitudes in this kinematic region.  Many
attempts have been made to clarify the physics of this regime, as well
as the related physics of small-$x$ structure functions in deep
inelastic scattering, but the situation remains murky.

QCD is an especially difficult theory in which to investigate this
issue, and were it not for the data we would have no good intuition
for the physics.  An important simplification is expected to occur
when the number of colors $N$ is taken very large.  In the limit $N\to
\infty$ followed by $s\to\infty$, scattering amplitudes in the Regge
regime are dominated by what is known as single-Pomeron exchange.  The
Pomeron is a coherent color-singlet object, built from gluons,
whose properties are universal; it is the object which is exchanged by
any pair of hadrons that scatter at high energy and large impact
parameter.\footnote{The Pomeron is a Regge singularity initially
proposed by Chew and Frautschi \cite{ChewFrautschi} and independently
by Gribov \cite{VNGribov:1961}, in honor of I. Ia. Pomeranchuk, who
first addressed the general question of the possible equality of total
cross-sections for particle-particle and particle-antiparticle
interactions at high energies.  Even before the theory of
QCD was introduced, it was recognized that the Pomeron propagator
should be endowed with the topology of a cylinder in a $1/N$
expansion, i.e., it represents
the exchange of a closed-string-like structure. 
See \cite{Veneziano:1973nw, Capella:1984tj}.
This topological feature was explored extensively in the 1980's, 
through the optical theorem, to understand patterns of 
particle production \cite{Capella:1992yb}.  Interest in the
phenomenological importance of the Pomeron  was rekindled in the
1990's; see \cite{DL}.}  In string theory, this is the object which is
exchanged in tree-level scattering in the Regge regime; it is not the
graviton but the graviton's Regge trajectory.  In real QCD at fixed
$N$, and in string theory at finite string coupling, multi-Pomeron
exchange eventually comes to dominate as $s\to\infty$.  We will not
address this regime in the present paper (aside from a few comments in
the conclusions), focusing instead on clarifying the properties of
single Pomeron exchange.

Even the single Pomeron is very subtle in QCD.  At positive $t$, the
notion of a ``soft'' Pomeron --- a Regge trajectory on which lies the
lightest $2^{++}$ glueball state ---is generally accepted.\footnote{
Due to mixing with ordinary mesons, experimental identification of
glueball states has been challenging.  The best evidence for their
existence has been through lattice gauge
theory~\cite{Morningstar:1999rf}.  For inferring the property of the
Pomeron trajectory from lattice data, see \cite{Meyer:2004jc}.  The
relevant tensor glueball state was first studied
in~\cite{Brower:1999nj,Constable:1999gb,Brower:2000rp} from an AdS/CFT
duality perspective. For first attempts at calculating glueball masses
using AdS/CFT, following work of \cite{Witten:1998zw},
see \cite{Csaki:1998qr, deMelloKoch:1998qs}.}  It
generalizes the observed ``soft'' charge-carrying Reggeons, such as
the rho trajectory for the $\rho$ meson and its higher spin
recurrences \cite{rho}.  All glueball states are expected to become
stable as $N\to\infty$.  At negative $t$, the notion of a ``hard''
Pomeron has emerged from perturbative resummation of Feynman diagrams,
as pioneered by Balitsky, Fadin, Lipatov, and Kuraev
\cite{Lipatov:1976zz,Kuraev:1977fs,BL}, referred to as ``BFKL"; for a
modern introduction and a more complete list of references, see
\cite{FR}.  The original calculation is at leading order in $\alpha$
and resums all terms of order $[\alpha\ln(s/t)]^n$, $n\geq 0$.  The
BFKL approach has been controversial, especially following the
understanding that in QCD the next-order correction to BFKL is large
and of opposite sign to the leading-order answer \cite{Fadin:1998py,
Camici:1997ij}.  Meanwhile, no existing calculational method, or
experimental data, can simultaneously address the physics at both
positive and negative $t$.  All in all, the relation between the two
Pomerons, the theoretical status of the BFKL method, and the physics
of the $|t|<\Lambda_{QCD}^2$ region have never been made entirely
clear.

Importantly, the large size of the correction to the leading-order
BFKL result is in part due to the large size of the beta function in
QCD.  For this reason, a second significant simplification for an
analysis of the Regge regime involves specializing to large-$N$ gauge
theories whose beta function is either zero or small, in particular of
order $1/N$.  If the beta function vanishes, the theory is strictly
conformally invariant, and the BFKL computation can be carried out
without confusing the effects arising from the running coupling with those
from other sources.  Indeed, it has been shown that the next-order
corrections to the BFKL result in \nfour\ Yang-Mills are a third as
large as those in QCD, making the analysis much more reasonable and
interpretable.  We will also see that our analysis is especially
simple in this case.  Theories with a small beta function can then be
understood as a small perturbation on the conformally invariant case.

Although the resummation calculation of BFKL applies at $s\gg |t|$, it is
only valid in regimes where confinement effects can be completely neglected.
At best, these include (1) computations in the regime $s\gg -t\gg
\Lambda^2_{QCD}$, where the large momentum transfer implies that the
scattering takes place on scales small compared to the confinement scale, and
(2) computations, for any $t\leq 0$, but with $s$ not exponentially large,
concerning hadrons whose size $\rho$ is sufficiently small compared to
$\Lambda_{QCD}^{-1}$, as would be the case for quarkonium states with quarks
of mass $M\gg \Lambda_{QCD}$.  For this reason, the cleanest application of
BFKL is to quarkonium-quarkonium scattering \cite{Mueller:1994gb}, or to
deep-inelastic scattering off a quarkonium state, or to off-shell
photon-photon scattering \cite{BL}.  But it cannot be used to study the
classic Regge regime, for which the physics of confinement is dominant.

In this paper, we aim to show, in certain large-$N$ QCD-like theories
with beta functions that are vanishing or small in the ultraviolet,
how the BFKL regime (which disagrees with flat-space string theory)
and the classic Regge regime (which roughly agrees with it) can both
simultaneously be described using curved-space string theory.  We will
extend this to all values of $s\gg \Lambda_{QCD}^2$ and $t$, obtaining
thereby the full analytic structure of the single-Pomeron exchange
kernel, including both the soft Pomeron at positive $t$ and the hard
Pomeron at negative $t$.  This is technically impossible in QCD, where
computations in lattice gauge theory and perturbative gauge theory
calculations are separated by a kinematic range where the physics is
both strongly coupled and Lorentzian in character.

As an illustration of the form of our results, we briefly summarize our
investigation of the simplest case: the scattering of two objects by
conformally-invariant dynamics. In conformally-invariant theories there are,
of course, no hadrons, but we may either consider four-point functions of
operators that are functions of nonzero momentum $p_i$, with
$s=-(p_1+p_3)^2\gg -t= (p_1+p_2)^2$, or we may add massive quarks as a
probe of the theory, at the cost of only a $1/N$ violation of conformal
invariance, and consider the scattering of quarkonium bound states.  For
$t=0$, the single-Pomeron-exchange amplitude for scattering of two such
objects $A$ and $B$, with center-of-mass energy $\sqrt s$, is of the form
\bel{BFKLamp}
\int {d\pperp\over \pperp} \int {d\pperp'\over \pperp'}
\Phi_A(\pperp) {\cal K}(\pperp,\pperp',s) \Phi_B(\pperp')
\ee
where $\pperp$ ($\pperp'$)
is the magnitude of the transverse momentum with
which the first (second)  object is probed by the Pomeron. 
The two functions $\Phi_i$, called ``impact
factors'', describe the transverse structure of the objects undergoing
the scattering. These impact factors are convolved together with the
BFKL kernel ${\cal K}$.

When the 't Hooft coupling $g^2N$ is very small and $N$ is very large,
the computation can be done using the methods of BFKL, according to
which, at leading nontrivial order in $\alpha$, the kernel can be
written exactly as an inverse Mellin transform in the spin $j$
\bel{exactkernel}
 {{\cal K}} (\pperp,\pperp',s) = \int^{C+i\infty}_{{C-i \infty}} {\frac{dj}{2i \pi }  s^j} \int_{-\infty}^\infty
d\nu\ e^{%2
i\nu\ln(\pperp/\pperp')}{ {1\over j -\hat j(\nu)}}
\ee
with the $j$-plane contour to the right of all $j$-plane singularities, and
\bel{BFKLj}
\hat j(\nu) = 1 + {\alpha N\over\pi}\left[-2\gamma_E 
- \Psi\left(\half + i\ %\nu
\frac\nu 2\right)- 
\Psi\left(\half - i\ %\nu
\frac \nu 2 \right)\right] 
+ O(\alpha^2)
\ 
\ee
where $\gamma_E$ is Euler's constant and
$\Psi(z)$ is the Digamma function.\footnote{Note that in many
papers a different normalization of $\nu$ is used; in particular
our variable $\nu$ is twice as large as that used in \cite{FR}.}  
(Here we limit ourselves
to the term with conformal spin equal to zero.)
A good approximation to this kernel can be found by expanding
the function $\hat j$ to second order in $\nu$
\bel{omegaquad}
 \hat j (\nu) = j_0  - {\cal D}\nu^2 + {\rm order}(\nu^4) \ ,
\ee
where
\bel{coeffs1}
 j_0 = 1 +  {4\ln 2\over \pi}\alpha N \ , \quad
{\cal D} = 
{7\zeta(3)\over 2\pi}\alpha N \ .
\ee
~From this one obtains
\bel{quadkernel}
{\cal K} (\pperp,\pperp',s)\approx 
{s^{{j_0} } \over \sqrt{{ 4 
\pi{\cal D} }\ln s}}
e^{-(\ln \pperp'-\ln \pperp)^2/4 
{\cal D}\ln s}
\ee
Strictly speaking, $s$ must be replaced with $s/s_0$, where
$\sqrt{s_0}$ is a characteristic energy scale which we will discuss
later.  We may recognize ${\cal K}$, in this
approximation, as a power of $s$ times a diffusion kernel, with the
diffusion occurring in the variable $\ln\pperp$ over a diffusion time
$\tau \sim\ln s$.

This is a very curious result.  Ordinary
Regge behavior ${\cal A}\sim s^{\alpha(t)}$, in flat-space string
theory or in the classic Regge regime, is related
to diffusion in {\it transverse  position} space.  Around the
``intercept'' at $t=0$, the Regge trajectories are initially linear,
with $\alpha(t) = \alpha_0+\alpha' t + \dots$; the higher order
corrections are zero in ordinary flat-space string theory and are
apparently small in QCD. The Regge amplitudes, Fourier transformed
into position space, take a diffusive form.  Suppose the scattering
particles are traveling initially along the $x^{1}$ axis, with the
momentum transfer $k^\mu$ completely transverse to the $x^0, x^1$
coordinates, so that $t=-{\bf \kperp}^2$; then 
\bel{Regisdiff} 
\int \ d^{d-2}\kperp e^{i{\bf\kperp}\cdot {\bf
\xperp}} s^{\alpha(t)} \approx s^{\alpha_0} \int \ d^{d-2}\kperp
e^{i{\bf\kperp}\cdot {\bf \xperp}} e^{-\alpha'\kperp^2 \ln s} =
{ \frac{s^{\alpha_0 }  e^{-\xperp^2/4\alpha'\ln s}}{(4 \pi\alpha' \ln s)^{(d-2)/2}}}
 \ .  
\ee
~From the point of view of one string (or hadron), the other string
(or hadron) grows, with $\langle \xperp^2\rangle \sim \alpha'\ln s$,
via diffusion, with a diffusion time $\propto\ln s$.  This can also be
viewed as due to a time-resolution effect; for a modern discussion,
see Ref. \cite{susskind}.  The time-dilation of the boosted string, as
viewed by the ``target'', resolves more and more of its quantum
fluctuations.  This tends to make the string appear longer, and
consequently larger, by an amount that grows like a random walk in the
dimensions transverse to the motion of the string.  This is explicit
in our later discussion of Regge physics in the light-cone frame, in
Sec.~\ref{sec:light-cone}.

The similarity between these two types of diffusion, in two different
variables $(\ln \pperp$ versus $\xperp$) and in two different regimes (far from
confinement for BFKL, deep within confinement for the classic Regge regime),
is not accidental.
This can be seen by considering the same scattering problem at large
$g^2N$, where BFKL methods cannot be applied but where string
theoretic methods, which resum the expansion in $g^2N$ to all orders,
can be used.  The scattering of two states in a conformal field theory
translates into the scattering of strings on a curved background of
the form $AdS_5\times W$.  The coordinates on the $AdS_5$ space are
$x^\mu$, the usual Minkowski coordinates, and $r$, the fifth
coordinate that runs from $r=0$ at the horizon of the Poincar\'{e} patch
of $AdS_5$ to $r=\infty$ at its boundary.  The coordinate $r$ is
related to the energy scale $\mu$ in the quantum field theory; $r\to
0$ corresponds to the infrared and $r\to\infty$ to the ultraviolet.
We will show the resulting kernel at $t=0$ for the scattering of two strings on
an $AdS_5$ space via single-Pomeron exchange is 
\bel{quadkernel2}
{\cal K}(r,r',s)= 
{s^{j_0} \over \sqrt{4 \pi  {\cal D}\ln s}}
e^{-({\ln r - \ln r'})^2/4{\cal D}\ln s}
\ee
where 
\bel{coeffs2}
j_0=2-{2\over\sqrt\lambda} 
+ O(1/\lambda)\ 
, \quad {\cal D}= {1\over 2\sqrt\lambda} + O(1/\lambda)\ \ .
\ee
Here $\lambda\equiv R^4/\alpha'^2$, where $R$ is the curvature radius
of the $AdS_5$ space and $2\pi\alpha'$ is the inverse of the string
tension.  Note that $\lambda= g^2_{YM} N= 4\pi \alpha N $ in \nfour\
supersymmetric Yang-Mills theory --- the numerical coefficient can
differ in other theories but the proportionality always holds --- so
large $\lambda$ is large 't Hooft coupling.  Comparing this with
\Eref{quadkernel}, one sees that the fifth coordinate $r$ of the
string theory should be identified in this context with $\kperp$ of
the gauge theory.  The identification of $r$ and $\kperp$ has its
source in the UV/IR correspondence \cite{malda} and has been suggested
in numerous contexts (see for example \cite{hardscat,Brower:2002er,
brodsky} for related applications.)  
Note that the effective diffusion time $\tau$ is
of order $\ln s$ for both the BFKL and the Regge diffusion, at both
large and small $\lambda$.

This success is consistent with others that have emerged in recent
years.  The duality of gauge theory and string theory
has led us to expect that many of the
failures of string theory as a good model of the physics of QCD are
due not to having the wrong string theory, but to putting the right
string theory on the wrong space-time background, namely flat
Minkowski space.  It is now known that much better phenomenological
string models for QCD are given by string theory on certain curved
spaces.  The result just described implies that at large $N$,
vanishing beta function, and $t=0$, {\it the form of the BFKL result
may be interpreted as the Regge physics of a string theory
compactified to $AdS_5$, in the form of diffusion along the curved
fifth dimension.}
The only substantial difference between the small 't Hooft coupling result
\eref{quadkernel} and the large 't Hooft coupling result
\eref{quadkernel2} lies in the coefficients; compare
Eqs.~\eref{coeffs1} and \eref{coeffs2}.  This is strongly suggestive
that the kernel is always of the form \eref{quadkernel}, with the
overall power $j_0$ and the diffusion constant ${\cal D}$ being
continuously varying functions of $\alpha N$.  Indeed, 
such a result follows from the constraints of conformal
invariance.

The extension of the single-Pomeron kernel to nonzero $t<0$ (physical
scattering at small angles) is quite involved in the BFKL regime
\cite{finitetBFKL,KirschLipat}.  By contrast, using string-theory
methods, the kernel in the strong-coupling approximation is easily
derived.  We will argue that aspects of our result are necessarily
true in an asymptotically conformally-invariant theory.  As before,
the result depends on coefficients that are functions of the 't Hooft
coupling.  The weak- and strong-coupling results at finite $t$ have
the same formal structure, and qualitative similarities.  They differ
in that at weak coupling the Pomeron can couple to individual partons,
while at strong coupling it couples only to the entire hadron, in
analogy with the physics of deep inelastic scattering in the two
regimes \cite{jpmsDIS}.

In sections 2, 3 and 4, we will derive the result \eref{quadkernel2},
and its extension to $t<0$, in three ways.  The first uses a low-brow
approach which returns to the results of earlier work \cite{jpmsDIS}.  The
second carefully obtains the result from string theory in conformal
gauge, introduces Pomeron vertex operators as computational tools in
string theory, and discusses a number of underlying theoretical
aspects of the calculation.  The final derivation uses light-cone
gauge.  

Next we will generalize our results to include effects of confinement,
first considering theories whose ultraviolet physics is conformally
invariant (Sec.~5), and then considering theories with a
logarithmically running coupling (Sec.~6).  We use string theory to
study the full analytic structure of the single-Pomeron exchange
amplitude, for all values of $t$.  For $t\ll -\Lambda^2$, with $\Lambda$ of
order the confinement scale, the kernel is nearly independent of
confinement, and our results from the conformal case require no
modification if the beta function vanishes in the ultraviolet, and a
more substantial but model-independent modification if the coupling
runs.  For $t\gg +\Lambda^2$ the Regge trajectories on which the
hadronic resonances sit can be identified and studied.  It is
straightforward to compute the hadron spectrum using differential
equations which match directly to the equations governing the
diffusion at $t<0$.  This makes it possible to answer long-standing
questions concerning the behavior, as $t$ is taken from positive to
negative, of the Regge trajectories $\alpha(t)$.  Details of the
Regge-trajectories are model-dependent, but their presence and their
general $t$-dependence are not.  The region $|t|\sim\Lambda^2$, which
dominates total cross-sections and near-forward scattering, is the
most model-dependent. 
In addition to computing the kernel's
asymptotic form at large $s$, we also note various transient
effects which are present for $s$ not asymptotically large, some of
which are also model-independent.

Our all-$t$ results for the analytic behavior of the single-Pomeron
exchange amplitude are consistent with what is known from a
combination of data at positive $t$ and analytic work in field theory
at negative $t$.  The structure of our formalism makes concrete the
intuitive approach suggested in \cite{levintan}, which suggested
a unified treatment of the Pomeron and developed an intuitive
picture of diffusion in both hard and soft regimes.\footnote{See also
\cite{FR}, Section 5.6.}
We believe our result is
the first, however, to connect the positive and negative $t$ behavior
in a reliable and consistent theoretical framework.\footnote{Our
results contradict the earliest attempts \cite{IntegerPoles} to
connect positive and negative $t$, where it was proposed that Regge
trajectories at positive $t$ flatten out and extend to negative
integer values of $j$ as $t\to -\infty$.  Our approach and conclusions 
differ as well from refs.~\cite{Janik:1999zk,Janik:2000pp};
in particular~\cite{Janik:2000pp} finds a Pomeron intercept that is 
independent of $\lambda$. The conjectures of
\cite{Andreev:2004sy} also differ from our results, though by not as
great a degree; we will see below there is a certain commonality of
viewpoint at the point of departure, though in the end our approach
and conclusions are distinct from theirs.}

\section{A heuristic derivation}
\label{sec:heuristic}

In ref.~\cite{hardscat}, it was argued that scattering amplitudes in gauge
theories with good string dual descriptions 
can be expressed in a simple general form, in
which the underlying ten-dimensional amplitude is essentially local and the
four-dimensional amplitude is given by a coherent sum over scattering
anywhere in the six transverse dimensions.  In this section we will apply
this to scattering in the Regge limit.  We will see that at a certain point
the approximation breaks down, and we will need to make an educated guess as
to the correct amplitude.  In the next section we will make a more systematic
world-sheet analysis, seeing why the local approximation breaks down and how
to correct it.

For conformally invariant gauge theories, the metric of the dual
string theory is a product\footnote{More generally
it could be a warped product, where the $AdS$ metric is multiplied by
a function of the coordinates on $W$.} $AdS_5 \times W$,
\begin{equation}
\label{AdSWmetric}
ds^2 = \frac{r^2}{R^2} \eta_{\mu\nu} dx^\mu dx^\nu + \frac{R^2}{r^2} dr^2 + ds^2_W\ ,
\label{product}
\end{equation}
where $0 < r < \infty$.  We use $x^M$ for the ten-dimensional
coordinates, or $(x^\mu, r, \theta)$ with $\theta$ being the five
coordinates on $W$, or $(x^\mu, y^m)$ when we discuss all six
transverse coordinates $y$ together.  Our convention is that the
metric signature is spacelike-positive.  For the dual to ${\cal N}=4$
supersymmetric Yang-Mills theory~\cite{malda} the AdS radius
$R$ is
\begin{equation}
R^2 \equiv\sqrt{\lambda} \alpha'
= (4\pi g_{\rm string} N)^{1/2} \alpha' = 
(g_{\rm YM}^2 N)^{1/2} \alpha' \ ,
\end{equation}
and $W$ is a 5-sphere of this same radius.  By a ``good string dual
description'' we mean that $\lambda \gg 1$, so that the spacetime
curvature is small on the string scale, and $g_{\rm string} \ll 1$ so
that we can use string perturbation theory.

We are interested in gauge theories that are conformally invariant at high
energy but with the invariance broken at low energy, resulting in a mass gap
and confinement.  Roughly speaking, this means that the dual string metric is
of the $AdS$ form but with a lower cutoff on the coordinate $r$, so that $r_0
< r < \infty$.  More precisely, the product structure breaks down in the
infrared, and we must use a general warped product
\begin{equation}
ds^2 = G_{MN} dx^M dx^N = e^{2 A(y)} \eta_{\mu\nu} dx^\mu dx^\nu + G_{\perp mn} dy^m dy^n\ .
\end{equation} 
To simplify the discussion we define the radial coordinate by
$r^2/R^2 = e^{2A}$, so that $r_0^2/R^2 = \min(e^{2A})$.  (In
models where $e^{2A}$ has more than one local minimum, $r$ is not a good
coordinate, and one should everywhere replace $r/R$ with $e^A$.)
The precise metric
depends on the details of the conformal symmetry breaking.  Most of the
physics that we will study takes place in the conformal region where the
metric is the approximate $AdS$ product~(\ref{product}).  Even here we might
generalize to geometries that evolve slowly with $r$, as in cascading gauge
theories and in the running coupling example to be 
studied in Sec.~\ref{sec:runcplg}.

Glueballs arise as discrete modes in the six-dimensional transverse `cavity.'
For example, a scalar glueball created by the operator $F_{\mu\nu}F^{\mu\nu}$
would have a dilaton wavefunction
\begin{equation}
\Phi(x,y) = e^{i p \cdot x} \psi(y)\ . \label{cavity}
\end{equation}
Local operators in the gauge theory also translate into bulk
excitations, the difference being their boundary conditions
(non-normalizable rather than normalizable) as $r \to
\infty$~\cite{GKPAdS,WittenAdS,Balasubramanian:1998sn}.  The Pomeron is a closed string state, and at leading
large-$N$ order its properties do not depend on the open string
spectrum.  Thus we can study its physics without introducing branes
and open strings, though it will later be useful to do so in order to
model quarkonium scattering and to discuss open string trajectories.
The open string wavefunctions are of the same form~(\ref{cavity}) but
with support restricted to whatever branes have been embedded in the
transverse space.

In the string picture, scattering amplitudes are given as usual by
path integrals over string world-sheets embedded in the deformed
$AdS_5\times W$, with appropriate vertex operators for the external
states.  Aside from a few remarks at the end, we will consider only
the leading $1/N$ approximation, corresponding to spherical
world-sheets.  In general this would still be a forbidding
calculation, but a great simplification occurs at $\lambda
\gg 1$.  The string world-sheet action contains factors of $R^2$ from
the spacetime metric and $1/\alpha'$ from the string tension, which
combine into $R^2/\alpha' = \sqrt{\lambda}$.  Thus the string
world-sheet coupling $1/\sqrt{\lambda}$ is small, and the world-sheet
path integral is almost Gaussian.  It is not exactly Gaussian because
the constant mode on the world-sheet has no quadratic term.  We must
therefore separate the fields on the string world-sheet into 
their zero modes and the remaining
nonzero-mode parts,
\begin{equation}
X^M(\sigma^1,\sigma^2) = x^M + X'^M(\sigma^1,\sigma^2)\ .
\end{equation}
At fixed $x^M$, the Gaussian integral over the
nonzero modes is exactly as one would do in flat
spacetime,\footnote{The RR backgrounds have scaled away with the
curvature, so we are spared dealing with them.} thus producing the
ten-dimensional flat spacetime S-matrix
that would be seen by a local
observer (except that the momentum delta function from the zero modes
is missing.)  We denote this amplitude $i {\cal A}_{\rm
local}(x,y)$.  Integrating over the zero mode then gives the S-matrix
\begin{equation}
{\cal S} = i\int d^4x\,d^6y\,\sqrt{-G}{\cal A}_{\rm local}(x,y) \ . \label{local}
\end{equation}
We define ${\cal A}_{\rm local}$ to be a scalar; the $\sqrt{-G}$ then
arises from the path integral measure, which respects coordinate
invariance.

The dependence on the external states is implicit, but for high energy
scattering it can be put in a more explicit form.  In the
wavefunction~(\ref{cavity}), the gradients in the transverse
directions are generally of order $1/R$, much less than string scale.
In the noncompact directions we will be considering momenta at least
of order the string scale, and in fact much larger.  It follows that
the local inertial observer sees momenta directed essentially in the
noncompact directions.  In terms of inertial coordinates parallel to
the $\mu$ axes, the momenta are
\begin{equation}
\tilde p^\mu = \frac{R}{r} p^\mu\ . \label{red}
\end{equation} 
The $p^\mu$, which appear in the wavefunction~(\ref{cavity}), are the
conserved Noether momenta and so are identified with the gauge theory
momenta, while the $\tilde p^\mu$ are the momenta seen by a local
inertial observer.  Thus,
\begin{equation}
{\cal A}_{\rm local}(x,y) \to {\cal T}_{10} (\tilde p) \prod_{\rm ext.\atop states} e^{i p_i \cdot x} \psi_i(y)\ ,
\end{equation}
the flat-spacetime scattering amplitude with momenta $\tilde p$ times the product of the wavefunctions at $(x,y)$.  The integral over $x$ produces the four-dimensional momentum delta function.  Then, for scattering of scalars, the general local superposition~(\ref{local}) becomes
\begin{equation}
{\cal T}_4 = \int d^6 {y}\, \sqrt{-G}\, {\cal T}_{10}(\tilde p) \prod_{\rm ext.\atop states} \psi_i(y) \ ,\label{amp}
\end{equation}
where ${\cal S} = i(2\pi)^4 \delta^4 ({\textstyle \Sigma p}) {\cal T}_4$.
Note that $G$ is the determinant of the full ten-dimensional metric.
For external states with spin, ${\cal T}_{10}$ is naturally written
with tangent space indices, and so would be contracted with the
external wavefunctions in tangent space form.

The final results~(\ref{local}), (\ref{amp}) are simple and intuitive,
just a coherent superposition over all possible scattering locations.
The scattering is effectively local because the scale of fluctuations
of the string world-sheet is set by $\alpha'$, which is small compared
to the variations of the geometry.  The 
Gaussian approximation that leads to this
local expression is rather robust.  It will break down later in this
section, but only because we introduce a competing large parameter.
In the next section we will analyze this breakdown at the world-sheet
level, and see how to correct it.

Now let us apply this to Regge scattering, focusing for simplicity on
$2\to2$ scattering of scalars.  Thus $s = -(p_1 + p_3)^2$ is taken
large, with $t = -(p_1 + p_2)^2$  fixed.  The local inertial quantities
are
\begin{equation}
\tilde s = \frac{R^2}{r^2} s\ , \quad \tilde t = \frac{R^2}{r^2} t\ ,
\end{equation}
and so for scattering at a given value of $r$ the ten-dimensional process is also in the Regge regime.  Thus at fixed $r$ we have
\bel{regten}
{\cal T}_{10}(\tilde s, \tilde t) \to f(\alpha' \tilde t) (\alpha' \tilde s)^{2+\alpha' \tilde t /2}\ ,
\ee
with $f$ a process-dependent function.  In fact the relevant value of
$r$ in the superposition~(\ref{amp}) evolves with $s$ but does so only
slowly, so we remain well within the regime where the
form~(\ref{regten}) is valid.  Thus we have
\begin{equation}
{\cal T}_4(s,t) = \int d^6 {y}\, \sqrt{-G}\, 
f(\alpha' R^2 t/r^2) (\alpha' R^2 s/r^2)^{2+\alpha' R^2 t/ 2r^2}
\prod_{i=1}^4 \psi_i(y) \ .\label{ramp}
\end{equation}
Examining the exponent of $s$, we see that the intercept, the exponent at $t= 0$,
is 2 just as in flat spacetime.  We also see that the slope, the coefficient
of $t$, depends on $r$ as 
\be \frac{\alpha'_{\rm eff}(r)}{2} = \frac{R^2
  \alpha'}{2r^2} \; .
\label{alfeff} 
\ee 
The 2's appears in the denominator because this is a closed string
trajectory.  It is as though, in this ``ultralocal'' approximation, the
five-dimensional Pomeron gives rise to a continuum of four-dimensional
Pomerons, one for each value of $r$ and each with a different slope.\footnote{
The notion of a tension depending on a fifth dimension
dates to \cite{Polyakov:1997tj}. 
The idea of superposing many four-dimensional Pomerons is conceptually
anticipated in the work of \cite{Brower:2002er},
and more technically in that of \cite{Andreev:2004sy}, where an idea
similar to that of the next paragraph is  considered.}
This is illustrated schematically in Fig.~\ref{fig:ultralocal}.

\FIGURE[ht]{
%\begin{center}
\includegraphics[width = 3.5 in]{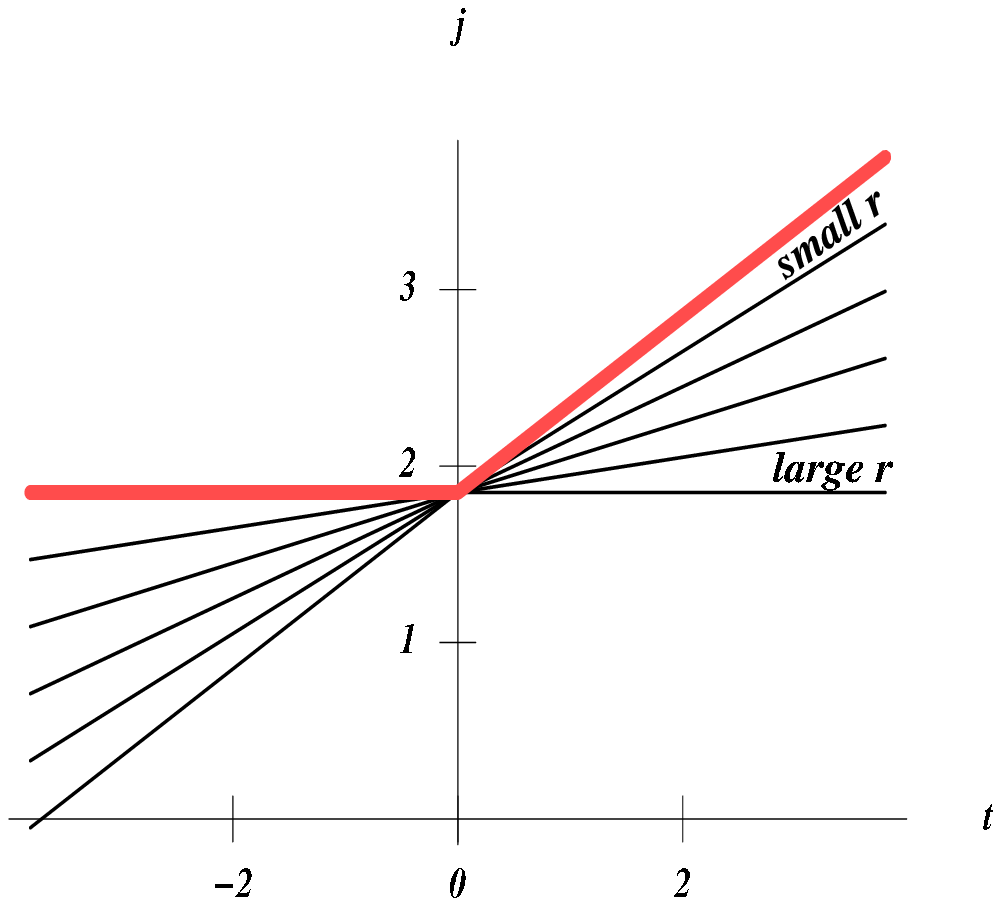}
%%\end{center}
\caption{In the ultralocal approximation, the slope of the leading
trajectory redshifts from order one at small $r$ down to zero as
$r\to\infty$; trajectories at several different $r$ values, with
their different slopes, are shown.
The leading singularity (the singularity with largest $j$ at fixed
$t$) is the usual linear Regge trajectory for $t>0$ but is a constant
near 2 for $t<0$.  The full story is more elaborate, as will be
shown.}
\label{fig:ultralocal}
}
%\end{figure}
%

At large $s$, the highest trajectory will dominate.  For
positive $t$, this would be the one at the minimum value $r_{0}$:
\be
\alpha'_{\rm eff} (t > 0) = \frac{R^2 \alpha'}{ r_0^2} \equiv \alpha'_0
\ . \label{alf0} 
\ee
For negative $t$, it would be the trajectories at large $r$.  The
wavefunctions in the superposition~(\ref{ramp}) make the integral
converge at large $r$, so at any given $s$ the dominant $r$ is finite,
but as $s$ increases the dominant $r$ moves slowly toward infinity and
so we have 
\be 
\alpha'_{\rm eff} (t < 0) = 0 \ .  
\ee 
We see that, in this approximation, the
dominant trajectory has a kink at $t = 0$, similar to (though more
abrupt than) the behavior in QCD.  Moreover, as also happens
in QCD, the nature of the
Pomeron changes.  For positive $t$
it sits at small $r$ and so its properties are determined by the
confining dynamics: it is a glueball.  At negative $t$ it sits at
large $r$ and so is effectively a very small object, analogous to the
tiny (and therefore perturbative) two-gluon Pomeron of
BFKL \cite{Brower:2002er}.

However, we can go well beyond the ultralocal approximation.
Note that we have two large quantities, $\lambda$ and $s$.  In the
discussion thus far, we have taken $\lambda$ large first, making the
Gaussian approximation, and then considered large $s$ within this
approximation.  But in order to expose a wider range of QCD-like
physics, we would like to keep values of $s$ which are large compared
to $\lambda$.  More precisely, the interesting physics arises in the
limit
\begin{equation}
\lambda, s \to \infty\ , \quad \frac{\ln s}{\sqrt{\lambda}}\ {\rm fixed}\ . 
\label{larges}
\end{equation}
Thus $s$ is exponentially large in $\sqrt{\lambda}$.  This is an enormous
scale from the point of view of AdS/CFT physics, but to reach real QCD we
must continue to small $\lambda$ and so the physics that we find in this
regime can become important.

In the regime~(\ref{larges}) it is necessary to retain terms of order
$1/\sqrt{\lambda}$ in the exponent $2 + \alpha' t/2$ of $s$.  Thus in the
ten-dimensional momentum transfer $\tilde t$ we must keep a term previously
dropped, coming from the momentum transfer in the six transverse directions:
\bel{kinemt}
\alpha' \tilde t \to  {\alpha'} \nabla_{P}^2 \equiv
\frac{\alpha' R^2 t}{r^2} + {\alpha'} \nabla_{\perp P}^2\ . 
\ee
The transverse Laplacian is proportional to $R^{-2}$, so that the added term
is indeed of order ${\alpha'}/{R^2} = 1/\sqrt{\lambda}$.  The Laplacian acts
in the $t$-channel, on the product of the wavefunctions of states 1 and 2 (or
3 and 4).  The subscript $P$ indicates that we must use the appropriate
curved spacetime Laplacian for the Pomeron being exchanged in the
$t$-channel; we will determine this shortly.

The inclusion of the transverse momentum transfer leads to several new
effects.  First, we will  now be able to determine the first strong coupling
correction, $O(1/\sqrt{\lambda})$, to the intercept 2.  Second, we will see
that $s^{\alpha' \tilde t /2}$ is now a diffusion operator in all
eight transverse dimensions, not just the Minkowski
directions, and that this leads to BFKL-like physics.  Third, to
obtain the Regge exponents we will now have to diagonalize the differential
operator~(\ref{kinemt}), so that instead of the `ultralocal'
Pomerons~(\ref{alfeff}) that arose in the earlier regime, we will
have a
more normal spectral problem.\footnote{Ref.~\cite{Andreev:2004sy}
proposes a different modification of the ultralocal Pomeron spectrum,
an Ansatz based on straight trajectories with a discrete set of
slopes.  The structure that we find in AdS/CFT, where the trajectories
are given by the eigenvalues of an effective Hamiltonian, is more
closely parallel to that found by BFKL in perturbation theory.}  

Now let
us determine the form of $\nabla_{\perp P}^2$.  We first make some
definitions.  The scalar Laplacian is
\begin{eqnarray}
\nabla_0^2 \phi &=& (-G)^{-1/2} \partial_M \!\left[ (-G)^{1/2} G^{MN} \partial_N \phi \right] \nonumber\\
&=& (r/R)^{-4} G_\perp^{-1/2} \partial_M \!\left[ (r/R)^4 G_\perp^{1/2} G^{MN} \partial_N \phi \right]\ .
\end{eqnarray}
To study the contribution of the exchange of a transverse traceless
  field of spin $j$ at high energies, it is useful to define the Laplacian in
  light-cone coordinates,
\begin{equation}
x^{\pm} = \frac{1}{\sqrt{2}} (x^0 \pm x^1)\ .
\end{equation}
Specifically, consider the tensor component $\phi_{+^j}$, with $j$ indices $+$, in a frame where the Pomeron momentum components $(p_1 + p_2)^\pm$ vanish.  The
covariant Laplacian reduces to 
\begin{equation}
\nabla_j^2 \phi_{+^j} = (r/R)^j \nabla_0^2 
\! \left[ (r/R)^{-j} \phi_{+^j}  \right] + \frac{j}{4} {\cal R}_+\!^+ \phi_{+^j}\ ,
\end{equation}
where $\cal R$ is the Ricci tensor.  It is convenient to define also
\begin{eqnarray}
\label{Deltajdefn}
\Delta_j &=& \nabla_j^2 - \frac{j}{4} {\cal R}_+\!^+ 
\nonumber\\
\Delta_j \phi_{+^j} &=& (r/R)^j \nabla_0^2 
\! \left[ (r/R)^{-j} \phi_{+^j}  \right] \ . \label{deltaj}
\end{eqnarray}
As we will see these expressions make sense even for noninteger
 $j$; they define an  on-shell Regge exchange process.

The difference between $\Delta_j$ and $\nabla_j^2$ is a curvature term and so
covariance alone does not determine the Laplacian to this order.  We must
resolve this ambiguity in order to find the shift in the Regge intercept, and
so we must look at the dynamics.  For transverse traceless fluctuations of
the metric, $h_{++}$, one finds from the supergravity field equations that
the wave equation in a warped background is
\be
\Delta_2 h_{++} = 0\ . \label{wave}
\ee
A simple way to check this is to note that in the long-wavelength limit the
transverse traceless perturbation becomes $h_{++} = (r/R)^2$; this is a
linear reparameterization of the background metric $(r/R)^2 \eta_{\mu\nu}$
and so must satisfy the correct wave equation.  From the explicit
form~(\ref{deltaj}) it follows that this is the case for the
equation~(\ref{wave}), and would not be true with an added curvature term.
Now, we have noted that the Regge intercept differs from 2 by an amount of
order $1/\sqrt{\lambda}$.  If it were exactly 2 we would be sitting on the
graviton pole, so we conclude that $\Delta_P = \Delta_2$ up to a correction
of order $1/\sqrt{\lambda}$.  The Laplacian term is already of order
$1/\sqrt{\lambda}$, so the shift is second order and can be neglected:
\begin{equation}
\nabla_{P}^2 \to \Delta_2 + O(j-2)\ .
\end{equation}
Thus
\begin{equation}
{\cal T}_4(s,t) = \int d^6 {y}\, \sqrt{-G}\,  \psi_3(y) \psi_4(y)
f(\alpha' R^2 t/r^2) (\alpha' R^2 s/r^2)^{2+\alpha' \Delta_2/2}
 \psi_1(y) \psi_2(y) \ .\label{ramp2}
\end{equation}
  This somewhat heuristic argument will be supported by the more formal treatment in the next section~\footnote{In flat space, the function $f$ has a pole at zero,
from massless $t$-channel exchange.  In a more precise treatment, we need to replace  the argument  of 
$f$ with the curved-spacetime propagator $\alpha'\Delta_2$. See Sec. \ref{sec:reggewarpedspacetime}.
In most cases we are interested
in the imaginary part of $\cal T$, and ${\rm Im}\, f(0)$ is finite.
}.

Now let us work out the explicit form of the amplitude for processes
that take place in the high-energy $AdS$ product region (or in an
exactly conformal theory, such as discussed above Eq.~(\ref{BFKLamp}) in the
introduction.)  In
terms of the coordinate $u = \ln(r/r_0)$, which is $\gg 1$ in the high
energy region, the metric is
\be
ds^2 = \frac{r_0^2}{R^2} e^{2u} \eta_{\mu\nu} dx^\mu dx^\nu + R^2 du^2 + ds^2_W\ .
\ee
The Laplacian is
\begin{eqnarray}
\alpha' \Delta_2 &=& \frac{1}{\sqrt{\lambda}} 
( \partial_u^2  - 4 ) + \alpha'_0 t e^{-2u} + \alpha' \nabla_W^2 \nonumber\\
&=& \frac{1}{\sqrt\lambda} \left[ \partial_u^2  - 4 + z_0^2 t e^{-2u} \right]
+ \alpha' \nabla_W^2 \ , \label{alDel2}
\end{eqnarray}
where $\alpha'_0 = \alpha'_{\rm eff}(r_0) = R^2 \alpha' / r_0^2$ is the
infrared slope~(\ref{alf0}), and $1/z_0 = r_0/R^2$ is the mass scale of the
lightest glueballs (Kaluza-Klein excitations).  Note that
$1/\sqrt{\alpha'_0} = {\lambda}^{1/4}/z_0$.

For simplicity we assume that the wavefunctions are independent of
$W$, and consider first the case $t=0$.  Then Eq.~(\ref{ramp2})
becomes
\begin{equation}
{\cal I}m\;  {\cal T}_4 = {\rm const} \times  s^2 \int du\ \psi_3(u)\psi_4(u)  /\sqrt{\lambda})
\!\left[ \alpha'_{\rm eff}(r) s \right]^{(\partial_u^2 -
4)/2\sqrt{\lambda}} \psi_1(u)\psi_2(u) \ . \label{t4diff}
\end{equation}
We have
canceled a factor of $r^4$ from $\sqrt{-G}$ against a factor of
$r^{-4}$ from the Regge amplitude.  In doing this we have assumed that
the latter factor acts to the left of the diffusion operator, rather
than being symmetrized with it (the difference is of the same order as
the effects that we are retaining).  This follows automatically from
the more systematic analysis of the next section, but in fact one can
already deduce it from the symmetry of the diffusion kernel, which
would not hold with any other ordering.

The amplitude~(\ref{t4diff}) will indeed be dominated by large $u$
provided that one or more wavefunctions are strongly peaked at large
$u$ (as is the case for quarkonium states or external operators of
large momentum), and also provided that $s$ is not too large (else the
diffusion will reach the confinement scale).  The latter effect will
be considered in more detail in Sec.~5.

The $s$-dependence now takes the form
\begin{equation}
s^{2 + (\partial_u^2 - 4)/2\sqrt{\lambda}}\ .
\end{equation}
This gives diffusion in $u$ over a time
\begin{equation}
\tau = \frac{1}{2\sqrt\lambda} \ln 
\Bigl(\left[\alpha'_{\rm eff}(\bar r)\right]s\Bigr)\ .
\label{tau}
\end{equation}
where $\bar r$ is an averaged value of $r$ appropriate to
the particular physical process.\footnote{While the precise choice
of $\bar r$ is a subleading effect, confusion might arise were it
not addressed here.  In particular, $\bar r$ is not, in general,
$r_0$.  In this and the earlier
equations we have retained
$\alpha'_{\rm eff}(r)$ to a small power.  This is necessary for good
form with units, but there is also some physics in it.  We could
imagine that at the values of $r$ relevant to the physics,
$\ln( \bar r/r_0)$ may be large enough that we would wish to retain
effects of order $\ln(\bar r/r_0)/\sqrt{\lambda}$; 
keeping $\alpha'_{\rm eff}(\bar r)$ does this.  If $\psi_1\psi_2$ and
$\psi_3\psi_4$ are peaked at a common scale $\bar r$, we can replace
$\alpha'_{\rm eff}(r)$ with $\alpha'_{\rm eff}(\bar r)$; any correction
is higher order in $1/\sqrt{\lambda}$.  If they are peaked at two
different scales $r_1$ and $r_3$ then we can replace it with the
geometric mean of $\alpha'_{\rm eff}(r_1)$ and $\alpha'_{\rm
eff}(r_3)$.  Thus the diffusion time appearing in $e^{\tau
\partial_u^2}$ is, more generally, to be taken as
\begin{eqnarray}
\tau = \frac{1}{2\sqrt\lambda} \ln \Bigl({s}{\sqrt{\alpha'_{\rm eff}(r_1) \alpha'_{\rm eff}(r_3)}}\Bigr)\ .
\nonumber
\end{eqnarray}
}

Recall that a diffusion operator resembles a Schr\"odinger operator in
imaginary time, and can be similarly treated.  We need to solve
\begin{equation}\label{SchrodingerEq}
H\Psi_E(u)=[-\partial_u^2 + V(u)]\Psi_E(u) = E\Psi_E(u)
\end{equation}
with $V(u)=4$.  This has the same eigenstates as a
free particle; its delta-function normalizable eigenstates are
$e^{i\nu u}$ for $-\infty < \nu < \infty$, with eigenvalues $E= 4+
\nu^2$.  That is, this operator has a continuum of states with a
continuum of energies --- a cut in the $E$ plane starting at $4$.  The
operator $2 + (\partial_u^2 - 4)/2\sqrt{\lambda}$ therefore has
eigenvalues
\bel{jeq}
j = j_0- {\cal D}\nu^2
\ee
with
\bel{j0eq}
j_0= 2 - {2\over \sqrt\lambda} \ , \quad {\cal D}={1\over 2\sqrt\lambda} \ .
\ee
Thus there is a cut in the $j$ plane ending at $j_0$.  This is the
same behavior as found by BFKL in perturbative contexts, and we can
identify $j_0$ as the strong coupling limit of the BFKL exponent.
This exponent has also been derived recently in ref.~\cite{klv5}; we
will discuss this work more fully in Sec.~3.3.

The high-$s$ behavior of amplitudes is roughly of
the form $s^{j_0}$, but this would be strictly true only
if the leading eigenvalue of $\Delta_2$ were discrete, so that the
leading singularity in the $j$ plane would be a pole rather
than a cut.  The precise form is
\begin{equation}
{\cal I}m\; {\cal A}(s,t=0) \propto  \int du\,du'\,
\psi_3(u)\psi_4(u) {\cal K}(u,u',\tau) \psi_1(u') \psi_2(u')
\end{equation}
where
\be
{\cal K}(u,u',\tau) = s^{j_0} {\cal K}_0(u,u',\tau)
\ee
and ${\cal K}_0(u,u',\tau)$ is the diffusion (heat) kernel
\beq
\label{Kzero}
{\cal K}_0(u,u',\tau) &=& e^{\tau\partial_u^2} \delta(u-u')
%\nonumber\\
%&=&
=\int_{-\infty}^\infty \frac{d\nu}{2\pi} \, e^{i\nu(u-u')} 
e^{- \nu^2\tau}  \nonumber\\
&=&
\frac{1}{2\sqrt{\pi\tau} } e^{-(u-u')^2/4\tau}\ ,
\eeq
where again $\tau$ is defined in \eref{tau}.
This result is of the form \eref{quadkernel2}, and, as discussed
in the introduction, is similar in form
to that of \cite{Kuraev:1977fs,BL}.  
Note that it is consistent
with the ultraviolet conformal symmetry of the field theory: scale
invariance (translation of the $u$ coordinate) 
and inversion symmetry (reflection of the $u$ coordinate)
require dependence only on $(u-u')^2$.

Let us now consider the kernel at nonzero $t<0$, working in the regime
$u,u'$ large and positive ($r\gg r_0$), and with $|t|\gg \Lambda^2$.
In this regime the kernel is determined only by the
conformally-invariant region of the gauge theory, and is independent
of confinement effects, which we will treat later.  This problem was
first solved in the BFKL context in position space \cite{finitetBFKL}.
The result we obtain here bears some resemblance to the form
anticipated in \cite{levintan} and recently reconsidered in
\cite{Bondarenko:2003xb}.  We obtain it directly in momentum space.

From \eref{ramp2} and \eref{tau}, the exponential operator appearing in
the kernel at $t\leq 0$ is (at leading order in $\ln s$)
of the form $e^{-H\tau}$, where
$H$ is the Hamiltonian for Liouville quantum mechanics
\bel{Hdefn}
-\sqrt{\lambda}\alpha' \Delta_2 
\equiv H =  -\partial_u^2  + V(u)
\ee
\bel{potential}
V(u) = 4 - {z_0^2 t} e^{-2u} .
\ee
Since $t<0$ here, this is an exponentially growing potential as $u$
decreases.  
If $z_0^2 |t| = |t|/\Lambda^2 \gg 1$ the particle is
repelled from the small-$u$ region, so confinement effects are highly
suppressed and the calculation becomes identical to the
computation in a conformally invariant theory.  
The eigenvalue
spectrum for $H$ determines the location of the singularities in the
$j$-plane, where $j = 2 - E/2\sqrt{\lambda}$.  There is a continuum
beginning at $E=4$, with eigenvalues $E=4 + \nu^2$, and corresponding
eigenfunctions,
\begin{equation}
\psi(\nu,u)= \sqrt{2/\pi}\
K_{i\nu}(z_0\sqrt{|t|}e^{-u})/\Gamma(i\nu) \ ,\quad \nu>0\ .
\end{equation}
(For
$t=0$, this spectrum reduces to the free particle momentum states
$e^{i\nu u}$.)  Thus, independent of $t$, 
there is a cut in the $j$ plane
beginning at
$E=4$, that is, 
$j = j_0=2-2/\sqrt{\lambda}$.  The kernel itself,
however, has nontrivial $t$-dependence, since the exponential
potential suppresses the eigenfunctions in regions where $u\ll u_t \equiv
\ln ( z_0 \sqrt{|t|})$.  The kernel can be written \cite{D'Hoker:1982er}
\bel{Knonzerot}
{\cal K}_0(u,u',\tau, t) = {2\over \pi^2}\int_0^\infty\ d\nu\,
\nu\, \sinh \pi\nu \
K_{i\nu} (z_0 |t|^{1/2}e^{-u}) K_{-i\nu} (z_0 |t|^{1/2}e^{-u'}) 
e^{-\nu^2 \tau}\ .
\ee
This kernel is exponentially suppressed if either $u$ or $u'$ is much
less than $u_t$.  

For $u,u'\gg u_t$, the behavior of the kernel at
small $\tau$ resembles the $t=0$ kernel, while for $\tau \gg uu'$ the
kernel falls\footnote{More
precisely, at fixed $\tau$ and large $u,u'\gg u_t$, perturbation
theory around the $t=0$ case yields
\begin{eqnarray}
{\cal K}_0(u,u',\tau,t) = \frac{ 1 }{2\sqrt {\pi \tau}}
{e^{-\frac{(u-u')^2}{4\tau}}} e^{ \tau (z_0^2 t) \frac{(
e^{-2u}-e^{-2u'} )}{2(u-u')} }
% \; \theta(\tau)
\nonumber
\end{eqnarray}
At large $\tau$ (for fixed $u,u'\gg u_t$) the kernel is dominated by the
lowest mode,  so 
for $\sqrt{\tau}\gg u,u'\gg u_t$
\begin{eqnarray}
%\nonumber
 {\cal K}_0(u,u',\tau,t)\simeq   \frac{ 1}{2 \sqrt {\pi\tau^3}}  \; 
K_{0}(z_0 \sqrt{-t} e^{-u} )  K_{0} (z_0\sqrt{-t} e^{-u'}) %\; \theta(\tau)
\ . \nonumber
\end{eqnarray}
}  
to zero as $\tau^{-3/2}$. 
This is due to the reflection of the diffusion off the exponentially
rising potential.  Indeed, the kernel is moderately well-approximated
by the $t=0$ kernel combined with reflection off a wall at $u=u_t$
with a Dirichlet boundary condition.
To see this, it is useful to write $
K_{i\nu}(x)=(i\pi/2)[I_{i\nu}(x)-I_{-i\nu}(x)]/\sinh\pi\nu$, so that
incoming (``L'', for left-moving) and reflected waves (``R''-moving)
are more explicit.  The kernel now is a sum of $LL$, $LR$, $RL$ and
$RR$ subkernels. Interference between the $LL, RR$ terms and the
$LR,RL$ terms removes the $1/\nu$ pole as $\nu\to0$, so the integrand
goes as $\nu^2$ near $\nu=0$ and gives $\tau^{-3/2}$ behavior.
Meanwhile, the $t=0$ kernel is obtained by going to large positive
$u$, so that the barrier at $u=0$ is infinitely distant and the
$I_{i\nu}$ become plane waves; the $LR$ and $RL$ interference
terms are exponentially suppressed after the $\nu$ integration, while
the $LL$ and $RR$ terms can be seen as the $\nu>0$ and $\nu<0$ parts
of the integral in \eref{Kzero}.  The $1/\nu$ pole multiplying the
$I_{i\nu}(x)$ is no longer cancelled, and the standard $\tau^{-1/2}$
behavior of a diffusion kernel is obtained.

This $\tau^{-3/2}$ feature is very general and survives even when
conformal symmetry breaking is introduced in the infrared, as we will
see later.  It corresponds to having a softened branch point in the
$j$ plane compared to the $t=0$ conformal kernel.  More precisely, the
``partial-wave'' amplitude, obtained by a Mellin transform of ${\cal
K}(s)$, has a vanishing square-root type singularity,
\begin{equation}
\tilde {\cal K} (j, t, u, u')\sim \sqrt {j-j_0} + {\rm regular}.
\end{equation}
By contrast, at $t=0$, 
where as we just saw the $\nu^2$ factor is cancelled 
giving a non-vanishing
contribution at $\nu=0$, there is a divergent singularity
in $j$,
\begin{equation}
\tilde {\cal K} (j, t, u, u')\sim \frac{1}{\sqrt {j-j_0}} + {\rm regular}.
\end{equation}
Correspondingly, the large $\tau$ behavior is $s^{j_0}/\tau^{1/2}$ at
$t=0$.  Thus, the branch cut in the $j$-plane begins at a
$t$-independent point $j_0$, but the nature of the cut differs at
negative $t$ from its form at $t=0$.

The same physics --- the reflection of the diffusion off an effective
barrier at $e^{2u}\sim k_\perp^2\sim |t|$ --- leads to identical
powers of $\tau$ at weak coupling. The barrier is absent at $t=0$, so
the eigenmodes are just plane waves in $\ln k_\perp$, giving the
$\tau^{-1/2}$ behavior of \Eref{quadkernel}.  For $t<0$, however, the
eigenmodes are a combination of incoming and reflected modes, with the
leading $\tau^{-1/2}$ behavior removed by destructive interference
between them.

To see this, consider the standard weak-coupling BFKL calculation at
finite $t$ \cite{finitetBFKL,KirschLipat}.  The amplitude is expressed
in terms of a $t<0$ kernel and impact factors $\Psi_A,
\Psi_B$,
\begin{equation}
\label{weakkernel1}
\frac{{\cal I}m\; A(s,t)}{s} 
=
\frac {\cal G}{(2\pi)^4} \int d^2{p_\perp} 
d^2 p'_\perp \; \Psi_A( p_\perp,k_\perp) \; 
{\cal K} (s,  k_\perp,  p_\perp,  p'_\perp)\; \Psi_B( p'_\perp, k_\perp) 
\end{equation}
where
\begin{equation}
{\cal K} (s,  k_\perp,  p_\perp,  p'_\perp)
=  
\frac{1}{(2\pi)^6} \int d\nu 
\left[\frac {\nu}{\nu^2 + 1/4}\right]^2
e^{\alpha_s \chi(\nu) \tau} 
\psi_\nu(p_\perp,  k_\perp)\; \psi^*_\nu( p'_\perp, k_\perp) \ .
\label{weakkernel2}
\end{equation}
Here $\psi_\nu$ is a conformal wave function,
an eigenfunction of the homogeneous BFKL 
equation.\footnote{Our notation corresponds to a BFKL kernel that, as a
four-point function, carries external transverse momenta $p_\perp\pm
k_\perp/2$ and $p'_\perp\pm k_\perp/2$, with $t=-k_\perp^2$; it has a
dimension of $[{\it length}]^6$. We have retained only the leading order
term, i.e., the so-called $n=0$ term, for the BFKL kernel, with $\chi
$ in Eq.~(\ref{weakkernel2}) related to $\hat j$ of \eref{BFKLj} by $\hat j
= 1+ \alpha_s\chi$. To match the formulas in the BFKL literature,
e.g., \cite{FR}, we have normalized the dummy variable $\nu$ in a way
{\it which differs from that used in the rest of the paper} by a
factor of 2, as noted earlier. We find that the conformal wave
function can be expressed, defining the functions $a_1^2\equiv
{x(1-x)}(k^2_\perp/p_\perp^2)$, and $a_2^2\equiv | p_\perp + (x-\half)
k_\perp|^2/p_\perp^2$, as
\begin{eqnarray}
\psi_\nu(p_\perp, k_\perp) &=&\frac{(2\pi) (|k_\perp|/2)^{2i\nu}}{\Gamma^2(\half
+ i\nu)} \int_0^1 \frac {dx}{\sqrt{x(1-x)}} \int_0^{\infty} dr \; r^2
J_0( a_2 |p_\perp| r) K_{-2i\nu}(a_1 |p_\perp| r ) \nonumber \cr 
&=& \frac{(2 \pi)^2
(\nu^2 +1/4) (k_\perp/2)^{2i\nu}}{\cosh \pi\nu\ \Gamma^2(\half + i\nu)}
\int_0^1 \frac {dx}{p_\perp^3[x(1-x)]^2} 
F(-i\nu+3/2, i\nu+3/2; 1; -a_2^2/a_1^2)
\, .\nonumber
\end{eqnarray}
}

Care must be taken in comparing the weak- and strong-coupling
representations since the effective degrees of freedom used are
different. However, since conformal invariance plays an essential role
here, it is not surprising that they share many qualitative and some
quantitative similarities.  In particular, the Bessel function in
$\psi_\nu(p_\perp,k_\perp)$, while not identical in form to the one
appearing in \eref{Knonzerot}, shares the feature that it can be
written as an incoming and an outgoing wave, and that the $t=0$ kernel
is obtained by effectively dropping the interference
between them, resulting in a change in the
$\nu$-dependence at small $\nu$.  Thus the analytic structure in the
$j$-plane, and the $\tau$ dependence, of the weak- and strong-coupling
kernels agree.

We have not attempted 
to obtain more precise connections with
weak coupling calculations --- as is clear from the above formulas,
the weak-coupling kernel at finite $t$ is structurally more
complicated than the $t=0$ case --- but we believe it should be
possible to do so.

\section{A systematic derivation}
\label{sec:systematic}

The result in the previous section is simple and intuitive, but it is
useful to present a more systematic derivation.  For one thing
we have begun with an expression~(\ref{amp}) in which the scattering
is local in the bulk, and then (when $s$ is taken large with
$\lambda$) we have found a diffusive effect that makes the scattering
arbitrarily nonlocal in the bulk.  As a result we have had to guess
about such things as operator ordering.  For another, we are retaining all
orders in $\ln s /\sqrt{\lambda}$.  From the point of view of the
world-sheet theory this is a resummation of perturbation theory, and
we would like to determine its exact nature.  We will see that it is
something familiar: that the large $s$ amplitudes will involve
world-sheet distances of order $1/s$, so this is simply a
renormalization group improvement in the world-sheet field theory.

\subsection{Regge behavior in flat spacetime}

We first analyze Regge scattering in flat spacetime in a rather
general way which may have other applications.  For any process in
which the external particles (which may include D-branes as well as
strings) can be divided into two sets that are at large relative
boost, we derive the leading behavior of the amplitude.

\subsubsection{Example: Bosonic string tachyons}

Consider the Virasoro-Shapiro amplitude, for bosonic string tachyon scattering:\footnote{In order to keep equations uncluttered, we adopt simple overall normalizations in Eqs.~(\ref{vsamp}), (\ref{regvert}), but keep all later equations normalized with respect to these.  For the same reason we omit the momentum delta functions in the translationally invariant directions.}
\begin{equation}
{\cal A} =  \int d^2w\, |w|^{-4 - \apm t/2} |1 - w|^{-4-\apm s/2} \ .\label{vsamp}
\end{equation}
When $s$ and $t$ are both large, the integrand has a saddle point at $w = t/(s+t)$.  When the integral is appropriately defined by analytic continuation, this point indeed dominates~\cite{GrossMende}.  If this were to continue to apply for large $s$ at fixed $t$, then $w$ of order $s^{-1}$ would dominate in the Regge regime.  This is true but the integral can no longer be evaluated by the saddle point method, rather we must integrate it explicitly in the small-$w$ region:
\begin{eqnarray}
{\cal A} &\sim&  \int d^2w\, |w|^{-4-\apm t/2} e^{\apm s(w + \bar w)/4}\nonumber\\
&=& 2 \pi \frac{\Gamma(-1-\alpha't/4)}{\Gamma(2+\alpha't/4)} (e^{-i\pi/2} \apm s/4)^{2 + \apm t/2}\ . \label{regamp}
\end{eqnarray}
The integral is defined by continuation from $-4 > \apm t > -8$ and positive imaginary $s$.

Since $w$ is small we should be able to reproduce the above Regge behavior via the OPE,
\begin{equation}
e^{i p_1 \cdot X(w,\bar w)}\, e^{i p_2 \cdot X(0)}
\sim |w|^{-4 - \apm t/2} e^{ik \cdot X(0)}\ , \quad k = p_1 + p_2\ .
\end{equation}
However, this reproduces only the first term in the integrand~(\ref{vsamp}): it gives the tachyon pole but not the Regge behavior.  The point is that because $sw$ is of order one we must retain additional terms in the OPE, which are normally subleading.  The result is still quite simple:
\begin{equation}
e^{i p_1 \cdot X(w,\bar w)}\, e^{i p_2 \cdot X(0)}
\stackrel{\rm Regge}{\sim} |w|^{-4 - \apm t/2} e^{ik \cdot X(0)
+ i p_1 \cdot (w\partial + \bar w \bar\partial) X(0)}\ . \label{rope}
\end{equation}
Contractions involving $p_1 \cdot (w\partial + \bar w \bar\partial) X(0)$ will generate a factor of $s$ for each factor of $w$.  We insert the OPE into the vertex operator amplitude,
\begin{eqnarray}
\langle e^{i p_1\cdot X(w,\bar w)}  e^{i p_2\cdot X(0)}  e^{i p_3\cdot X(1)}  e^{i p_4\cdot X(\infty)}
 \rangle 
\hskip 3.3 in & \nonumber \\ 
\sim |w|^{-4 - \apm t/2}
\left
\langle  e^{ik \cdot X(0)
+ i p_1 \cdot (w\partial + \bar w \bar\partial) X(0)}\,
e^{i  p_3 \cdot X(1)}\, e^{i p_4 \cdot X(\infty)}
\right
\rangle\ . \nonumber \\
\label{4vert}
\end{eqnarray}
Evaluating the contractions reproduces the integrand in eq.~(\ref{regamp}).
It is also interesting to instead carry out the $w$ integral first, at the operator level:
\begin{equation}
\int d^2w\, e^{i p_1 \cdot X(w,\bar w)}\, e^{i p_2 \cdot X(0)}
\stackrel{\rm Regge}{\sim} \Pi(\alpha' t)
e^{ik \cdot X(0)}
\!\left[ p_{1} {\cdot} \partial X \, p_{1} {\cdot} \bar\partial X(0)\right]^{1 + \apm t/4}
 \ ,  \label{regvert}
\end{equation}
where we have defined
\be
\Pi(\alpha' t) = 2\pi \frac{\Gamma(-1-\alpha't/4)}{\Gamma(2+\alpha't/4)}  e^{-i\pi -i\pi \apm t/4}\ ,
\ee
with $t=-k^2$. Inserting this matrix element into the expectation value~(\ref{4vert})  immediately gives the Regge amplitude~(\ref{regamp}).

The result~(\ref{regvert}) displays the essential idea that we will need for analyzing Regge behavior in curved spacetime as well.
We can think of the operator on the right as a Pomeron vertex operator. For $t=-4/\apm$ it is the tachyon, for $t=0$ it is the graviton, and so on.  Note that it is always on shell in the sense of satisfying the physical state conditions,\footnote{To be precise, we should replace $2 p_1$ with $p_1 - p_2  = p_{12}$; these are equivalent in the Regge limit but $p_{12}$ is exactly orthogonal to $k$.}
 even when $t$ is not the mass-squared of a physical particle, but it is outside the normal Hilbert space because of the fractional power.
In spite of the importance of Regge physics in the history of string theory, we are unaware of any previous introduction of such a vertex operator.

\subsubsection{Generalizations}

The result~(\ref{regvert}), derived for four tachyons, can be broadly generalized.  First, let us note that the OPE~(\ref{rope}) is essentially the same for any pair of vertex operators,
\begin{equation}
{\cal V}_1(w,\bar w) \, {\cal V}_2(0)
\sim C_{12} |w|^{-4 - \apm t/2} e^{ik \cdot X(0)
+ i p_1 \cdot (w\partial + \bar w \bar\partial) X(0)}\ . \label{rope2}
\end{equation}
The tensor terms in the vertex operators contract to give a constant $C_{12}$ times a power of $|w|$, and then the rest is as for the tachyons; the final power of $|w|$ depends only on the momentum transfer.  We can again integrate this directly in operator form,
\be
\int d^2w\, {\cal V}_1(w,\bar w) \, {\cal V}_2(0)
\sim  C_{12} \,\Pi(\alpha' t) e^{ik \cdot X(0)}
\!\left[p_{1} {\cdot} \partial X \, p_{1} {\cdot} \bar\partial X(0)\right]^{1 + \apm t/4}
\ . \label{rope3}
\ee
The constant $C_{12}$ can be interpreted as the coupling of the Pomeron to states 1 and 2; we will express this coupling in another form in eq.~(\ref{pomfact}) below.

Having the relation~(\ref{rope3}) in operator form immediately allows a broad generalization: we can replace the tachyon vertex operators~${\cal V}_{3}, {\cal V}_{4}$ with any number $l \geq 2$ of vertex operators, for any collection of string states.  The M\"obius group fixes ${\cal V}_2$ plus two of the $l$ additonal vertex operators.  Let ${\cal W}$ denote the product of the $l$ vertex operators and their $l-2$ position integrations.  Then the standard string amplitude is
\be
{\cal A}_{12\cal W} \sim C_{12}\, \Pi(\alpha' t) \left\langle e^{ik \cdot X(0)}
\!\left[p_{1} {\cdot} \partial X \, p_{1} {\cdot} \bar\partial X(0)\right]^{1 + \apm t/4}
{\cal W}  \right\rangle \ .  \label{regvert2}
\ee
This captures the asymptotic behavior as the vertex operators in ${\cal W}$ are boosted to some large rapidity $y$, relative to ${\cal V}_{1}$ and $ {\cal V}_{2}$.  Then
\be
e^{y} = s/s_0 \ ,  \quad y =  2 \sqrt{\lambda}\, \tau + \mbox{const}\ ,
\ee
where $s_0$ is the center of mass energy-squared when the relative boost is zero and $\tau$ is the diffusion time introduced earlier.  Contractions of
$\partial X(0)$ or $\bar\partial X(0)$ in the Pomeron vertex operator with fields in $\cal W$ are of
order $e^{y}$, producing a factor of $e^{(2 + \alpha' t/2)y}$.  We will
refer to ${\cal V}_{1, 2}$ as right-moving and $\cal W$ as left-moving along
the direction of the boost.

The results so far are still special, in that we have an arbitrary
 left-moving process but on the right-moving side there is just 1-to-1
 scattering.  To derive the Regge amplitude in full generality, let ${\cal
   W}_R$ and ${\cal W}_L$ denote arbitrary sets of $l_R$ and $l_L$ vertex operators together with their
 associated $l_R - 2$ and $l_L-2$ world-sheet integrations.  This leaves a single 
 integration over world-sheet coordinates, corresponding to an overall scaling of the coordinates in ${\cal W}_L$.  The string amplitude is then
\bel{xlxr}
 A_{{\cal W}_L {\cal W}_R} = \int d^2w \left\langle {\cal W}_R w^{L_0 - 2}
   \bar w^{\tilde L_0 - 2} {\cal W}_L \right\rangle\ , \ee
 where $L_0$ and $\tilde L_0$ are the right- and left-moving Virasoro
 operators which generate the world-sheet scale transformations,
\bel{vira}
L_0 = \frac{\alpha'}{4} k^2 + N\ ,\quad
\tilde L_0 = \frac{\alpha'}{4} k^2 + \tilde N\ ,
\ee
and $N$ and $\tilde N$ are the right- and left-moving excitation levels.
This is a standard way of organizing string amplitudes for purposes of discussing unitarity: the integral over the region $|w|<1$ produces the closed string propagator  in the $t$-channel,
\be
\frac{\delta_{L_0-\tilde L_0}}{L_0 + \tilde L_0 - 2}\ .
\ee
Note also that the OPE~(\ref{rope}) can similarly be written
\begin{eqnarray}
e^{i p_1 \cdot X(w,\bar w)}\, e^{i p_2 \cdot X(0)}
&\stackrel{\rm Regge}{\sim}& |w|^{-4 - \apm t/2} e^{ik \cdot X(0)
+ i p_1 \cdot (w\partial + \bar w \bar\partial) X(0)}\nonumber\\
&=& w^{L_0 - 2} \bar w^{\tilde L_0 - 2}  e^{ik \cdot X(0)
+ i p_1 \cdot (\partial + \bar\partial) X(0)}\ ; \label{rope4}
\end{eqnarray}
for the bosonic string $N$ and $\tilde N$ just count the number of $\partial$ and $\bar \partial$ in the vertex operator.

We wish to study the amplitude in
 the limit that the two sets differ by a large boost in the $\pm$ plane, so that the
 momenta in ${\cal W}_R$ have large $+$ components and the momenta in ${\cal
   W}_L$ have large $-$ components; the exchanged momentum $k$ is orthogonal to the $\pm$ plane.  The generalization of the OPE is to
 insert a complete set of string states in the matrix element~(\ref{xlxr})
 but again to retain only those that survive in the Regge limit $sw \sim 1$:
\beq A_{{\cal W}_L {\cal
     W}_R} &\sim& \sum_{m,n=0}^\infty \int d^2w\, \frac{2^{m+n} w^{m - 2 -
     \alpha' t/4} \bar w^{n - 2 - \alpha' t / 4} }
 {\alpha'^{m+n} m! n! } \nonumber\\
 &&\qquad
\left\langle {\cal W}_R\, e^{i k \cdot X} (\partial X^-)^m
   (\bar\partial X^-)^n \right\rangle \left\langle e^{-i k \cdot X} (\partial
   X^+)^m (\bar\partial X^+)^n {\cal W}_L \right\rangle\ .  \eeq
Note that $\partial X, \bar\partial X$ are the vertex operator factors for the string excitations $\alpha_{-1}, \tilde\alpha_{-1}$.
After
 inserting the states, we have explicitly evaluated $w^{L_0 - 2} \bar
 w^{\tilde L_0 - 2}$ acting on the intermediate vertex operator.  The next
 step is typographically tricky.  We note that all the $m$-dependent terms
 combine to form an exponential, and similarly all the $n$-dependent terms:
 \bel{expo} \exp\!\left( 2 w \partial X^-_R \partial X^+_L/\alpha' + 2 \bar w
   \bar\partial X^-_R \bar\partial X^+_L/\alpha' \right)\ , \ee
where the
subscripts indicate whether the operator appears in the expectation value
 with ${\cal W}_R$ or the one with ${\cal W}_L$.  Acting now with $\int d^2
 w\, |w|^{-4 - \alpha' t / 2}$ (where the integral is defined by
 continuation as before), the exponential~(\ref{expo}) becomes
\be
 \Pi(\alpha' t)(2 \partial X^-_R \bar\partial X^-_R/\alpha')^{1 + \alpha'
   t/4} (2 \partial X^+_L \bar\partial X^+_L/\alpha')^{1 + \alpha' t/4} \ee
Inserting this back in the amplitude gives
\begin{eqnarray}\label{pomfact} A_{{\cal W}_L {\cal
     W}_R} \sim  
\Pi(\alpha' t) \times \hskip 4.5 in &\nonumber \\
\left\langle {\cal W}_R\, e^{i k \cdot X} (2
   \partial X^- \bar\partial X^-/\alpha')^{1 + \alpha' t/4} \right\rangle
 \left\langle e^{-i k \cdot X}(2 \partial X^+ \bar\partial X^+/\alpha')^{1 +
     \alpha' t/4} {\cal W}_L \right\rangle\ .  \nonumber \\
\end{eqnarray}
We could have guessed this by symmetrizing the earlier
result~(\ref{regvert2}).  We can also make the Regge behavior more explicit
by boosting the states ${\cal W}_L$ and ${\cal W}_R$ back to their
approximate rest frames (denoted by a subscript 0), so that the large boost
$e^y = s/s_0$ enters explicitly:
\be A_{{\cal W}_L
   {\cal W}_R} \sim \Pi(\alpha' t) (s/s_0)^{2 + \alpha' t/2} \left\langle
   {\cal W}_{R0}\, {\cal V}_P^- \right\rangle \left\langle {\cal V}_P^+ \,
   {\cal W}_{L0} \right\rangle \; . 
\label{pomgen} \ee
The result~(\ref{pomgen}) has a simple interpretation as a Pomeron propagator,
of the form
$\Pi(\alpha' t) (s/s_0)^{2 + \alpha' t/2}$, times the couplings of the Pomeron
to the two sets of vertex operators, with Pomeron vertex operator
\bel{pomvertex}
  {\cal V}_P^{\pm} = (2 \partial X^{\pm} \bar\partial
X^{\pm}/\alpha')^{1 + \alpha' t/4}\; e^{\mp i k \cdot X} \; .
\ee

Note that this formalism works equally well for Regge scattering of strings
and D-branes or of D-branes and D-branes.  For coupling to a D-brane one
simply replaces the vertex operators in ${\cal W}$ with a world-sheet hole
with appropriate boundary conditions, and the factorization analysis goes
through unchanged.  Thus, scattering processes involving ultrarelativistic
D-branes will also display Regge behavior~\cite{Bachas}. 

This analysis extends readily to the superstring.  Let us start for
simplicity with the OPE of two type II tachyons; these have the wrong GSO
projection but their product is then GSO-allowed.  We work in the 0 picture
because this is most closely analogous to the bosonic string and to other
formulations of the superstring.  Then
\beq
&& \int d^2w\, d^2\theta\, d^2\theta'\,e^{i p_1\cdot \mbox{\boldmath\scriptsize $X$}(w,\theta)}\, e^{i p_2\cdot \mbox{\boldmath\scriptsize $X$}(0,\theta')} \nonumber\\
&&\qquad \sim \int d^2w\, d^2\theta\, d^2\theta'\, |w - \theta\theta'|^{-2 -
  \alpha' t/2}
e^{i k\cdot \mbox{\boldmath\scriptsize $X$}(0,\theta') + i p_1\cdot (w\partial_w + [\theta-\theta'] \partial_{\theta'}) \mbox{\boldmath\scriptsize $X$}(0,\theta')} \nonumber\\
&&\qquad= \hat\Pi(\alpha' t) \int d^2\theta'\, e^{i k {\cdot}
  \mbox{\boldmath\scriptsize $X$}} p_{1} {\cdot} D_{\theta'} \mbox{\boldmath
  $X$} \, p_{1} {\cdot} D_{\bar\theta'} \mbox{\boldmath $X$} (p_{1} {\cdot}
\partial \mbox{\boldmath $X$} p_{1} {\cdot} \partial \mbox{\boldmath $X$}
)^{\alpha' t/4} |_{0,\theta'}
\nonumber\\
&&\qquad\equiv \hat\Pi(\alpha' t) \hat{\cal V}_P\ .  \eeq
The Pomeron vertex operator has the same bosonic part as in the bosonic
string, together with fermionic terms as required by world-sheet
supersymmetry.  The Pomeron propagator no longer has a tachyon pole:
\be
\hat\Pi(\alpha' t) = 2\pi \frac{\Gamma(-\alpha't/4)}{\Gamma(1+\alpha't/4)} e^{ -i\pi \apm t/4}\ .
\ee
This result, derived for the simplest vertex operators, can then be generalized broadly as in the bosonic case:
\be
{\cal A}_{{\cal W}_L {\cal W}_R} \sim
\hat\Pi(\alpha' t) (s/s_0)^{2 + \alpha' t/2} \left\langle
{\cal W}_{R0}\, \hat{\cal V}_P^- \right\rangle \left\langle
\hat{\cal V}_P^+ \,
{\cal W}_{L0} \right\rangle\ .
\ee

\subsection{Regge behavior in warped spacetime}
\label{sec:reggewarpedspacetime}
Now let us try to repeat these steps in a warped metric
\be
ds^2 = e^{2A(y)}\eta_{\mu\nu} dX^\mu dX^\nu + ds_\perp^2\ .
\ee
We again start with the bosonic string to avoid tensor
complications.\footnote{We do not know of any $AdS_5$ solutions for
the bosonic string, so this is slightly formal.  However, there is an
$AdS_3 \times S^3 \times T^{20}$ solution, to which the analysis at
$t=0$ applies.  This solution might be interesting to explore further
because the world-sheet CFT is exact.}  
In the Gaussian limit the string wavefunctions~(\ref{cavity}) translate directly into vertex operators
$e^{ip\cdot X} \psi(Y)$, where the capital letters denote world-sheet fields.  We start with the OPE
\beq
&&e^{i p_1\cdot X(w,\bar w)}\psi_1(Y(w,\bar w))\,
e^{i p_2\cdot X(0)}\psi_2(Y(0)) \nonumber\\
&&\qquad\qquad\qquad \stackrel{\rm Regge}{\sim} w^{L_0 - 2} \bar w^{\tilde L_0 - 2} e^{ik \cdot X(0)
+ i p_1 \cdot (\partial + \bar\partial) X(0)} \psi_1(Y(0)) \psi_2(Y(0)) \ . \label{wrope}
\eeq
This is identical in form to the flat spacetime OPE~(\ref{rope4}): we naively multiply the vertex operators, keeping only those terms that survive in the Regge limit, and the $w$-dependence arises from the Virasoro generators.  Now, however, we need to go beyond the Gaussian approximation to include terms of order $(\ln s)/\sqrt{\lambda} \sim |\ln w|/\sqrt{\lambda}$.  In the form~(\ref{wrope}) it is clear how this is done: we must keep terms of order
$1/\sqrt{\lambda}$ in the world-sheet dimension $L_0$.  This is
precisely the renormalization group improvement referred to at the
beginning of the section.

In order to diagonalize $L_0$ we must go to a basis of definite spin,
\begin{equation}
{\cal V}(j) = (\partial X^+ \bar\partial X^+)^{j/2} e^{ik \cdot X} \phi_{+^{j}}(Y)\ ,
\label{vofj}
\end{equation}
working again in the frame where the large momentum is $p_+$.  The
operator on the right of the OPE~(\ref{wrope}) can be expanded in such
a basis.  The one-loop world-sheet
dimension of ${\cal V}(j)$ (which includes all
terms up to second derivatives in space-time) must be of the form
\begin{equation}
L_0 {\cal V}(j) = \tilde L_0 {\cal V}(j) = (\partial X^+ \bar\partial X^+)^{j/2}
\Biggl[ \frac{j}{2} - \frac{\alpha'}{4} (\Delta_j + \delta_j) \Biggr] e^{ik \cdot X} \phi_{+^{j}}(Y)\ ,
\end{equation}
where $\Delta_j$ is the covariant Laplacian defined before and
$\delta_j$ is an unknown shift of order $R^{-2}$.  The calculation of
$\delta_j$ is an interesting exercise in string theory, which perhaps
already exists in some form in the literature, but for our present
purposes we can argue as in the previous section.  That is, for $j=2$
we know from the low energy field equations that $\delta_2 = 0$, and
the relevant $j$ are close to 2, and so
\begin{equation}
L_0 {\cal V}(j) = \tilde L_0 {\cal V}(j) = (\partial X^+ \bar\partial
X^+)^{j/2}
\Biggl[ \frac{j}{2} - \frac{\alpha'}{4} \Delta_j \Biggr] e^{ik \cdot X} \phi_{+^j}(Y)
+ O(1/\lambda)\ .
\end{equation}
Note that we are keeping $\Delta_j = e^{(j-2)A} \Delta_2 e^{-(j-2)A}$ rather than the simpler $\Delta_2$ for good form, because it has the correct covariance properties. 

Applying this to the OPE, Eq. (\ref{wrope}), the $j$-dependent factors, $e^{\pm jA}$ in $\Delta_j$ can be combined as $(e^{2[A(Y_L) - A(Y_R)]} \partial X^+ \bar\partial X^+)^{j/2}$, where we have again introduced an ordering notation, such that $A(Y_L)$ and $A(Y_R)$ are understood to act to the left or right of the Laplacian, respectively.
Then
\begin{eqnarray}
&&e^{i p_1\cdot X(w,\bar w)}\psi_1(Y(w,\bar w))\,
e^{i p_2\cdot X}\psi_2(Y)
\nonumber\\[3pt]
% &&\qquad
% = e^{ik \cdot X + i p_1 \cdot (w \partial + \bar w \bar\partial) X}
% |w\bar w|^{-2 - \alpha'\Delta_j/4} \psi_1(Y) \psi_2(Y)
% \nonumber\\[3pt]
&&\qquad
= e^{\{ik \cdot X + i e^{[A(Y_L) - A(Y_R)]}
p_1 \cdot (w \partial + \bar w \bar\partial) X\}}  e^{-2A(Y)}
|w\bar w|^{-2 - \alpha'\Delta_2/4} e^{2A(Y)} \psi_1(Y) \psi_2(Y)
\nonumber\\[3pt]
&&\qquad
=e^{\{ik \cdot X + i e^{[A(Y_L) - A(Y_R)]}
p_1 \cdot (w \partial + \bar w \bar\partial) X\}}
F(Y)
\ ,\label{wrope2}
\end{eqnarray}
where
$$
F(y) = e^{-2A(y)} |w\bar w|^{-2 - \alpha'\Delta_2/4} e^{2A(y)} \psi_1(y) \psi_2(y) \ . \label{p1p3}
$$
Fields without arguments are understood to be at the origin, for compactness.

Now consider the matrix element
\beq
\left\langle  {\cal V}_1(w,\bar w) \, {\cal V}_2(0)  {\cal V}_3(1) {\cal V}_4(\infty) \right\rangle\hskip 3.7in
\nonumber\\[3pt] 
 = \Bigl\langle
e^{ik \cdot X + i e^{A(Y_L) - A(Y_R)} p_1 \cdot (w \partial + \bar w \bar\partial) X} F(Y)
e^{i p_3\cdot X(1)}\psi_3(Y(1))\,
e^{i p_4\cdot X(\infty)}\psi_4(Y(\infty))
\Bigr\rangle\ . \ \ \nonumber \\
\eeq
We next evaluate this in the semiclassical approximation.  At the
saddle point, $X^M(\sigma^1,\sigma^2) = x^M$ is constant
on the world-sheet: to first approximation we just replace the fields
with the zero modes.  The leading contribution of the $p_1 \cdot
\partial_a X$ in the exponent comes from their contraction with
$p_3 \cdot X(1)$, where the propagator is evaluated at fixed values of
the zero modes. 
The action for the $X^\mu$ is multiplied by $e^{2A(y)}/\alpha'$, so this
gives 
\be
-\frac{\alpha'}{2} e^{-A(Y_L) - A(Y_R)} p_1 \cdot p_3 = [\alpha'_{\rm eff}(Y_L)\alpha'_{\rm eff}(Y_R)]^{1/2} \frac{s}{4} \equiv \bar\alpha' \frac{s}{4}
\ee
for each contraction.  Integrating the zero mode with weight $\sqrt{-G} = e^{4A}\sqrt{G_\perp}$, the matrix element becomes
\be
{\rm const.} \times
\int d^6y\, \sqrt{G_\perp}\, e^{2A(y)} \psi_3(y) \psi_4(y)
e^{-\bar\alpha' s (w + \bar w) /4 }  |w\bar w|^{-2 - \alpha'\Delta_2/4} e^{2A(y)}\psi_1(y) \psi_2(y) \ . \label{semimat}
\ee
Finally, taking $\int d^2w$ gives
\be
{\cal T}_4 = {\rm const.} \times
\int d^6y\, \sqrt{G_\perp}\, e^{2A(y)} \psi_3(y) \psi_4(y) \Pi(\alpha'\Delta_2)
(\bar\alpha' s )^{2 + \alpha'\Delta_2/2}  e^{2A(y)}\psi_1(y) \psi_2(y) \ . \label{ramp3}
\ee
Noting that $\alpha'\Delta_2 \approx \alpha'(r) t$, this
reproduces the earlier result~(\ref{ramp2}).  The operator ordering
issue raised there has been resolved, and in particular the kernel is symmetric (note that $\Delta_2 = e^{2A}( \sqrt{G_\perp}^{-1}) \partial_M e^{-4A} \sqrt{G_\perp} G^{MN} \partial_N e^{2A}$).  Also, the appearance of $\bar\alpha'$ confirms the assertions about the diffusion time that were made in footnote 11.

The logic of the world-sheet calculation is exactly as in any weakly
coupled field theory for a correlation function with a large hierarchy
of separations: we evaluate the OPE in lowest order, renormalize the
resulting operators with the one loop anomalous dimension, and
evaluate the final matrix element to lowest order.
The result~(\ref{ramp3}) is extended readily to the superstring and to
any external scalars. 

\subsection{BFKL and anomalous dimensions}
\label{subsec:BFKLDGLAP}

Our result on $j_0-2$ is situated in a wider context.  The study
of the relationship between Regge singularities and anomalous dimensions of certain
operators has a long history \cite{Jaroszewicz:1982gr}.  In \nfour\
supersymmetric Yang-Mills theory, it has been argued that properties
of BFKL and DGLAP operators are related by analyticity
\cite{Lipatov:1996ts,Kotikov:2000pm, Kotikov:2002ab,Kotikov:2004er}.
We can understand this connection also from the string theory side of
the duality.  In this subsection we do so in the large-$\lambda$
approximation that we have used throughout; in the next we argue that
it extends to all values of $\lambda$.

This discussion takes place in the conformal limit, where the
spacetime is $AdS_5 \times W$.  The AdS/CFT dictionary relates string
states to local operators in the gauge theory~\cite{GKPAdS,WittenAdS}.
Consider the vertex operator
\begin{equation}
(\partial X^+ \bar\partial X^+)^{j/2} e^{ik \cdot X} \phi_{+^{j}}(Y)\ ,\quad
\phi_{+^{j}}(Y) = e^{\zeta U}\ . \label{dglapvert}
\end{equation}
That is, the transverse part of the vertex operator depends only on the radial coordinate
$U$.  Under scale transformations $\delta U = \epsilon$, $\delta X^\mu
= - \epsilon X^\mu$, this has weight $\zeta - j$.  It corresponds to a
perturbation
\begin{equation}
\int d^4x\, {\cal O}\ ,
\end{equation}
where $\cal O$ is an operator of spin $j$ and dimension $\Delta$; the scale transformation determines that 
\be
\Delta - 4 = \zeta - j\ . \label{dimen}
\ee
The physical state conditions $L_0 = \tilde L_0 = 1$ determine
$\zeta$ and hence $\Delta$ as a function of $j$.  This is the
`nonnormalizable mode'~\cite{GKPAdS,WittenAdS}; the space-time
inversion symmetry implies a second solution $\Delta \to 4 - \Delta$.
The vertex operators~(\ref{dglapvert}) for integer $j \geq 2$ (in the
IIB theory) correspond to the lightest string states of given spin,
and so are dual to the lowest dimension operators of those spins.
These would be the leading twist operators, whose gauge part is
tr$(F_{+\mu} D_+^{j-2} F_{+}{}^{\mu})$.  Thus the physical state
conditions determine the dimensions of the leading twist operators.

On the other hand, the vertex operators~(\ref{dglapvert}) are of the
same form as the Pomeron vertex operator that controls the Regge
behavior.  As in Eqs.~(\ref{SchrodingerEq})--(\ref{Kzero}), 
we take plane wave
normalizable states.  
In the invariant inner product
\begin{equation}
\int d^4x\,du\, \sqrt{-G} (G_{+-})^{-j}  \phi_{+^{j}} \phi_{-^{j}} =
\int d^4x\,du\, e^{(4 - 2j) u}  \phi_{+^{j}} \phi_{-^{j}}\ ,
\end{equation} 
the plane wave states would be 
$e^{(j-2)u} e^{i\nu u}$, that is, $\zeta= j-2+i\nu$.
The dimension~(\ref{dimen}) is then
\begin{equation}
\Delta = 2 + i\nu\ .
\end{equation}
We have seen that the Pomeron vertex operator gets extended to
general $j$ in the analysis of the Regge limit.  Similarly the gauge
theory operators can be extended to general $j$~\cite{Balitsky:1987bk}
(they are no longer local).  Thus it is natural to think of $\Delta$ as a function of complex $j$, or more precisely that the physical state conditions define a curve in the $\Delta$-$j$ plane.  The operator dimensions (which enter into DGLAP evolution) are given by $\Delta(j)$ at $j=2,3,4,\ldots\ $.  The BFKL exponents are given by the inverse function $j(\Delta)$ at $\Delta = 2+i\nu$, and in particular $j_0 = j(2)$.  

Let us now repeat the large-$\lambda$ calculation of $j_0$ in light of the above discussion.  We have the physical state condition
\begin{equation}
1 = L_0 = \frac{j}{2} - \frac{1}{4\sqrt\lambda} (\Delta-2)^2 + \frac{1}{\sqrt\lambda}\ .
\label{jeq2}
\end{equation}
The first term on the right is the oscillator level.  The form of the second term is determined by inversion symmetry; its coefficient follows by lifting the Virasoro generators~(\ref{vira}) to curved space and matching the term $-(\alpha'/4R^2)\partial_u^2$.  The value of the final constant follows from the solution $j=2, \Delta = 4$, corresponding to the energy-momentum tensor.  Finally, $j_0 = 2 - 2/\sqrt\lambda$ is obtained by setting $\Delta = 2$.  This is similar to the discussion in ref.~\cite{klv5}.

\subsection{Extension to general $\lambda$}

Thus far we have leaned heavily on semiclassical calculations on the
world-sheet.  In this section we would like to set up the framework as
much as possible without this assumption.  We might hope in the future
to extend our results to higher order in $1/\sqrt\lambda$, and also to
attempt to make some contact with the strongly coupled world-sheet
theories that would be dual to perturbative gauge theories.  In order
to have some constraint on the structure we will focus on the high
energy conformal regime.

The calculations thus far indicate the general strategy, factorizing
in the $t$-channel in terms of a sum over Pomeron vertex operators.
We keep only the leading Regge trajectory, meaning vertex operators
constructed from the undifferentiated $X^{3,4}$ and $U$ fields, and 
from $\partial_{w,\bar w} X^{\pm}$.  That is, we are assuming that as 
$\lambda$ is varied the dominant Pomeron states are in one-to-one 
correspondence with those at large $\lambda$; this appears to be 
consistent with what is known at small $\lambda$, as we note below.  At 
$t = 0$, a complete set
of vertex operators is of the form
\begin{equation}
{\cal V}_{\pm j,\nu,0} = e^{(j-2 + i \nu )U} (\partial X^{\pm} 
\bar\partial X^{\pm})^{j/2}.
\label{vjn}
\end{equation}
In the previous section we used a semiclassical argument to justify 
this, but we believe that it is simply a consequence of symmetry: these 
are the principal continuous representations of the conformal 
group~\cite{Bargmann:1946me}.
Similarly for negative $t=-k^2$,
\begin{equation}
{\cal V}_{\pm j,\nu,k} \propto e^{i k \cdot X} e^{(j-2 )U} K_{i\nu}
(z_0 |t|^{1/2}e^{-U}) (\partial X^{\pm} \bar\partial X^{\pm})^{j/2}\ ;
\label{Kvert}
\end{equation}
for positive $t$ the Bessel function becomes $J_{i\nu}$.  Again these 
forms
should be completely determined by conformal symmetry.

The quantum numbers $(j,\nu,k)$ commute with $L_0, \tilde L_0$, so
\begin{equation}
L_0 {\cal V}_{j,\nu,k} = \tilde L_0 {\cal V}_{j,\nu,k} = h_{j,\nu} 
{\cal V}_{j,\nu,k} \ ,
\end{equation}
where the weight $h$ is a function of the spin and the $SO(2,2)$ 
Casimir; for example, at strong coupling eq.~(\ref{jeq2}) gives 
$h_{j,\nu} = \frac{1}{2} j
+ \frac{1}{4}(\nu^2 + 4)/\sqrt{\lambda}$.
   The coefficient of the unit operator in the OPE of two such operators 
is then
\begin{equation}
{\cal V}_{j,\nu,k}(w,\bar w) {\cal V}_{j',\nu',k'}(0,0) \sim 
\frac{c_{j,\nu}}{(w\bar w)^{h_{j,\nu}}}
(2\pi)^4 \delta(j+j') \delta(\nu + \nu') \delta^2(k+k')\ .
\end{equation}
General principles of CFT (see 
Chapter 6.7 of \cite{Polchinski:1998rq})
then give the factorization
\begin{equation}
\left\langle {\cal W}_R w^{L_0 - 2} \bar w^{\tilde L_0 - 2}
{\cal W}_L \right\rangle = \int \frac{dj\,d\nu}{(2\pi)^2}
\frac{(w\bar w)^{h_{j,\nu} - 2}}{c_{j,\nu}}
\left\langle {\cal W}_R {\cal V}_{j,\nu,k} \right\rangle
\left\langle {\cal V}_{-j,-\nu,-k} {\cal W}_L \right\rangle
\ .
\end{equation}
We are using a non-standard Hilbert space for the $\pm$ oscillators,
so we must determine the path of the $j$-integral.  We can think of
$j$ as being introduced through a Mellin transformation with respect
to $s$, and so the $j$ integral is an inverse Mellin transformation.
Thus the $j$ integral runs parallel to the imaginary axis, Re$(j) =
j_* $, and to the right of all singularities.

Now take the integral $d^2 w$ in a neighborhood of the origin;
in order for this to converge at 0 we first deform the contour to large 
$j_*$.
Integrating and also boosting back to the approximate rest frames, we 
have
\begin{equation}
{\cal T} \sim \int \frac{dj\,d\nu}{(2\pi)^2}
\frac{(s/s_0)^j}{c_{j,\nu} (h_{j,\nu} - 1) }
\left\langle {\cal W}_{R0} {\cal V}_{j,\nu,k} \right\rangle
\left\langle {\cal V}_{-j,-\nu,-k} {\cal W}_{L0} \right\rangle\ .
\end{equation}
In order to obtain the large-$s$ asymptotics we now deform the contour 
back
toward negative $j_*$, picking up the pole in $h_{j,\nu} - 1$ having
the largest $j$; call this $j(\nu)$.  Then
\begin{equation}
{\cal T} \sim \int \frac{d\nu}{(2\pi)}
\frac{(s/s_0)^{j(\nu)} {\rm Res}(\nu) }{c_{j(\nu),\nu} }
\left\langle {\cal W}_{R0} {\cal V}_{j(\nu),\nu,k} \right\rangle
\left\langle {\cal V}_{-j(\nu),-\nu,-k} {\cal W}_{L0} \right\rangle\ . 
\label{formal}
\end{equation}

 From this formal argument we can anticipate that much of the
qualitative physics at large $\lambda$ will persist to smaller values
of $\lambda$.  The inversion symmetry of the conformal group, which
takes $u\to u_0-u$ for any $u_0$, implies that $j(\nu)$ can depend
only on $\nu^2$.  Therefore $\nu=0$ is an extremum, presumably a
maximum,\footnote{At weak coupling, 
Ref.~\cite{Kotikov:2004er} gives $j(\nu)$
in \nfour\ Yang-Mills through three loops.}
 of $j(\nu)$. 
This ensures there is a cut
extending from some value $j=j_0$ at $\nu=0$ down to smaller values of
$j$.  The leading behavior at large $s$ will
always be given by expanding
\begin{equation}
j(\nu)\approx
j_0 - {\cal D}\nu^2 + {\rm order}(\nu^4)\ .
\end{equation}
Evaluating the $\nu$ integral~(\ref{formal}) by saddle point again
gives, at $t=0$, a diffusion kernel in $U$.  
For negative $t$, the
vertex operators~(\ref{Kvert}) are damped at small $U$, 
generalizing the result found in the effective quantum mechanics at large
$\lambda$; there
is effectively a repulsive potential in this region. Meanwhile,  
the end of the cut, $j_0$, is $t$-independent for all $\lambda$,
because the eigenvalue $h_{j,\nu}$ can be obtained at large $U$, where the
vertex operator is asymptotically independent of $t$.
See our discussion surrounding \Eref{Knonzerot}.

The DGLAP dimensions and BFKL exponent are still determined by the same 
vertex operator, so the relation discussed in the previous section 
should hold for all $\lambda$, consistent with the weak coupling 
result.  That is, $j_0$ is determined by $\Delta(j_0) = 2$.  In the 
large-$\lambda$
limit we have studied the $\Delta$-$j$ relation only near $j = 2$,
where it takes the form~(\ref{jeq2}).

  In the weakly coupled limit the $(\Delta,j)$ locus has a
complicated structure \cite{Jaroszewicz:1982gr}.
\FIGURE[ht]{
%\begin{center}
\includegraphics[width = 3.7in]{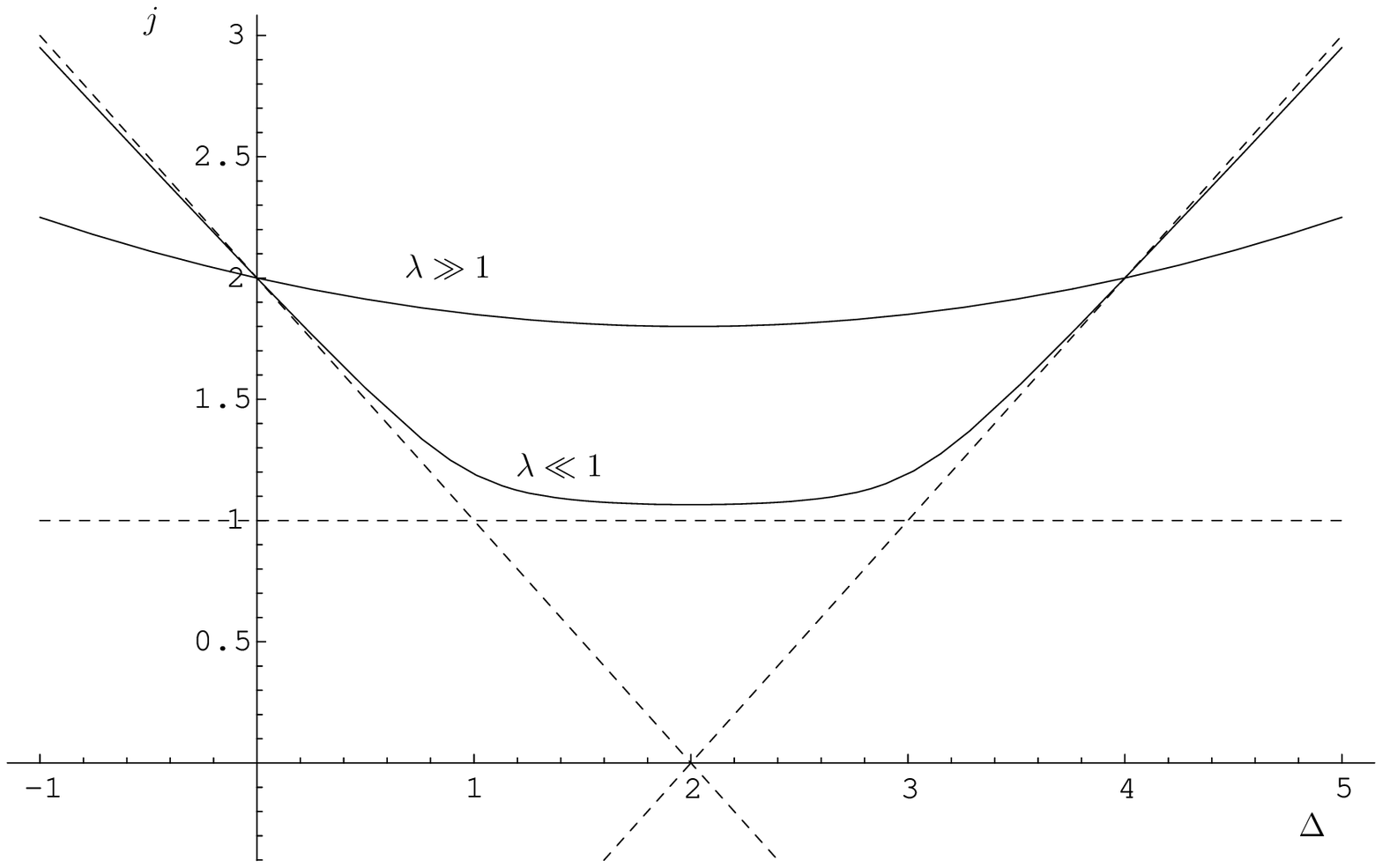}
%\end{center}
\caption{Schematic form of the $\Delta-j$ relation for $\lambda\ll 1$ 
and $\lambda\gg 1$.  The dashed lines show the $\lambda =0$ DGLAP 
branch (slope 1), BFKL branch (slope 0), and inverted DGLAP branch 
(slope $-1$).  Note that the curves pass through the points (4,2) and 
(0,2) where the anomalous dimension must vanish.  This curve is often 
plotted in terms of $\Delta - j$ instead of $\Delta$, but this obscures 
the inversion symmetry $\Delta \to 4-\Delta$.}
\label{fig:BFKLDGLAP}
}%\end{figure}
In the normal operator analysis one has $\Delta
= 2 + j + O(g^2)$, i.e. twist two in the free limit.  BFKL identify
another branch to the solution, where $j = 1 + O(g^2)$.  The inversion 
symmetry $\Delta \to 4-\Delta$ implies a third branch $\Delta = 2-j + 
O(g^2)$.  It follows that at zero coupling the Pomeron physical state 
condition must be
\begin{equation}
(\Delta - 2 - j)(\Delta - 2 + j)(j-1) = 0
\end{equation}
in order to capture all three branches of the solution.  At one loop we 
would expect a correction
\begin{equation}
(\Delta - 2 - j)(\Delta - 2 + j)(j-1) = a(\Delta,j) g^2\ . \label{phys1}
\end{equation}
Let us analyze the solution near the point $(\Delta,j) = (3,1)$ where 
the BFKL and DGLAP lines meet.  We assume that $a(\Delta,j)$ is 
nonsingular there and approximate it by a constant $a \equiv a(3,1)$; 
we can also approximate $\Delta - 2 + j = 2$.  The intersection of the 
BFKL and DGLAP lines is then resolved into a smooth hyperbola, one 
branch of which is shown in Fig.~\ref{fig:BFKLDGLAP}.

If we approach this point along the BFKL branch, the physical state 
condition~(\ref{phys1}) becomes
\begin{equation}
j = 1 + \frac{a g^2}{2(\Delta - j - 2)} =  1 + \frac{a g^2}{2(\Delta - 
3)} + O(g^4)\ . \label{jpert}
\end{equation}
If we approach it along the DGLAP branch, then
\begin{equation}
\Delta = 2 + j + \frac{a g^2}{2(j - 1)}\ . \label{dpert}
\end{equation}
Thus the physical state condition~(\ref{phys1}) reproduces the known
perturbative poles in the BFKL exponent and anomalous dimensions; the
common value $a = -N/\pi^2$ gives the correct coefficient in both
(\ref{jpert}) and~(\ref{dpert}).  We emphasize that this result, that
the BFKL calculation determines the $j=1$ singularity in the anomalous
dimensions, is well-known~\cite{Jaroszewicz:1982gr,
Lipatov:1996ts,Kotikov:2000pm,
Kotikov:2002ab,Kotikov:2004er}; we are simply giving a different
perspective on it.\footnote{The full structure is even richer than we
have indicated, because the poles that we have found are just the
first of infinite sets, arising from $\Psi$ functions.  Thus there are
evidently an infinite number of additional branches to the solution of
the physical state condition at $\Delta = 2 + j + m$, $\Delta = 2 - j
- m$, and $j = 1 - m$ for positive integer $m$; these do not intersect
with the branch we are focussing on.  There are also additional 
contributions from states with nonzero spin in the 2-3 plane.  
Presumably all these additional contributions correspond to the 
exchange of string states with higher oscillators excited.}

For the $\rho$ trajectory~\cite{Kwiecinski:1981yk,Kirschner:1982qf,McGuigan:1992bi,Bartels:2003dj,Bartels:2004mu}
 the BFKL exponent goes to $0$ at weak
coupling, so the BFKL and DGLAP curves all meet at the point $(2,0)$
that determines the BFKL exponent.
That the weak-coupling BFKL exponent in
this case is of order $\sqrt\lambda$, rather than the perturbative
$\lambda$, is presumably connected with this fact.

\section{Derivation in light-cone gauge}
\label{sec:light-cone}

The light-cone has proven to be a natural formalism for studying the high
energy limits of quantum field theories, leading to a vivid physical picture
in Feynman's parton language. It is interesting to re-interpret
our results for the BFKL singularity in this framework.

The light-cone gauge for a superstring~\cite{Metsaev:2000yu} in $AdS_5 \times
W$ eliminates all spurious degrees of freedom by fixing the bosonic
co-ordinates $X^+(\sigma,\tau) = (X^0+X^1)/\sqrt{2} = \tau$,
$P^+(\sigma,\tau) = \mbox{const}$, and the first derivatives of
$X^-(\sigma,\tau) = (X^0-X^1)/\sqrt{2}$ as quadratic functions of the
transverse fields via the Virasoro constraints.\footnote{ It is convenient in
  this section to use $\tau$ to denote worldsheet time.  In other sections
  $\tau$ denotes the Regge diffusion ``time''. }  (See
Refs.~\cite{Polchinski:1998rq,Zwiebach:2004tj} for details.) The remaining
physical bosonic degrees of freedom are two transverse co-ordinates
$X_\perp(\sigma,\tau) = (X_2,X_3)$, the radial field $Z(\sigma,\tau)$ in
$AdS^5$, (its zero-mode being $z=R^2/r$), and five fields
$\Theta(\sigma,\tau)$ in $W$, plus the center of mass coordinates $x^- =
(x^0-x^1)/\sqrt{2}$ and conjugate momentum $p^+ = (p^0 + p^1)/\sqrt{2}$.  The
fermionic sector is treated analogously, but we can safely ignore it in our
present discussion.

We will first discuss how to extract the Regge limit for an
elastic open-string scattering amplitude ($p_1,p_3 \rightarrow - p_2, - p_4$)
in a light-cone setting, before generalizing the analysis to closed-string
scattering. After treating scattering in flat space, we will deal
with the case of warped spacetime.

The elastic scattering amplitude for external states with momentum vectors 
$p_i^M =
(p^+_i,p^-_i,p^\perp_i)$, and corresponding vertex operators $V_i$, can be
expressed as a Euclidean  Polyakov path integral in light-cone gauge,
\be
{\cal A}(s,t)\ \delta^2\left(\sum
 p^\perp_i 
%+p^\perp_2 +
%p^\perp_3 + p^\perp_4
\right)  = \int dT \int  {\cal D}X_\perp {\cal D}Z\   \; 
% V_2 V_3 V_4 
\; e^{\textstyle - \int d\tau  \int^{ p^+}_0 d\sigma \ {\cal
    L}[X_\perp,Z]} \ \prod_{1}^4 V_i\
\; ,
\label{eq:lcpath}
\ee 
where we ignore the bosonic modes in $W$ as well as all fermionic
modes, since they don't contribute to the leading $j$-plane
singularity.  Scattering can be shown to take place on a worldsheet of
fixed width given by the total $p^+ = p^+_1 + p^+_3$ in the
$s$-channel, as illustrated in Fig.~\ref{fig:worldsheet}. Open strings
obey Neumann boundary condition on the boundaries of the worldsheet
(the horizontal solid line segments in Fig.~\ref{fig:worldsheet}).  The
worldsheet has a single modulus $T$ specifying the duration of the
interaction in worldsheet time $\tau$.  (Closed strings obey
periodic boundary conditions and have an additional modulus to enforce
level matching.)
\FIGURE[ht]{
%\begin{center}
%
\setlength{\unitlength}{0.7 mm}
\begin{picture}(125,60)
\linethickness{.25mm}
\put(0,10){\line(1,0){125}}  
\put(-20,25){$p^+_3$} \put(135,25){$-p^+_4$}
\put(0,40){\line(1,0){40}}
\put(27,3){$\tau = 0 \; \leftrightarrow \; T$}
\put(40,10) {\dashbox{2}(15,40)}
\put(55,40){\line(1,0){70}}
\put(-20,45){$p^+_1$} \put(135,45){$-p^+_2$}
\put(0,50){\line(1,0){125}} 
\put(-35,10){\vector(0,1){10}} \put(-40,18){$\sigma$}
\put(-35,10){\vector(1,0){10}} \put(-33,5){$\tau $}
\end{picture}
\caption{The light-cone worldsheet domain, $X^+ = \tau \in [-\infty,\infty]$, 
  $\sigma \in [0,p^+]$,with $p^+=p^+_1 + p^+_3$ for elastic scattering in the
  brickwall frame.}
\label{fig:worldsheet}
%\end{center}
}%\end{figure}

\subsection{Open string scattering in flat spacetime}

The open string tachyon elastic scattering
amplitude in flat space has the well-known  Veneziano form,  
\be 
{\cal A}(s,t) =  \int^1_0 dw\; (1-w)^{ - 2 - \alpha's}
\;w^{ -2 -  \alpha't }  \simeq   \Gamma(-1 - \alpha't  ) ( e^{- i\pi}
\alpha' s)^{1 + \alpha't } \; ,  
\ee
where the integral in the Regge limit is dominated by $w = O(1/s)$. In
Sec.~\ref{sec:systematic} this observation for the closed string led
naturally to the use of the OPE for the $p_1$-$p_2$ vertex operators. 
The methods of Sec.~\ref{sec:systematic} analogously give 
\be
{\cal V}^\pm_R(k,w) =  (\dd_w X^\pm(w))^{1+ \alpha' t} e^{ \mp i k X(w)} \quad \mbox{and}
\quad \Pi_R(\alpha't) = \Gamma(-1-\alpha't) e^{- i \pi - i \pi \alpha' t}\; .
\label{eq:Rvertex}
\ee
for the open string Reggeon vertex operator and
propagator in flat space.~\footnote{We also
note in passing that using the vertex operator~(\ref{eq:Rvertex})
considerably reduces the labor of earlier methods used in extracting
the asymptotic multi-Regge behavior for general $n$-point
functions~\cite{Detar:1971gn,Detar:1971dj,Brower:1974yv}.}

We shall explain in this subsection how we can arrive at these results
starting from the flat-space light-cone path integral, given
by~(\ref{eq:lcpath}) without the $Z$ coordinate. In conformal gauge,
we integrate over the position $w$ of one of the vertex operators.  In
a light-cone approach, all external vertices are fixed, while the
modulus $T$ is integrated over.  The conformal Christoffel
transformation~\cite{Zwiebach:2004tj} which maps the upper half
complex $w$-plane into the light-cone worldstrip in
Fig.~\ref{fig:worldsheet}, takes the region $w = O(1/s)$ that
dominates the integral in the Regge limit into the regime $T =
O(1/p^+)$.  This difference notwithstanding, we will see that we can
still identify factors in the light-cone derivation closely related to
${\cal V}^\pm_R(k,w)$ and $\Pi_R(\alpha't)$ above.

\subsubsection{Brief  discussion for light-cone gauge at high energies}

Since the density of $P^+(\sigma)$ is conserved in the light-cone gauge, it
is traditional, for scattering processes, to label each string segment (or
string bit) $\Delta \sigma$ by equal quanta $\Delta p^+$, and to choose the
string length\footnote{We have set the world sheet speed of propagation  to be $c=1/(2\pi \alpha')$. More generally, $c= {\it l_s}/(2\pi \alpha' p^+)$~\cite{Polchinski:1998rq}.} to   be ${\it l_s} = p^+$. Since this gauge is not manifestly
Lorentz invariant, it is helpful to pick a convenient frame.  We have
chosen the {\it brickwall frame} in the center of momentum, where the
transverse momenta are reflected by the collision: $p^\perp_r = \pm
k^\perp/2$. In this frame~\footnote{To be precise, we define the brickwall
  frame with transverse momenta $ p^\perp_1 = p^\perp_2 =  k^\perp /2 $, and
  rapidities $\exp[\pm y_i] = \sqrt{2}p^\pm_i/\sqrt{(M^2_i + k^2_\perp/4)}$,
  so that the invariants are $t = - k^2_\perp$ and $s = M^2_1 + M^2_3 +
  \sqrt{M^2_1 + k^2_\perp/4} \; \sqrt{M^2_3 + k^2_\perp/4}\; \cosh(y_1 -
  y_3)$. Boosting to the center of longitudinal momentum frame sets $y_1 = -
  y_3$. } the two strings joining at $(\sigma,\tau) =
(\sigma_{int},0)$, split at $(\sigma_{int},T)$ with exactly the same value of
$\sigma$ or string bit, as illustrated in  Fig.~\ref{fig:boundaries}.
  
A major simplification of this frame is the fact that the $t$-channel
worldsheet diagram ($T <0$) vanishes identically, leaving only the
$s$-channel contribution ($T\ge 0$).
\FIGURE[ht]{
%\begin{center}
%
\setlength{\unitlength}{0.7 mm}
\begin{picture}(125,60)
\linethickness{.25mm}
\put(0,10) {\dashbox{2}(125,40)}
\put(0,10){\line(1,0){125}}  
\put(-20,25){$B_3$} \put(135,25){$B_4$}
%\put(0,42){\line(1,0){40}}
\put(0,40){\line(1,0){40}}
\put(27,3){$\tau = 0 \; \leftrightarrow \; T$}
%\put(40,10) {\dashbox{2}(15,40)}
\put(40,40) {\dashbox{2}(15,0)}
%\put(55,42){\line(1,0){70}}
\put(55,40){\line(1,0){70}}
\put(-20,45){$B_1$} \put(135,45){$B_2$}
\put(0,50){\line(1,0){125}} 
\put(-12,39){$\sigma_{int}$}
\put(-35,10){\vector(0,1){10}} \put(-40,18){$\sigma$}
\put(-35,10){\vector(1,0){10}} \put(-33,5){$\tau $}
\end{picture}
\caption{The light-cone worldsheet domain splits
  into two parallel sheets which join only at the dashed line
  at $\tau \in [0,T]$ and
  $\sigma =\sigma_{int}$. Solid (dashed) lines have Neumann (Dirichlet) 
  boundary conditions
  respectively.}
\label{fig:boundaries}
%\end{center}
}%\end{figure}
In this case the path integral can be evaluated by first cutting the
worldsheet along the horizontal dashed line in
Fig.~\ref{fig:boundaries} at fixed $\sigma = \sigma_{int}$, so that it
forms two independent strips of 
fixed width $p^+_1=-p^+_2$ and $p^+_3=-p^+_4$ respectively. The two
independent strings must then be rejoined along the dashed line by
imposing there a Dirichlet boundary condition, with a delta-functional
constraint,
\be
\int {\cal D}k^\perp(\tau)\; \exp[i\int^T_0 d\tau \; 
  k^\perp(\tau) (X^{3-4}_\perp(\sigma_{int},\tau) -
X^{1-2}_\perp(\sigma_{int},\tau))] \; .
\label{eq:delta}
\ee
Effectively what we have done is insert a non-local vertex operator 
\be
\exp[-i\int^T_0 d\tau \; k^\perp(\tau)
X^{1-2}_\perp(\sigma_{int},\tau)]
\ee
on the $\sigma=\sigma_{int}$ 
boundary of the 
$1$-$2$ string,
and a corresponding vertex 
%$\exp[i\int^T_0 d\tau \; k^\perp(\tau)
%X^{3-4}_\perp(\sigma_{int},\tau))]$ 
on the boundary of the $3$-$4$
string.  The path
integrals over the worldsheets of 
the $1$-$2$
and $3$-$4$ strings can now be performed separately, followed
by the integral over $\int d T \int {\cal D}k^\perp(\tau)$.

The Regge limit represents a collinear boost into what is sometimes
referred to as an {\it infinite momentum} frame, in which the boosted
$3$-$4$ string grows in length ($p^+_3 = - p^+_4 \sim O(\sqrt{s})$)
and the $1$-$2$ string decreases in length ($p^+_1 = - p^+_2 =
O(1/\sqrt{s})$). Therefore, in the Regge limit, it is convenient to
refer to these as the ``long" and ``short" strings respectively.  They
propagate in $\tau$ independently, except in the interaction region.
As a consequence, the non-trivial physics involves a very small area of
the worldsheet, $\Delta\tau = T \sim 1/p^+_3 \sim 1/\sqrt{s}$ and
$\Delta \sigma \sim p^+_1 \sim 1/\sqrt{s}$, so one should be able to
associate the Regge mechanism with a ``local'' conformal worldsheet
transformation near the interaction region.  The sole impact of this
brief interaction is to constrain the ends of the long and short
strings to coincide.  As we shall show below, the Regge behavior comes
entirely from the growth of the long string with the total
center-of-mass energy $\sqrt s$.

Before proceeding to this analysis, it is useful to first fix the
normalization for the amplitude.  We do this by
considering the Regge limit of the scattering amplitude
at $t=0$, {\it i.e.}, for $p^\perp_r \equiv 0 $.  
In this limit, the path integral in
(\ref{eq:lcpath})
can be carried out (recall we are not including a warped coordinate $Z$ at
this point), and the forward amplitude reduces to a
single integral over $T$~\cite{Mandelstam:1973jk} 
\be 
  {\cal A}(s,0)  \simeq \; \int \frac{ dT}{T^2} \; 2 \alpha' p^+ e^{ 
    p^- T}   = - \alpha' s \int^\infty_\epsilon \frac{d\zeta}{ \zeta^2 } \; e^{ -\zeta}
\simeq   -   \frac{\alpha' s}{\epsilon} \; ,
\label{eq:normalize}
\ee
with  $\zeta = - p^- T$.  The divergence at $T = 0$ corresponds to the pole in
the propagator, $\Pi_R(\alpha't)$, in Eq.~(\ref{eq:Rvertex}) from ``photon''
exchange at $t = - k_\perp^2 = 0$. The divergence at $T=\infty$ is due to $s$-channel poles; we evaluate the integral by analytic continuation to ${\cal R}e \;\; p^-<0$. This is consistent with our expectation
that the leading Regge singularity is dominated by contributions from the
small $T$ region. For later reference we define the measure
$d\mu(T) =  [ 2 \alpha' p^+  \exp( p^- T) /T^{2}] dT$.

\subsubsection{Regge behavior}

In flat space, the light-cone
Lagrangian density is
\be
{\cal L} = \frac{1}{2}(\dd_\tau X_\perp)^2 +\frac{1}{2} \frac{\textstyle
  1}{(\textstyle 2 \pi   \alpha')^2}{(\dd_\sigma X_\perp)^2}  \; .
\ee
The light-cone Hamiltonian $H = p^-$ is the generator of translations
in $x^+$. The Virasoro constraints set $-\dd_\tau X^-$ equal to
the Lagrangian: ${\cal L}= -\dd_\tau X^-$. (In what follows, we shall
use the notation $\dot X_\perp \equiv \dd_\tau X_\perp$ and
$X'_\perp\equiv \dd_\sigma X_\perp$.) Next we introduce the
vertex functions for ground state tachyons on the vertical boundaries
$B_r$ in Fig.~\ref{fig:boundaries},
\be 
V_r[p_r,X] = \exp\left[(1/p_r^+) 
  \;\int_{B_r} d \sigma   [ i p^\perp_r X_\perp(\sigma,\tau_r) +    p^-_r
  X^+(\sigma,\tau_r)] \right] \; , 
\ee
with center of mass coordinates: $\tau_r =(1/p_r^+) \int_{B_r} d \sigma X^+$
and $x_\perp^{(r)} = (1/p_r^+) \int_{B_r} d \sigma X_\perp$. The limits $\tau_r
\rightarrow \mp \infty$ for in/out states put the scattering amplitude on
shell.

We will compute the worldsheet path integral by first evaluating it in six
rectangular blocks, holding the worldsheet fields on the boundaries of the
blocks fixed, and then finally integrating over the boundary data of the
blocks.  The six rectangular regions, shown in Fig.~\ref{fig:six_pieces}, are
formed by cutting the worldsheet at $\sigma = \sigma_{int}$ into the $1$-$2$
and $3$-$4$ strips described above, and then dividing each strip into three
regions: incoming ($\tau < 0$), interacting ($\tau \in[0,T]$), and outgoing
($\tau>T$) segments. In Fig.~\ref{fig:six_pieces}, these six regions are
labeled by the contributions $\Phi^{(r)}$ to the path integral from the four
external states, and the Green's functions $G^{(1,2)}$ and $G^{(3,4)}$ for
the interactions regions.  On the vertical boundary of each region marked
$\Phi^{(r)}$, $r=1,2,3,4$, we hold the transverse fields fixed to
  $X^{(r)}_\perp(\sigma)$:
\beq
X_\perp(\sigma,0) &=& X^{(1)}_\perp(\sigma-\sigma_{int} ) \theta (\sigma-\sigma_{int} ) + X^{(3)}_\perp(\sigma ) \theta (\sigma_{int} -\sigma) \nn\\
X_\perp(\sigma,T) &=& X^{(2)}_\perp(\sigma-\sigma_{int}  ) \theta (\sigma-\sigma_{int} ) + X^{(4)}_\perp(\sigma ) \theta (\sigma_{int} -\sigma) 
\eeq
Note that $\sigma_{int}=p^+_3$. This leads to an
exactly factorized representation for the amplitude,
\be
{\cal A}(s,t)\delta^2(p^\perp_1 + p^\perp_2 + p^\perp_3 + p^\perp_4) \simeq
\int d\mu(T) \int {\cal D}k^\perp(\tau) \; {\cal V}_{12}[k^\perp(\tau)] \; \;
{\cal V}_{34}[-k^\perp(\tau)] \; . \label{eq:amplitude} 
\ee
Each ${\cal V}_{rs}$ is given by a {\it one-dimensional} path integral
over the  boundary fields $X^{(r)}_\perp(\sigma)$,
\be 
{\cal V}_{rs}[k^\perp(\tau)] = \int {\cal D}X^{(r)}_\perp \; \int {\cal D}X^{(s)}_\perp\;
\Phi^{(r)}[X^{(r)}_\perp]\;\; G^{(r,s)}[X^{(r)}_\perp, X^{(s)}_\perp,k^\perp(\tau)]\; \;\Phi^{(r)}[X^{(s)}_\perp]\; ,
\ee
where $\Phi^{r}$ and $G^{(r,s)}$ are the results of the {\it two-dimensional}
path integrals over the corresponding regions, with fixed boundary
values given by $X^{(r)}_\perp(\sigma)$, with $0\leq \sigma\leq |p^+_r|$.
\FIGURE[ht]{
%\begin{center}
%
\setlength{\unitlength}{0.7 mm}
\begin{picture}(125,60)
\linethickness{.25mm}
\put(0,10){\line(1,0){120}}  
\put(20,18){$\Phi^{(3)}$} \put(46,18){$G^{(3,4)}$}\put(75,18){$\Phi^{(4)}$}
\put(0,30){\line(1,0){40}}
\put(33,3){$\tau = 0 \quad \leftrightarrow \; T$}
\put(40,30) {\dashbox{2}(25,0)}
\put(40,10) {\dashbox{2}(25,40)}
\put(65,30){\line(1,0){55}}
\put(20,38){$\Phi^{(1)}$} \put(46,38){$G^{(1,2)}$} \put(75,38){$\Phi^{(2)}$}
\put(0,50){\line(1,0){120}} 
\put(-35,10){\vector(0,1){10}} \put(-40,18){$\sigma$}
\put(-35,10){\vector(1,0){10}} \put(-33,5){$\tau $}
\end{picture}
\caption{The six rectangular regions for evaluating the
elastic scattering path integral in the    brickwall frame.}
\label{fig:six_pieces}
%\end{center}
}%\end{figure}
Both the incoming and the outgoing regions involve free string propagation,
so the $\Phi^{(r)}$ are just the usual Gaussian wave functions for
propagating tachyon boundary states,
\bel{Phidef}
%V_r \rightarrow 
\Phi_{r}[X^{(r)}_\perp] = \exp[ i p^\perp_r
x^{(r)}_\perp \; - \Half \sum^\infty_{n=1} \omega^{(r)}_n X^{(r)}_n X^{(r)}_n ] \; .
\label{eq:Phi}
\ee
We have used a standard normal mode expansion for each boundary field,
\be 
X^{(r)}_\perp(\sigma) = x^{(r)}_\perp  + \sqrt{\frac{2}{p^+_r}} \sum^\infty_{n=1}
X^{(r)}_n \cos(  \omega^{(r)}_n \sigma/c) \; .
\label{eq:expansion}
\ee
 With our choice of Euclidean worldsheet parameters, the frequencies
of the modes, $\omega^{(r)}_n = n/(2\alpha' |p^+_r|)$, are scaled by
$1/ |p^+_r|$ on each string.

 $G^{(1,2)}$ and $G^{(3,4)}$ can be obtained explicitly by a
variety of methods~\cite{Kaku:1974zz}. They obey mixed boundary
  conditions: Dirichlet boundary conditions on the vertical (dashed lines at
  fixed $\tau$), Neumann boundary condition at the free end (solid line at
  fixed $\sigma$) and a fixed Fourier distribution in $k^\perp(\tau)$ for the
  interaction between the strings (dashed line at $\sigma_{int}$ as described
  above).  At this point our treatment is still general. However we now
choose to evaluate them approximately in the Regge limit, by
  making a semi-classical approximation which is more easily  generalizable to 
our subsequent warped background.

In the Regge limit, one string is much shorter than the other, and the
  interaction time $T$ goes to zero.  In analogy with the OPE
  expansion, we consider an expansion in $T$.  At $T=0$, we can
  identify the boundary values of the worldsheet fields at $\tau=0$
  and $\tau=T$.  It is therefore useful to distinguish between the
  differences, $[X^{(1)}_\perp(\sigma)-X^{(2)}_\perp(\sigma)] $ and
  $[X^{(3)}_\perp(\sigma)- X^{(4)}_\perp(\sigma)]$, that are being set
  to zero and the averages, $\overline X^{(12)}_\perp(\sigma) = [
  X^{(1)}_\perp(\sigma)+X^{(2)}_\perp(\sigma)]/2$ and $\overline
  X^{(34)}_\perp(\sigma) = [X^{(3)}_\perp(\sigma)+
  X^{(4)}_\perp(\sigma)]/2$.  Now the delta-functionals in
  Eq.~(\ref{eq:delta}) become ordinary delta-functions
\bel{ordinarydeltas}
{(2\pi)^2 } \delta^2(\overline X^{(12)}_\perp(0) - \overline X^{(34)}_\perp(p^+_3))= 
\int {d^2k^\perp}\;\;  e^{\textstyle -i k^\perp \overline
  X^{(12)}_\perp(0)} \; e^{\textstyle i k^\perp   \overline X^{(34)}_\perp(p^+_3)} \; ,
\ee
for the average coordinates:  
The only quantity we need to evaluate to first order in $T$ is the action,
$\Delta S= \Delta S_{12} + \Delta S_{34}$, in the $\tau\in[0,T]$ region,
\be
 \Delta S_{12}[\overline X_\perp^{(12)}] \simeq T \int^{p^+_1}_{0} d\sigma\;  {\cal
   L}   \quad \quad \mbox{and}
\quad\quad
 \Delta S_{34}[ \overline X_\perp^{(34)}] \simeq T\int^{p_3^+}_0 d\sigma\; {\cal L}\;.
 \ee
Consequently $G^{(1,2)}$ can  be approximated by
 \be
 G^{(1,2)}[X_\perp^{(1)},X^{(2)}_\perp, k^\perp]\sim
 \delta[X_\perp^{(1)}(\sigma) - X_\perp^{(2)}(\sigma) ] \;  \exp\left[{\textstyle-i k^\perp \overline
 X^{(12)}_\perp(0) -\Delta S_{12}[\overline X_\perp^{(12)}]}\right]\;,
 \ee
 and similarly  for $G^{(3,4)}$. The remaining
 path integral  over  $\overline X_\perp^{(12)}$ and $\overline X_\perp^{(34)}$ can be carried out in terms of sums over their respective normal modes.

Let us first examine the interaction region for the short string.  Since the
excitation frequencies in the wave functions $\Phi^{(1)}$ and $\Phi^{(2)}$
grow with $s$, ($\omega_n = n/2 \alpha' p^+_1 \sim n \sqrt{s}$), the short
string interacts like a rigid point-like object, and its center of mass
effectively coincides with the interaction point,
$x^{(1)}_\perp=x^{(2)}_\perp\simeq X_\perp^{(1)}(0)=X_\perp^{(2)}(0)$.
The action for the short string during the interaction time, $\Delta S_{12}$,
provides a UV cutoff in the mode sum, leading to an
approximate local point-like short string vertex,
\be
{\cal V}_{12}(k^\perp) \sim \int \frac{d^2x^{(1)}_\perp}{(2\pi)^2} 
e^{ix^{(1)}_\perp (p^\perp_1 + p^\perp_2 - k^\perp)} = 
\delta^2(p^\perp_1 + p^\perp_2 -k^\perp)\;.  
\ee

On the other hand, in the interaction region for the long  string, the
situation is reversed, with frequencies in the wave functions $\Phi^{(3)}$ and $\Phi^{(4)}$ 
becoming smaller
at high energy ($\omega_n = n/2 \alpha' p^+_3 \sim n/ \sqrt{s}$). When $s$ is
increased, higher modes become increasingly  important, and the long string also becomes
extended in the transverse space, $x_\perp$.  As we shall see, without an
effective cutoff in the mode sum, the transverse size of the string  would be logarithmic
divergent. The interaction to first order in $T$ can be written explicitly as
\be 
\Delta S_{34}=  \frac{T }{2} \int^{p^+_3}_0 d\sigma[
{\dot X_\perp}^2 + \frac{1}{(2 \pi \alpha')^2}{X'_\perp}^2 ] =- T
\int^{p^+_3}_0 d\sigma \dot X^-(\sigma,0) \; ,
\label{eq:DeltaS34}
\ee
using the
Virasoro constraints to express in terms of $\dot
X^-(\sigma,0)$. Inclusion of this interaction term will be shown next to 
render   ${\cal V}_{34}$ a finite function of $k_\perp$, $T$ and $p^\perp_3$, and will also directly lead us to the desired
Regge behavior.

Assembling all the factors,  one is led to  a long string form factor, 
\beq
F_{34}(-k^\perp,s)& \equiv& F_{34} \equiv \int d\mu(T) {\cal V}_{34}(-k^\perp)\\
& \simeq & \int d\mu(T) \; \int D X_\perp
\Phi_{3}(X_\perp) \Phi_{4}(X_\perp) \; e^{ \textstyle   T
  \int^{p^+_3}_0 d\sigma \dot X^-(\sigma,0) } \; e^{\textstyle i k^\perp
  X_\perp(p^+_3,0)} \; .\nonumber
  \label{eq:stringformfactor}
   \eeq
This  can be evaluated directly by  expanding in normal modes, so
that 
\be
F_{34}  \simeq  \delta^2(p^\perp_3 + p^\perp_4 +
k^\perp)\; (2 \alpha' p^+)  \;
 \int dT \; T^{-2} e^{  p^- T} \;\exp\left[\; 
-  \sum_n \frac{\alpha' k^2_\perp }{n +
  n^2 T/2 \alpha' p^+_3}\;\right] \; .
   \label{eq:modesum}
\ee
The sum in the exponent, at large $p^+_3 \simeq p^+ \simeq  s/2 p^-$, leads to  a logarithmic growth,
\beq
F_{34} &\simeq&   \delta^2(p^\perp_3 + p^\perp_4 +
k^\perp)\; (-\alpha' s) \;\int   d\zeta  \; \zeta^{-2} e^{ - \zeta} \exp[ - \alpha' k^2_\perp\log(-\alpha' s/\zeta) +
 O(1/s)] \nonumber \\
&\simeq&   \delta^2(p^\perp_3 + p^\perp_4 +
k^\perp)\; (-\alpha' s) \;\Gamma( -1 + \alpha' k^2_\perp) \;\exp[- \alpha'
 k^2_\perp  \log(- \alpha' s)] \; .
\eeq
Finally combining the vertices for both short and long strings
and performing the $k^\perp$ integral gives the final result, 
\be
 {\cal A}(s,t) \simeq  
\Gamma(- 1 - \alpha' t) \;(  e^{-i\pi} \alpha' s)^{\textstyle 1 + \alpha'
 t } \; .
 \label{eq:regge}
\ee

\FIGURE[ht]{
%\begin{center}
\includegraphics[width = 3.4in]{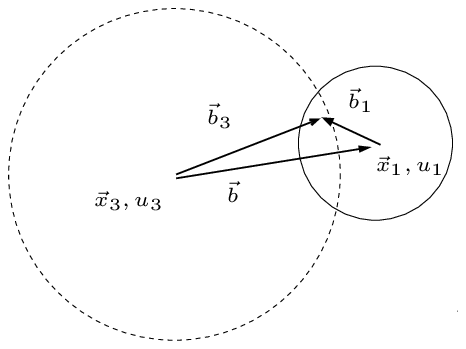}
%\end{center}
\caption{At the interaction the impact parameter is given by $\vec b =
  \vec b_3 - \vec b_1$ where $\vec b_i$ is the vector from the center
  of mass of each string to its end point. In AdS space the strings
  are separated by an additional transverse co-ordinate $u = u_1 -
  u_3=\ln (z_3/z_1)$ in the radial direction.}
\label{fig:Impact}
}%\end{figure}

\subsubsection{Diffusion in transverse space}

Looking back at this derivation,
we can see that Regge behavior is caused, in the infinite-longitudinal-momentum
transverse-brickwall frame, by the logarithmic growth in $s$ of the
average fluctuation of the end of the ``long'' string, $X^{(3)}_\perp(p^+_3)$,
relative to its center of mass, $x^{(3)}_\perp$.

It is instructive to examine further the physics of the interaction region
($\tau\simeq 0$) from the perspective of the transverse space. The size of
an incoming string can be characterized by the separation between  
its center of mass
 and its  end point where it interacts with the other strings: $\vec b_1=
X^{(1)}_\perp(0) - x_\perp^{(1)}$ and $\vec b_3=
X^{(3)}_\perp(p^+_3) - x_\perp^{(3)}$. 
With the constraint $X^{(1)}_\perp(0) = X^{(3)}_\perp(p^+_3)$
at the interaction, the conventional impact parameter is 
given by
\be
\vec b \equiv   x_\perp^{(1)}- x_\perp^{(3)}= \vec b_3 -\vec b_1 \; , 
\ee
as illustrated in Fig.~\ref{fig:Impact}. We could have chosen, 
in our light-cone analysis, to represent the scattering amplitude (\ref{eq:amplitude}) in impact-parameter space, Fourier transforming
from $k^\perp$ to $x_\perp$.   After integrating the fluctuations,
we would find the ``kernel",  ${\cal K}(y; x_\perp^{(1)},
x_\perp^{(3)})$, as the Fourier transform of the $s$-dependent factor
in Eq.~(\ref{eq:regge}).  The kernel satisfies a diffusion equation,
\be 
[\; \dd_y -1 - \alpha'
\dd^2_{x_\perp^{(1)}}\; ] \; {\cal K}(y; x_\perp^{(1)},x_\perp^{(3)}) =\delta
(x_\perp^{(1)}-x_\perp^{(3)})\; \delta(y)\; , 
\ee
where the evolution parameter is the rapidity, $y \sim \log(\alpha' s)$. 
Since one works only with physical degrees of freedom, 
the light-cone path integral has the advantage that
we can  follow the evolution of the physical transverse motion of the
string ``bits'', directly leading to a diffusion picture at high energies.  

In flat space, the diffusive growth in impact parameter is equivalent
to the ``Regge shrinkage'' for the Regge ``form factor'', $\exp[
\alpha' t \log(\alpha' s)]$. The amplitude decreases more rapidly in
$|t|$ at large values of the energy.  Historically, Regge behavior was
first exhibited in a relativistic setting by summing ``ladder
graphs'', or more generally, ``multiperipheral ladders''~\cite
{Amati:1962nv}. A crossing symmetric generalization led to the
consideration of ``fishnet
diagrams''~\cite{Nielsen:1970bc,Sakita:1970ep}, which played an
influential role in the construction of early string theories. Here,
we have reversed the argument, and have shown how a string-string interaction
in the Regge limit reproduces the underlying diffusion phenomenon.
This picture will be generalized in the next subsection to treat the
case of warped spacetime.

Our computation is related in an interesting way
to that of Ref. \cite{Polchinski:2001ju}.
Suppose we introduce a local conserved vector current coupled to charges at
the end of a single open string.  As emphasized in
Ref. \cite{Polchinski:2001ju}, in light-cone gauge such a form factor
would (naively) be obtained by introducing a local vertex $\dot
X_\mu(\sigma,\tau) \exp[i k^\perp X_\perp(\sigma,\tau)]$ at
$(\sigma,\tau)= (p^+,0)$ onto a straight worldsheet of width $p^+$ and
infinite length. The calculation is very similar to our Regge 
computation for the long string, but with $T$ strictly taken to zero,
which results in a logarithmic divergence, $\< {\Delta X}_\perp^2 \>
\sim\sum_n 1/n$, in the transverse size $\Delta X_\perp = X_\perp -
x_\perp$.  This is the well-known disease for local currents in flat
space~\cite{Polchinski:2001ju}. In contrast, the Regge ``form factor''
$F_{34}$ of the long string, Eqs.~\eref{eq:stringformfactor}--\eref{eq:modesum}, 
has its divergence cut off by the interaction
operator, $\exp[ T \int d\sigma \dot X^-]$, that we obtained working
to first order in $T$.

If we compare our light-cone Regge form factor $F_{34}$ in
Eq.~(\ref{eq:stringformfactor}), with the product of the conformal
Regge vertex and its propagator in Eq.~(\ref{eq:Rvertex}),
\be
\Pi_R(\alpha't) {\cal V}^\pm_R(k,w) = \int^\infty_0
d\zeta \; \zeta^{-2 - \alpha' t} e^{ \zeta \dd_w X^\pm(w)} e^{ \mp i k X(w)} \; ,
\ee
we note that they are clearly not the same. 
However we have checked by direct calculation that the light-cone-gauge-fixed
version of the conformal Reggeon vertex,
\be {\cal V}^{\pm(lc)}_R(k,w) = [\dd_w X^\pm(w)]^{1+ \alpha' t} e^{\mp i
  k^\perp X_\perp(w)} \; ,
\label{eq:lcRvertex}
\ee 
also reproduces the scattering amplitude in the Regge limit
using the Virasoro constraints.  A derivation of this form of the Regge vertex from the light-cone path
integral would be nice.

\subsection{Regge behavior in warped spacetime}

Now we consider the effect of adding a warped transverse direction, the
AdS radial direction, $Z$.  The light-cone action  in an AdS space of
curvature radius $R$
is~\cite{Metsaev:2000yu}
\beq \int_0^{p^+} d\sigma {\cal L} &=& \frac{1 }{2}\int^{ p^+}_0
d\sigma\ \left[ {\dot X_\perp}^2 + {\dot Z}^2 + \frac{1}{(2 \pi
\alpha' R^{-2}Z^2)^2}({X'_\perp}^2 + {Z'}^2) \right]\;, \nonumber\\
&=& \frac{1 }{2}\int^{ p^+}_0 d\sigma\ \left[ {\dot X_\perp}^2 +
R^2e^{-2U} {\dot U }^2 + \frac{1}{(2 \pi \alpha' )^2}(e^{4U}
{X'_\perp}^2 +R^2 e^{2U} {U'}^2) \right]\; ,\nonumber \\
 \eeq
 where have introduced $U(\sigma,\tau) = - \log(Z(\sigma,\tau)/R)$.  In the
 light-cone frame the conformal group $O(4,2)$ is restricted to the subgroup
 $SL(2,C)$: $Z \rightarrow \lambda Z, X_\perp \rightarrow \lambda X_\perp,
 \tau \rightarrow \lambda \tau, \sigma \rightarrow \sigma/\lambda $, which is the isometry of Euclidean $AdS_3$ for the three transverse coordinates, $(x_\perp, z)$. To
 exploit this invariance, we work with the $U$ variable and make a
 semi-classical expansion around the zero modes, $U = u$ and $X_\perp =
 x_\perp$ . The essential new feature is an effective string slope,
 $\alpha'_{\rm eff}(u) = \alpha' e^{-2 u}$, which leads to local dependence
 on $u$.  The dressed wavefunctions to Gaussian order become
\be
%V_r \rightarrow 
\Phi_r[X,U]=  \exp\left[ \textstyle i  p_r^\perp
  x^{(r)}_\perp  \;
 - \Half \sum_n \omega^{(r)}_n [ e^{2 u}
 X^{(r)}_n X^{(r)}_n +   R^2 U^{(r)}_n U^{(r)}_n ] \right] \; \psi_r(u_r)
\; ,
\label{eq:AdSwave}
\ee
to be compared with \Eref{Phidef}.

The calculation proceeds along the same line as that for the flat
background, with the addition of the one extra transverse coordinate $u$
(see Fig. \ref{fig:Impact}).  Factorization of the Dirichlet
constraint on $U$ in the interaction region requires that
\Eref{ordinarydeltas} be supplemented with an additional
delta-function,
\be
\delta \left(U^{(1)}_\perp(0) -
U^{(3)}_\perp(p^+_3)\right) = \int \frac{d\nu}{2\pi} \exp\left\{-i\nu [ U^{(1)}(0) -
U^{(3)}(p_3^+)]\right\} \; .
\ee
Again diffusion takes place only for the ``long'' boosted string, and
we must keep the interaction $\Delta S_{34}$ to first order in $T$.
The details are similar to the flat space derivation, except that
$\alpha'k^2_\perp$ is replaced by $\alpha'_{\rm eff}(u) k^2_\perp +
\nu^2$, in the new version of Eq.~(\ref{eq:modesum}). For
$k^2_\perp=0$, due to conformal invariance, the $\nu^2$ term
corresponds to flat space diffusion in the $u$-direction. For non-zero
$k^2_\perp \neq 0$, we must replace $\nu$ by an operator $i\dd_u $
conjugate to $u$ and take care with operator ordering.
  
In fact to deal rigorously with the operator ordering problem, one
must go beyond the Gaussian approximation to one loop
order~\cite{Callan:1986ja} for the worldsheet sigma model.  The result
of this calculation would be to introduce a shift $\nu^2 \rightarrow
\nu^2 +2 i\nu$, identical to that  in the
computation of the anomalous dimension of the on-shell photon vertex
operator~\cite{GKPAdS}.  We
choose an alternative approach, fixing this ambiguity by matching the
Regge spectrum with the on-mass-shell wave equation ($L_0=1$ at $j=1$)
for the vector field in AdS space.  Either way this results in the
following Hermitian differential equation,
\be
[ \dd_y  -1 -    \alpha'_0  \; e^{- 2 u}   \dd^2_{x_\perp}    
- \frac{1}{\sqrt \lambda } (\dd^2_u - 1)] \; 
{\cal K}_V(y; x_\perp, u, x'_\perp, u') =\delta (x_\perp-x'_\perp) \delta(u-u')  \delta(y) \; ,
\ee
which determines the leading $j$-plane singularity to leading order in
$\alpha'_0$. In momentum space, this is equivalent to a Euclidean Schr\"odinger equation
for the open string Reggeon kernel,
\be
[ \dd_y  -1 -    \alpha'_0  \;  t \; e^{ - 2 u}   
- \frac{1}{\sqrt \lambda } (\dd^2_u - 1)] \; 
{\cal K}_V(y; t, u, u') = \delta(u-u')  \delta(y) \; .
\label{eq:OpenStringkernel}
\ee
Since large $u$ suppresses the corresponding diffusion in $x_\perp$, 
diffusion in $u$ gives rise to the BFKL cut. This will be made more precise in the next section where we explore this  quantum mechanical analogy. 
This effect could have been anticipated qualitatively in terms of the
boosted incoming wave function (\ref{eq:AdSwave}).  The $U_n$ modes enter
like ordinary transverse modes in flat space.  The hadrons
are peaked at small $u$, but as $p^+$ increases, diffusion in $u$ pushes
the incoming hadron wave function into the large $u$ (UV) region.
This then acts to increase the
effective energies ($\omega_n e^{2u}$) of the $X^\perp_n$ modes, suppressing
diffusion in $x_\perp$.  It is interesting to compare this
with the physics in flat space, where increasing $p^+$ 
reduces effective energies ($\omega_n \sim n/p^+$) for the modes
of all transverse
directions; this effect
is responsible for the Regge shrinkage of the small-angle scattering peak.  
Eq.~({\ref{eq:OpenStringkernel})
leads to   a BFKL-like cut  for the open string Regge exchange~\cite{Kwiecinski:1981yk,Kirschner:1982qf,McGuigan:1992bi,Bartels:2003dj,   Bartels:2004mu} starting at $j = 1 - 1/\sqrt{\lambda}$.

To generalize this to the closed string, one must introduce periodic boundary
condition on the strings and an additional modulus $\theta$ that rotates the
Riemann surface around a cut on the $s$-channel intermediate closed string.
The integral over $\theta$ forces level matching between holomorphic and
anti-holomorphic modes. This has the  effect of replacing $\alpha'_0
\rightarrow \half \alpha'_0$.  Again, to get the full contribution to leading
order in strong coupling, we should do a one loop correction to the Gaussian
approximation, this time resulting in the anomalous shift
$\nu^2 \rightarrow \nu^2 + 4 i \nu$, for the graviton vertex, consistent with the on-shell linearized
graviton equation and general covariance in the $AdS^5$ background.  The
final result, when transformed back to a momentum representation, is
\be
[ \dd_y  -2 - \frac{\alpha'_0 }{2} \;   t\; e^{\textstyle - 2 u}   -  \frac{1}{2 \sqrt \lambda } (\dd^2_u - 4) ]\; {\cal K}(y; t, u, u') =\delta (u-u') \delta(y)  \; .
\ee
As noted before, the spectral decomposition for this Schr\"odinger
operator, Eq.  (\ref{Hdefn}), gives the leading BFKL singularity at $j= 2-
2/\sqrt \lambda$ in strong coupling. It is also interesting to note that the
same anomalous shift $\nu^2 \rightarrow \nu^2 + 4 i \nu$ for the graviton
vertex was found in Ref.~\cite{Polchinski:2001ju} to be essential in giving a
finite form factor at non-zero $k^2_\perp$.  The BFKL singularity, power
behavior at wide angles~\cite{hardscat,Brower:2002er} and finite
power-behaved form factors all have a common origin, at least in strongly
coupled ultraviolet-conformal theories.

\section{Regge Trajectories in UV-conformal Theories}
\label{sec:UVconf}
%Effects of confinement}

Our previous computations of the kernel were done in
conformally-invariant theories, or in kinematic regimes
where confinement played no role.  We now consider the effects of
confinement, while keeping the ultraviolet strictly conformal.  A simple
example of such a theory is the \nonestar\ model studied in
\cite{nonestar}.  This discussion is by necessity less precise than
the previous ones, simply because there is model-dependence in the
confining region.  Our goal in this section is to make as many
model-independent remarks as possible, and examine where
model-dependence is to be found.

If confinement sets in at a scale $\Lambda$ in the gauge theory, this
leads to a change in the metric away from $AdS_5\times W$ in the
region near $z = R^2/r \sim 1/\Lambda \equiv z_{0}$.  Typically
\cite{Witten:1998zw,nonestar, ks} the space is cut off, or rounded
off, in some natural way at $z=z_{0}$, or equivalently $u=u_0$.  This
leads to a theory with a discrete hadron spectrum, with mass
splittings of order $\Lambda$ among hadrons of spin $\leq 2$.  The
theory will also have confining flux tubes (assuming these are stable)
with tension $1/\alpha'_{0} = 2\sqrt{\lambda}\Lambda^2$; the same scale
sets the slope of the Regge trajectories for the high-spin hadrons of
the theory.  Note the separation of the two energy scales, by a factor
of $\lambda^{1/4}$; this is an important feature of the
large-$\lambda$ regime.

Since the metric is changed near $u_{0}$, the form of the differential
operator $\Delta_j\approx \Delta_2$, defined in \eref{Deltajdefn}, is
likewise changed in this region.  The effective potential $V(u)$ for
the Schr\"odinger problem assocated to the kernel approaches
\Eref{potential} only for $u\gg u_{0}$.  However, for $-t\gg
\Lambda^2$, the exponentially rising potential for $u\ll \ln
(\sqrt{|t|}\Lambda)$ implies the kernel is insensitive to the region
near $u_0$.  This is consistent with the expectation in the QCD
literature that the BFKL calculation is infrared-safe for
large negative $t$, while the effects of confinement become important as
$t\to 0^-$, and for any $t>0$.  The regime where
confinement-independent results are obtained will be discussed further
below.

\subsection{The hard-wall model}

To begin this discussion, it is instructive to work
our way through the ``hard-wall'' toy model.  While this model is 
not a fully consistent theory, it does capture key features
of confining theories with string theoretic dual
descriptions.

In the simplest form of the hard-wall model, the metric takes an
$AdS_5\times W$ form \eref{AdSWmetric} for $u>u_0$ and has a sharp
boundary at $u=u_0$; without loss of generality we take $u = -\ln
z/z_0$ and thus $u_0=0$.  This metric does not satisfy the
supergravity equations, but experience has shown
\cite{hardscat,jpmsDIS,rhouniv} that it captures much of the phenomenology
encoded in the metrics of consistent four-dimensional theories with
confining dynamics \cite{nonestar,ks}.  In particular, phenomena for
which the details of the metric in the confining region are not
important --- potentially universal features of gauge theory --- are
often visible in this model.  One can identify infrared-insensitive
quantities and general features of the hadronic spectrum, hadronic
couplings, etc, including, as we will see, aspects of Regge
trajectories and of the Pomeron.  Meanwhile, model-dependent aspects
of these and other phenomena also can be recognized, through their
sensitivity to small changes in the model.

\FIGURE[ht]{
%\begin{center}
\includegraphics[width = 4in]{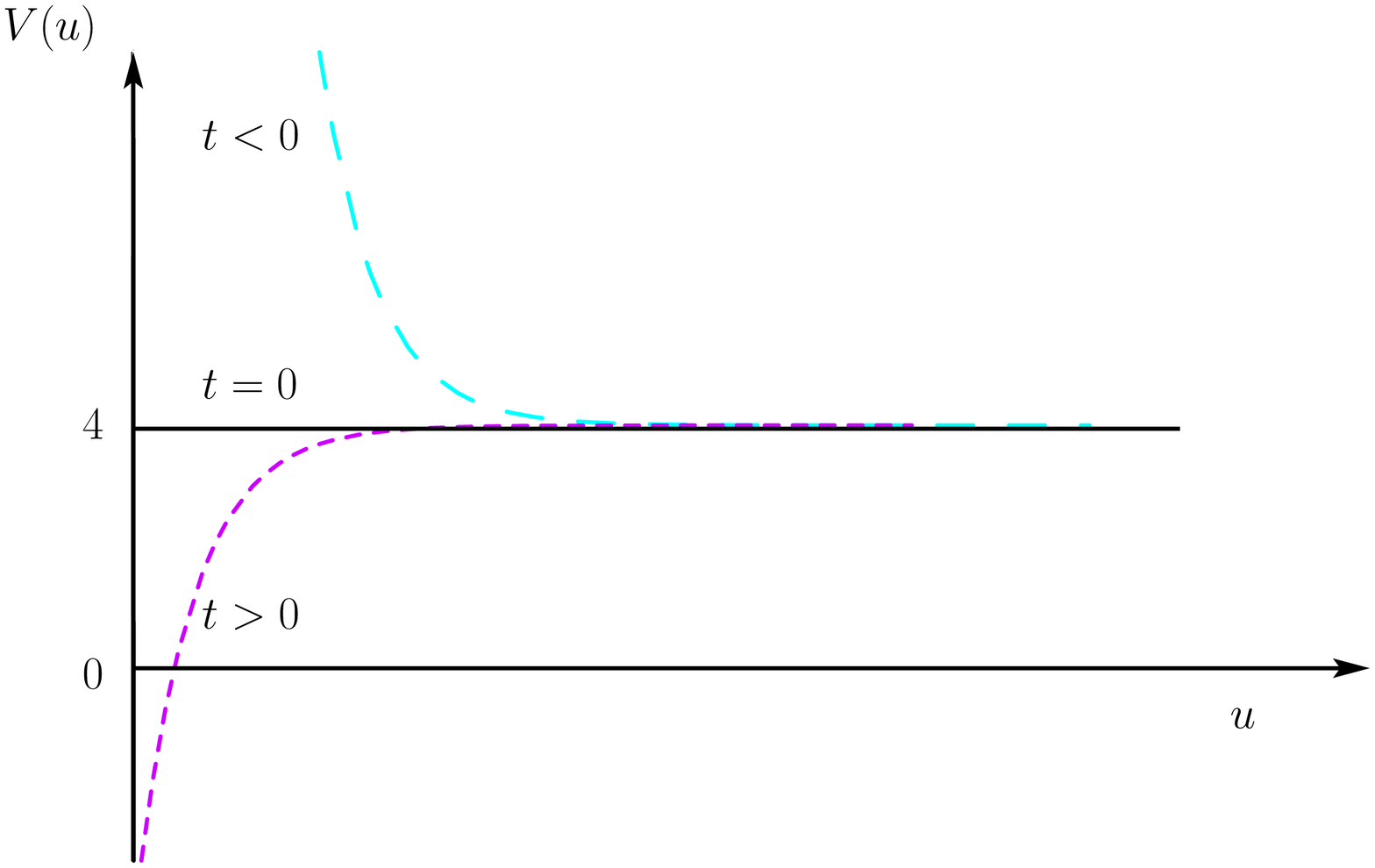}
%\end{center}
%
\caption{The potential for the effective Schr\"odinger problem in the
hard-wall model, for $t=0$ (solid), $t>0$ (short dash), and $t<0$
(long dash).  }
\label{fig:potential}
}%\end{figure}

The main advantage of the hard-wall model is that it can be treated
analytically, and the kernel can be written explicitly.  Since the
metric is still $AdS_5\times W$, we have the same quantum mechanics
problem to solve as in the conformal case, with potential $V(u)$ given
in \Eref{potential}, except for a cutoff on the space at $u=0$.
The boundary condition at the wall on the
five-dimensional graviton (and its trajectory for general $j$) is
constrained by energy-momentum conservation in the gauge theory.  We
must impose the boundary condition $\partial_r(r^{-2}\phi_{+^j})=0$
at $r=r_0$ for the analogue quantum mechanics system.  The logic is
the same as in deriving the wave equation~(\ref{wave}): the pure gauge
solution $h_{++} = r^2/R^2$ must be retained, else conservation of the
energy-momentum tensor will be violated. This condition extends to the
Pomeron for small $|j-2|$, which will be the regime we will mainly consider
below.

\subsubsection{Scattering of hadrons for $t<0$.}

For $t\leq 0$, the potential \eref{potential} has an exponential
growth at small $u$ and goes to a constant at large $u$.
Consequently, the spectrum of $\Delta_j$ is continous.  The kernel can
readily be expressed in terms of a set of delta-function-normalized
eigenfunctions,
\bel{K0usingIs}
{\cal K}_0(u, u',\tau,t)=  \int_0^\infty d \nu|c(\nu,t)|^2 
\{  I_{i\nu}(\xi) + R(\nu,t)  I_{-i\nu}(\xi) \}^* 
\{  I_{i\nu}(\xi') + R(\nu,t)  I_{-i\nu}(\xi') \} e^{-\tau \nu^2}
\ee
where $ I_{\nu}(\xi)$ is the modified Bessel function, $\tau\sim
{\ln s/2\sqrt{\lambda}}$ is defined in \Eref{tau},
$\xi=z\sqrt{-t}=(\sqrt {-t}/\Lambda) e^{-u}$, and $R(\nu,t)$ is fixed
by the boundary condition at $u=0$:
\be\label{Rdefn}
R(\nu,t) = -\frac{\partial_\xi[\xi^2 I_{i\nu}(\xi)] }{\partial_\xi[\xi^2 I_{-i\nu}(\xi)] }\Bigg|_{\xi= (\sqrt{-t}/\Lambda) } \ .
\ee
The parameter $\nu$ is related to the energy eigenvalue by $E=4+\nu^2$. 
The coefficients 
\bel{cdefn}
|c(\nu,t)|^2 = \frac{\nu}{2\sinh\pi\nu }
\ee
are normalization constants chosen so
that ${\cal K}_0(u, u',0,t)=\delta(u-u')$; because of conformal
invariance at large $u, u'$, the coefficients are actually 
$t$-independent.  

Since each Bessel function approaches a plane wave at $\xi\simeq 0$,
$R(\nu,t)$ is proportional to the reflection coefficient for a
plane-wave incident from $u=+ \infty$.  It will later be useful to
introduce a one-dimensional unitary S-matrix,
\be
S(\nu,t) \equiv e^{2i\delta(\nu,t)} =% -   
\left[\frac{\Gamma(1+i\nu) }{\Gamma(1-i\nu) }
\left(\frac{-t}{4\Lambda^2}\right)^{-i\nu}\right]   
R(\nu,t)\ .
%\frac{(\xi^2 I_{i\nu}(\xi))' }{(\xi^2 I_{-i\nu}(\xi))' }
%\Bigg|_{\xi= (\sqrt{-t}/\Lambda) }  
\label{SMatrice}
\ee

Let us first validate our expectations regarding scattering of
hadronic states.  At large $-t\gg \Lambda^2$, the scattering should be
model-independent, since the large momentum transfer shields the
scattering from the confinement region, as shown in \reffig{potential}.
Thus for large $-t$ and
large $u,u'$, the kernel should be almost identical to the kernel
\eref{Knonzerot} of a conformal theory.  This can straightforwardly
be verified by comparing \eref{Knonzerot} to \eref{K0usingIs},
with the use of \eref{Rdefn} and \eref{cdefn}.

As $t\to 0^-$, however, the effects of confinement become important.
This can most easily be seen in the $t=0$ kernel.  At $t=0$, the
Bessel function $I_{i\nu}$ in \eref{K0usingIs} reduces to a plane-wave
and the kernel becomes
\be
{\cal K}_0(u,u',\tau,t)= \int_{-\infty}^\infty \frac{d\nu }{4\pi} \{
e^{-i\nu u} + S(\nu,0) e^{i\nu u} \}^* \{ e^{-i\nu u'} + S(\nu,0)
e^{i\nu u'} \} e^{-\tau \nu^2} 
\ee
with
\be
S(\nu,0) = \frac{\nu-2i}{\nu+ 2i}
\ee
The integral can be performed, giving
\be
{\cal K}_0(u,u',\tau,t=0)
=\frac{e^{-(u-u')^2/4\tau} }{2\sqrt{\pi\tau} } + 
F(u,u',\tau)  \frac {e^{-(u+u')^2/4\tau}}   {2\sqrt{\pi\tau} } 
  \; , \label{Kreflects}
\ee
where
\be
F(u,u',\tau)=1-4 \sqrt{\pi \tau} e^{\eta^2} {\rm erfc} (\eta)\;,  \quad \eta=(u+u'+4\tau)/\sqrt{4\tau}\;.
\ee
Note $F=1$ for $\tau\to 0$ and $F=-1$ as $\tau\to \infty$, with
cross-over at $\tau \sim u+u'$ (for sufficiently large $u, u'$).

The formula \eref{Kreflects} is easy to interpret.  The first term is
the model-independent kernel \eref{Kzero} which describes diffusion
from $u'$ to $u$; the second term, which involves diffusion of the
image charge at $-u'$, is sensitive to the reflection off the wall at
$u=0$ and is thus model-dependent.  
Whether a given physical process
is model-dependent is determined by the relative importance of these
two terms.   For instance, the scattering
of two delta-function disturbances localized at $u_1 \gg u_2 \gg 0$
will be model-independent until $\ln s$ is large enough to permit
diffusion from $-u_2$ to $u_1$, and even then the second term will be
small compared to the first until still larger values of $\ln s$.

However, while hadrons of size $\rho$ typically peak at $u\sim -\ln
\rho\Lambda$, they also have power law tails extending out to large
$u$, of the form $r^{-\Delta}\sim e^{-\Delta u}$, where $\Delta$
(for a scalar hadron) is the dimension of the lowest dimension
interpolating operator for the hadron.  (More generally it is the
lowest-twist operator which determines the power.)  Consequently, 
the scattering
at $t=0$ of a small hadron, or off-shell photon, of size $\rho_1\ll
\Lambda^{-1}$ (with a wave function extending down to
$u_1\sim -\ln \rho_1\Lambda$) off of an ordinary hadron of size $\rho_2\sim
\Lambda^{-1}$ (with a wave function
peaking near the wall but sporting an $e^{-\Delta u}$ tail) has
several regimes.  Diffusion is unimportant for small $\ln s$,
model-independent but $\Delta$-dependent for moderate $\ln s$, and
model-dependent for $\ln s$ large compared to $\sqrt\lambda
u_1^2$.  These issues are addressed in the final calculations of
\cite{jpmsDIS} and will be revisited elsewhere.

\subsubsection{Regge trajectories at $t>0$}

The hard-wall model has a spectrum of hadrons typical of a confining
theory, including spin-two glueballs and their associated Regge
trajectories.  The spin-two glueballs are simply the discrete spectrum
of ``cavity modes'' of the Laplacian for a five-dimensional spin-two
field.  This Laplacian is the operator $\Delta_2$, slightly
reinterpreted, as we now discuss.

We have already explained that we may
view the operator $\Delta_j$ as
an effective Schr\"odinger operator $-\partial_u^2+V(u)$,
with $V(u)$ given in \eref{potential} (and shown in
\reffig{potential}) and
with energy eigenvalues $E$ that are
related to the corresponding spin $j$ by
\bel{EJreln}
  E = -2\sqrt{\lambda}(j - 2) \ ,
\ee
as we explained preceding \Eref{j0eq}.  For sufficiently positive $t$
this operator has a discrete set of normalizable ``bound-state'' modes
with eigenvalues $E=E_n<4$.  The bound state eigenvalues $E_n(t)$,
$n=1,2,\dots$, determine the Regge trajectories, where each trajectory
has $j_n(t) = 2 - E_n(t)/2\sqrt{\lambda}$.  The theory has a physical
spin-two glueball state of mass $m$ for each $t=m^2$ such that
$E_n(t)=0$ for some $n$
\cite{Brower:1999nj,Constable:1999gb,Brower:2000rp,Boschi-Filho:2005yh,Caceres:2005yx}. (Note
$V(u)$ goes to a positive constant as $u\to \infty$, so $E_n(t)=0$ can
only occur for $t > 0$.)  Higher spin hadrons on the trajectories lie
outside the supergravity regime.\footnote{For a review, see
\cite{Aharony:1999ti}.  } Meanwhile, since $V(u)$ goes asymptotically
to $4$ at large $u$, the spectrum of $\Delta_j$ also has a continuum
that extends over $E\geq 4$ ($j\leq j_0)$; this is the same continuum
as was present for $t<0$.

The bound states of the hard-wall model's
auxiliary quantum mechanics have wave
functions, in terms of $\chi=z\sqrt{t} = \sqrt{t} e^{-u}/\Lambda  = i 
\xi$,
proportional to
\bel{hardwallwf}
J_{\sqrt{4-E}}\left(\chi
%{\sqrt{t}\over\Lambda}e^{-u}
\right)
\ee
for those discrete values of $E$
where $\partial_\chi [\chi^2J_{\sqrt{4-E}}(\chi)]$ vanishes
at the wall $(\chi=\sqrt{t}/\Lambda)$. 
These values of $E$ correspond precisely to poles of the one-dimensional
S-matrix, Eq. (\ref{SMatrice}),
continued to $t>0$:
\be
S(\nu,t) \equiv e^{2i\delta(\nu,t)} \rightarrow  -
\left[\frac{\Gamma(1+i\nu) }{\Gamma(1- i\nu) }
\left(\frac{t}{4\Lambda^2}\right)^{-i\nu}\right]
\frac{(\chi^2 J_{i\nu}(\chi))' }{(\chi^2 J_{-i\nu}(\chi))' }
\Bigg|_{\chi= (\sqrt{t}/\Lambda) }
\ee
when $\nu$ lies on the positive imaginary axis.  The
glueball states,
found when $E_n(t)=0$,
have masses $m_n$ proportional
to the zeroes of
$4J_2(x)-xJ_3(x)$;
for $n\gg 1$ they are approximately equally
spaced, and by an amount $\Delta m\sim \pi \Lambda$.
The equal spacing in mass for large $n$ is
required by the WKB approximation applied to the potential
\Eref{potential}.

The continuum
states for $E>4$ can be read off from \eref{K0usingIs}, with
$I_{i\nu}(\xi)$ replaced by $J_{i\nu}(\chi)$,
$\chi=(\sqrt {t}/\Lambda) e^{-u}$. They take the form
\bel{hardwallcont}
J_{i\sqrt{E-4}}\left(\chi
%{\sqrt{t}\over\Lambda}e^{-u}
\right)
+ R\left(\sqrt{E-4},t\right)
J_{-i\sqrt{E-4}}\left(\chi
%{\sqrt{t}\over\Lambda}e^{-u}
\right)
  \ee
where the function $R$ is given in \Eref{Rdefn}.

We have plotted the spectrum as a function of $j-2$ [linearly related
to $-E$ by \Eref{EJreln}] in \reffig{hardwall}. The massive tensor
glueball states, marked with dots, are at $j-2=0$.  The familiar graph
of approximately linear Regge trajectories is supplemented by the
continuum of states (the BFKL-like cut) that begins at $j=
j_0=2-2/\sqrt{\lambda}$.  The unequal and nonconstant slopes of the
trajectories near $j=2$, like the equally-spaced hadron masses $m_n$,
are a model-independent feature of the supergravity limit $\lambda\gg
1$ for $j\approx 2$.  As $j$ increases the slopes gradually become
parallel and equal to $\alpha'_0 = (2\sqrt\lambda\Lambda^2)^{-1}$,
which is the reciprocal of $2\pi$ times the confining string tension.
\FIGURE[ht]{
%\begin{center}
%
\includegraphics[width=0.75\textwidth]{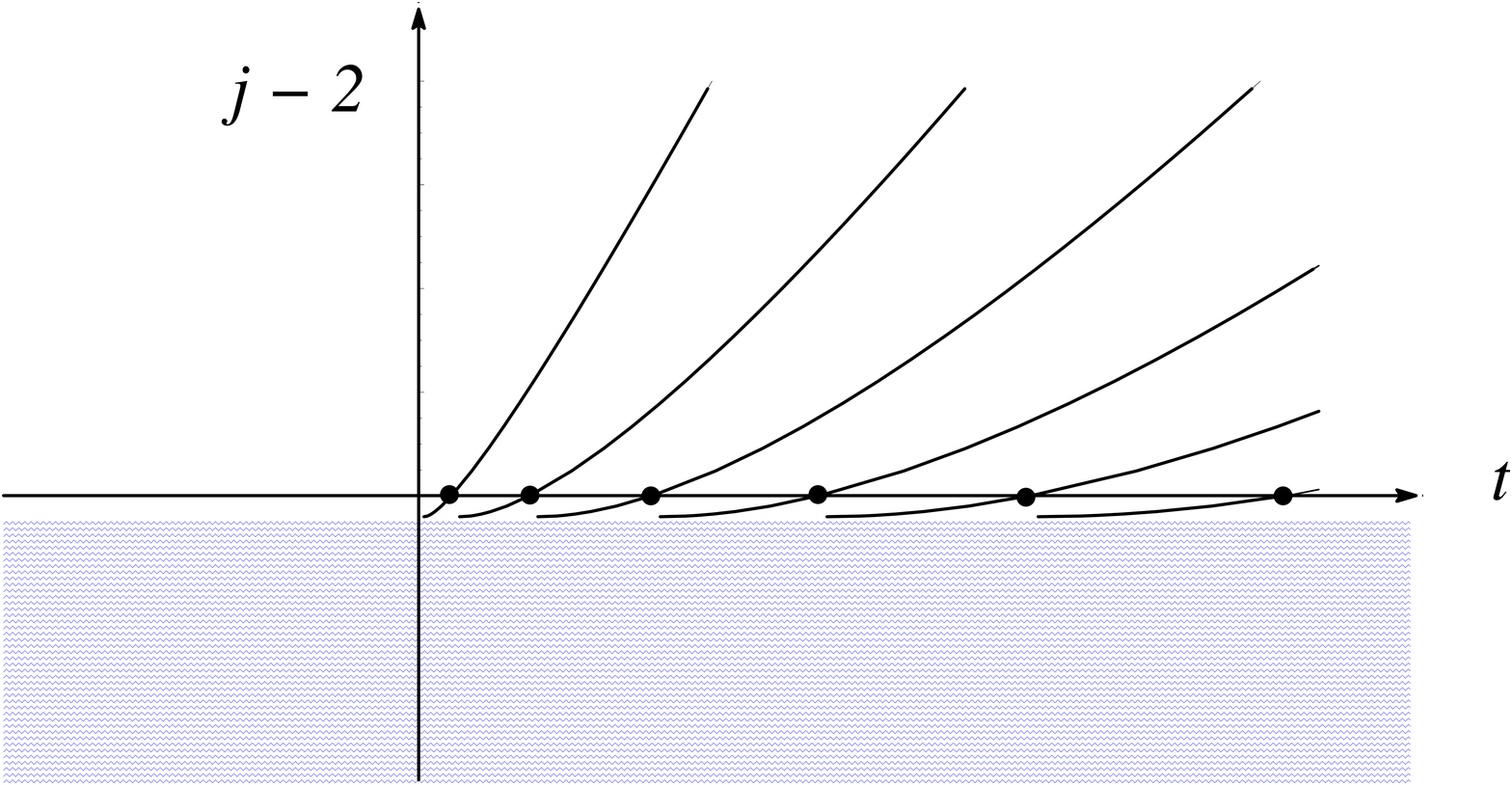}
\caption{The analytic behavior of  Regge trajectories in the hard-wall model,
  showing the location of the bound-state poles at $j = 2$ and the
  $t$-independent continuum cut (shaded) 
  at $j=j_0=2-2/\sqrt{\lambda}$ into which the Regge
  trajectories disappear.  The lowest Regge trajectory intersects the
  cut at a small positive value of $t$.  At sufficiently large $t$
  each trajectory attains a fixed slope, corresponding to the tension
  of the model's confining flux tubes.}
\label{fig:hardwall}
%\end{center}
}%\end{figure}
\FIGURE[ht]{
%\begin{center}
%
\includegraphics[width=0.65\textwidth]{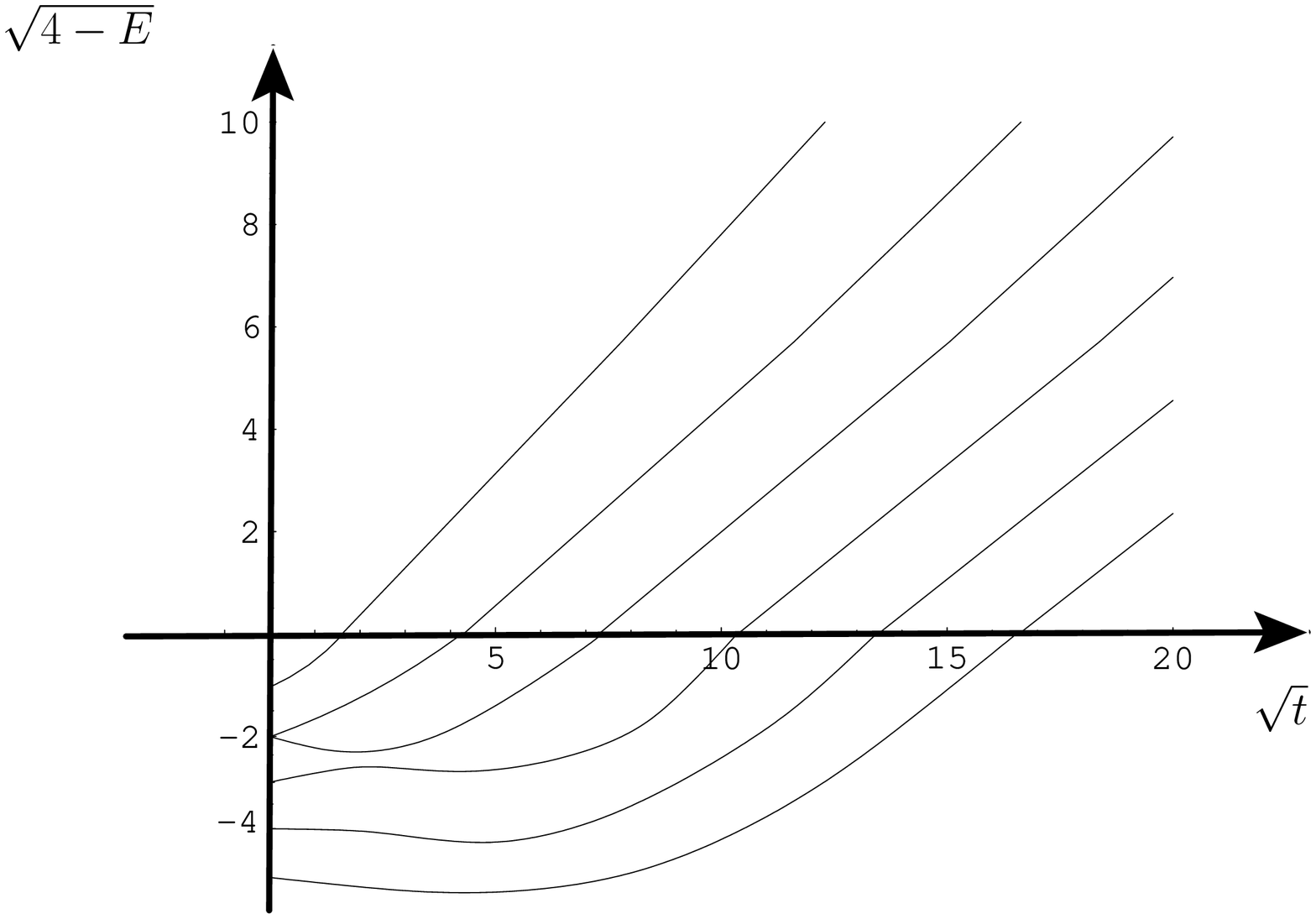}
\caption{The analytic behavior of  Regge trajectories in the hard-wall
  model plotting real $\sqrt{4-E}$ against $\sqrt{t}$. 
As $t$ decreases, each bound state pole moves from positive
to negative $\sqrt{4-E}$,
passing under the continuum cut in the $j$ plane and
 moving on the the second sheet.}
\label{fig:Ehardwall}
%\end{center}
}%\end{figure}
The transition to linear trajectories can be understood as follows.  As 
$t$ increases, the effective potential $4 - \Lambda^{-2} t e^{-2u}$ 
becomes deeper, and the states become more localized near the minimum 
at $u=0$.  The energy is then approximately $E\approx - \Lambda^{-2} t$, 
that is, $j \approx t/2\Lambda^2\sqrt\lambda$, giving 
the linear slope.  For a given trajectory $j_n(t)$, the WKB 
approximation shows that it reaches its
asymptotic slope for $(j-2)\gg n^2/\sqrt{\lambda}$.

The condition that the state be localized near the minimum, $\delta u 
\ll 1$, is consistent with the supergravity approximation, which 
requires that $R \delta u \gg \sqrt{\alpha'}$ or $\delta u \gg 
\lambda^{-1/4}$.  Thus the transition to linear trajectories occurs 
within the range of validity of supergravity.  However, to reach the 
excited states on the linear trajectory, where $j-2 = O(1)$, requires 
$\alpha'_0 t$ to be of order one.  This is outside the range of 
validity of supergravity: the momenta are of order the string scale.  
The supergravity regime thus describes the Pomeron trajectory for 
all negative $t$, and for positive $t \ll {\alpha'_0}^{-1}$.

In fact, our effective quantum mechanical description, obtained from 
the physical state condition $L_0 = 1$, extends to larger $t$.  Once 
$\alpha' t$ is of order one, higher derivative terms in $L_0$ are 
potentially unsuppressed.  However, because we know that the flat 
spacetime $L_0$ is exactly quadratic in derivatives,
these higher derivative terms are suppressed by at least one power of 
the curvature, a factor of $\lambda^{-1/2}$.  Hence to this accuracy we 
can continue to use
\begin{equation}
L_0 = \frac{j}{2} - \frac{\alpha'_0 t e^{-2u}}{4} - 
\frac{1}{4\sqrt\lambda} \frac{\partial^2}{\partial u^2} + 
O({\lambda}^{-1/2})\ .
\end{equation}

The hard-wall cutoff gives rise to an unwanted artifact, because the 
states become localized right at the wall.  A better model would have a 
smooth minimum.  For example, consider the variant
\begin{equation}
L_0 = \frac{j}{2} - \frac{\alpha'_0 t}{4(e^{2u} + e^{-2u} - 1)} - 
\frac{1}{4\sqrt\lambda} \frac{\partial^2}{\partial u^2} + 
O({\lambda}^{-1/2})\ ,
\end{equation}
combined with an orbifolding $u \to -u$.  
The potential has been 
designed to have the same large-$u$ behavior and the same minimum as 
before.  
Also, the boundary condition required for energy-momentum
conservation in this model
is consistent with the orbifold.  
This model gives a simple correction to the linearity of the 
trajectories.  For states near the quadratic minimum we can use the 
harmonic oscillator spectrum (even states only),
\begin{equation}
1 = L_0 = \frac{j}{2} - \frac{\alpha'_0 t}{4} + (2n + 
{\textstyle\frac{1}{2}}) (\alpha'_0 t)^{1/2} \lambda^{-1/4} + 
O({\lambda}^{-1/2})\ ,  \label{sqrtcorr}
\end{equation}
and so we obtain a $(\alpha'_0 t)^{1/2}$ correction to the slope, as in 
ref.~\cite{PandoZayas:2003yb}.
(The hard-wall model, where the potential is linear near the wall, 
gives an $(\alpha'_0 t)^{2/3}$ correction.)

We should note that the form~(\ref{sqrtcorr}) holds only for
$\alpha'_0 t = O(1)$, and breaks down at a yet higher energy
$\alpha'_0 t = O(\lambda^{1/2})$.  At this point the Pomeron becomes a
rotating string with a length of order the AdS radius, and we must
carry out a full string quantization~\cite{PandoZayas:2003yb}.  The
radial fluctuations (which are responsible for the $(\alpha'_0
t)^{1/2}$ term above) are ultimately exponentially suppressed, because
the radial fluctuations are massive on the world-sheet.  The leading
correction to the linearity of the trajectories is then the
L\"uscher term from the massless fluctuations~\cite{Luscher:1980ac},
which gives a negative constant shift.

\subsection{The analytic structure at all $t$ (constant coupling)}
\label{subsec:scathad}

Although it is a crude toy model, the hard-wall model may be
interpreted as capturing key universal features that will be true in
any quantum field theory with ultraviolet conformal invariance,
infrared confinement, large $\lambda$ and large $N$. Some of these
features, especially those that have to do with the analytic structure
of amplitudes, may be valid in any large-$N$ gauge theory at any
$\lambda$, and may in some cases survive to small $N$ as well.

Certain phenomena visible in \reffig{hardwall} will hold universally
in any similar theory.  Ultraviolet conformal invariance assures that
the auxiliary quantum mechanics problem in \Eref{Hdefn} will have a
potential which goes to the constant 4  for large $u$.  Conformal
invariance and analyticity assure that its $t$ dependence at large $u$
will be qualitatively similar; the deviations of $V(u)$ from 4 will
have a sign opposite to that of $t$ and a size that shrinks
exponentially at large $u$.  From this several consequences follow:
the spectrum will have
\begin{itemize}
\item  At all values of $t$, a continuum of states for $E\geq 4$ (the
BFKL-type cut at $j\leq 2-\frac{2}{\sqrt\lambda}$),
independent of the confinement physics at small $u$.
\item
For sufficiently large positive $t$, bound states with energies 
$E_n(t)$,
$n=1,2,\dots$ ({\it i.e.} Regge trajectories, with spins $j_n(t)$ and
positive slopes); the WKB approximation assures that for fixed
$j$ the trajectories at large $t$ are equally spaced in $\sqrt{t}$,
while for sufficiently large $j$
the slopes of the trajectories at low $t$ become
parallel, with slope
of order $(\lambda\Lambda^2)^{-1}$.  The trajectories remain linear, 
with a $(\alpha'_0 t)^{1/2} \lambda^{-1/4}$ correction, in the 
resonance regime $\alpha'_0 t = O(1)$.
\item
For sufficiently negative $t$, no bound states; as
$t$ decreases
the bound states move into the continuum, potentially
becoming resonances, and thus
the Regge trajectories 
disappear under the cut and move
onto the second sheet of the $j$ plane.
\item
Since $j_0<2$,
a spin-two glueball with mass
$m_n=\sqrt{t}$  at each $t$ for which
$j_n(t)=2$ (equivalently, $E_n(t)=0$); the WKB approximation ensures
that $m_n\propto n$ for large $n$, not $\sqrt{n}$ as
in flat-space string theory.
\end{itemize}

These quasi-universal phenomena are in contrast to certain important
model-dependent features.  The low-lying bound states of the quantum
mechanics problem, and the behavior of the trajectories $j_n(t)$,
$n\sim 1$, near $j=j_0$ --- the value of $t$ at which they touch the
end of the cut, and their behavior once they move onto the second
sheet of the $j$ plane --- are sensitive to the details of the
potential.  These aspects of the physics will be model-dependent even
at large $\lambda$ and large $N$.  The strongest model-dependence is
to be found where the leading trajectory disappears into the cut,
which unfortunately is a region of great physical importance.
Specifically, the value $t=t_1$ satisfying $E_1(t_1)=4$, where the
leading trajectory intersects the cut at $j=j_0$, is not strongly
constrained by general arguments.  It appears that one may vary the
potential $V(u)$ to obtain either sign for $t_1$, suggesting that
different confining models may lead to either sign.  Thus,
{\it whether the BFKL-type
cut at $j\leq j_0$ or the leading Regge pole $j_1(t)$ dominates the
large--$s$, $t=0$ behavior of the kernel appears to be
model-dependent.}  This is relevant for a number of processes whose
amplitudes are dominated by (or related through the optical theorem
to) forward scattering.

At smaller $\lambda$ (but still with large $N$ and a conformal
ultraviolet) certain aspects of \reffig{hardwall} will be modified.
From QCD data and BFKL calculations, we expect that the Regge
trajectories become steeper as $\lambda$ becomes smaller, and that
$j_0$ decreases from 2 toward 1.  Our results are consistent with
these expectations: the trajectories have slope $\sim
1/\sqrt{\lambda}\Lambda^2$, and $j_0 = 2-2/\sqrt{\lambda}$.  The
supergravity states and higher-spin string states begin to overlap as
$\lambda\to 1$, and the simple picture from the supergravity regime
must be supplemented; our auxiliary quantum mechanics problem becomes
non-local.  The analytic structure that we have found, however, may
remain intact.
This is because of the overall stability of the cut, whose presence at
all $t$ is required by conformal invariance, and of the trajectories,
which are required by confinement and the existence of glueballs at
positive $t$.

Conversely, however, our results support weak-coupling arguments
against applying BFKL to physics at $t=0$.  Single-Pomeron exchange
for $|t|\sim\Lambda^2$ is sensitive to the details of confinement, and
the dominant contribution from this regime need not be determined by
the physics of the model-independent cut --- the hard Pomeron ---
obtained at large negative $t$.  Different models with the same value
of $j_0$ can have different $t\to 0$ soft-Pomeron physics.  While we
have argued this in the regime $1\ll \lambda\ll N$, we see no reason
for it to change when $\lambda\ll 1\ll N$, or for smaller $N$.

In sum, the analytic structure realized in \reffig{hardwall} follows
at large $\lambda$ on very general grounds, with few assumptions, from
the constraints of ultraviolet conformal invariance and the physics of
confinement.  Its generality suggests that its
rough form survives to smaller $\lambda$.

\section{Effect of Running Coupling}
\label{sec:runcplg}

Up to this point, we have considered only theories for which the beta
function vanishes at high energies.  A logarithmically-running
coupling $\lambda$ makes a substantial qualitative change to the
kernel.

The effect on the differential 
operator appearing in the heat kernel is simple enough: as
long as the running is slow, one may view $\lambda$ as changing
adiabatically, and replace $\sqrt\lambda$ with $\sqrt{\lambda(u)}$, or
equivalently $R$ with $R(u)$, in
$\Delta_2$.  This reasoning is valid both in QCD and in
large-$\lambda$ theories such as the duality cascade.  Corrections to
this approximation are proportional to derivatives of $\sqrt\lambda$,
which, as we will show, are parametrically small in the region
relevant to our computation.

However, as is well-known from weak-coupling analyses, this small
change in the operator has a dramatic effect on the analytic structure
of the kernel \cite{finitetBFKL}.
 For a negative beta function, the continuous spectrum of
the operator is replaced with a discrete spectrum of bound
states, even for $t<0$; the BFKL cut is replaced with a dense set of poles,
the first of which is often called the ``hard Pomeron'' (in a shift
of the terminology formerly used to describe the cut.)  For a
positive beta function, the coupling in the ultraviolet is unbounded
and the cut simply begins at $j=2$, the infinite-$\lambda$
expectation.

\subsection{Effect in UV}

For a negative beta function (as in QCD)
results are most easily obtained at large negative $t$,
where the details of the ultra-strongly-coupled infrared physics
are unimportant.   The calculations are
dominated by the region where $1\ll \lambda(u)\ll N$,
that is, where the running coupling satisfies $g^2\ll 1\ll g^2N$.

 Examples of gauge theories with negative beta functions and
string-theoretic dual descriptions are known.  One such theory
\cite{AFM} has IIB strings on a space with an orientifold 7-plane, 4
D7 branes displaced far from the orientifold, and $N$ D3 branes on the
orientifold; this ${\cal N}=2$ $Sp(2N)$ gauge theory has a negative
beta function
\bel{betalambda}
\beta_\lambda=-{4\over N} {\lambda^2\over 4\pi^2} \ . 
\ee
It is important to note that $\beta_\lambda\sim 1/N$ in this model,
which is also true of the duality cascade (for which
$\beta_\lambda>0$.)  In such models the dual strings propagate on a
space which is approximately $AdS_5\times W$, with a slowly varying
metric and/or dilaton.\footnote{This statement need not be strictly
correct; in fact it is violated by the metric of the orientifold
model, at any $u$, in the region near the orientifold plane.  However,
the space-time region in which it is false decreases without limit as
$u$ becomes large.  This issue plays no role here and we will proceed
without examining it further.}

For the general $\beta_\lambda <0$ case,
we define
$$
0 < B\equiv -\beta_\lambda/\lambda^2\sim {1\over N} \ .
$$
The coupling varies slowly, as
\bel{runlambda}
{1\over \lambda(\mu)} = {1\over \lambda(\mu_0)}+B\ln(\mu/\mu_0)
\ee
Viewed from the ten-dimensional point of view, the
coupling depends on $z$, or equivalently $u$, as
\bel{runlambda2}
{1\over \lambda(u)} = {1\over \lambda(u_0)}+B (u-u_0)
\ee
since $u\propto -\ln z \propto \ln r$ and $r\sim \mu$ in the AdS/CFT
correspondence.  In this section it will be convenient to take
$\lambda(u_0)=\infty$ and $u_0=0$, so that $\lambda(u) = (Bu)^{-1}$;
note that in the previous section we set $u=0$ to be at the
confinement scale $\Lambda$, but we will not assume so here.

What is the effect of this running coupling?  The details are
model-dependent, but only in regions at small $u$.  At large $u$ 
the effect can only be an adiabatic alteration of the
$AdS_5\times W$ metric (except possibly, as noted
in the previous footnote, at isolated regions of small
measure on the
internal space $W$, which will have no effect on the calculation.)
Working in string frame, the metric will take the form, to leading
order at large $u$, 
\bel{runningmetric}
ds^2 = e^{2A(u)} dx^2 + R^2(u) \left[du^2 + ds^2_W(u)\right]
\ee
where, to leading order, $A\approx u$ and
\bel{Rofu}
R^4(u) = 4 \pi\lambda(u) \alpha'^2 = {4 \pi\alpha'^2\over Bu}\ .
\ee
It will prove inconvenient to have a running $g^{uu}$, so
for later use we may put the metric in the form 
\bel{runningmetric2}
ds^2 = e^{2A(w)} dx^2 + \alpha' dw^2 + R^2(w) ds_W^2(w) \ .
\ee
using a variable $w$ satisfying
\bel{wdefn}
dw = {R(u)\over\sqrt{\alpha'}} du \Rightarrow  u = C w^{4/3}
\ee
where $C\propto B^{1/3}$.

We now turn to the differential operator whose spectrum determines
the kernel.  This is
$\Delta_j\approx \Delta_2$,
as defined in \eref{Deltajdefn},
with the replacement of the factor $r/R$ by $e^{A}$:
\begin{eqnarray}
\Delta_2 &=& %\nabla_j^2 - \frac{j}{4} {\cal R}_+\!^+ 
%\nonumber\\
%\Delta_j \phi_{+^j} &=& 
e^{2A} \nabla_0^2 
\! \left[ e^{-2A} \phi_{++}  \right] \ . %\label{deltaj}
\end{eqnarray}
(In the hard-wall model below, the boundary
condition at the wall will now be
$\partial_w(e^{-2A}\phi_{++})=0$.) 
In the adiabatic regime, all terms in the metric vary slowly at large
$u$ except for the warp factor $e^{2A(u)}$.  We need therefore only
keep derivatives acting on $A(u)$, while dropping all derivatives
acting on $R(u)$ and on the slowly-varying metric on $W$.  Similarly,
if the theory has other varying fields, their slow variations need not
be retained at large $u$.

In the $w$ coordinate, the differential operator is of Schr\"odinger
form.  Diagonalizing the operator is equivalent to solving a
Schr\"odinger problem 
\be 
{\cal H}\Psi_{\cal E}(w)=[- \partial^2_w + V(w)]\Psi_{\cal E}(w)
= {\cal E} \Psi_{\cal E}(w) 
\ee 
with potential 
\bel{newVeff}
V(w) = \left({\displaystyle{\frac{8}{3}}} C\right)^2 w^{2/3} - h(w)
 {t}\  e^{- 2C w^{4/3}}
\ee
where $h(w)$ is a positive-definite function (of mass dimension $-2$)
whose form depends on details beyond the adiabatic approximation; its
effect is subleading, because it varies slowly compared to the
exponential that it multiples.  (To the same level of approximation,
the boundary condition in the hard wall is $\partial_w[e^{- 2C
w^{4/3}}\Psi_{\cal E}(w)]=0$.)  Here the operator ${\cal H}$ and
the eigenvalue ${\cal E}$ differ from our earlier conventions; in the
limit of a vanishing beta function, ${\cal H}=\sqrt\lambda H$ and
${\cal E} = \sqrt\lambda E$. With a running coupling, our earlier $H$
and $E$ are not defined, since they were {expressed} in terms of what
is now a running $R$; but ${\cal E} = 2(2-j)$ is well-defined.
Normal-ordering issues involving ${\cal H}$ and the running $\lambda$
are subleading within the adiabatic approximation.

The nonadiabatic corrections to these expressions are of order
$1/u\sim (R/\sqrt{\alpha'})/w$ 
$\sim B\lambda(u)$; since $B\sim 1/N$ this implies
corrections are of order $\lambda(u)/N$, which is small in the region
of interest.  Said another way, if $t$ is large and negative, but not
exponentially large, the repulsive exponential potential forces the
calculation to the region of large $u$, where $1\ll \lambda(u)\ll N$
and our adiabatic approximations and supergravity are both valid.
Thus the large negative $t$ region is, as in the conformal case,
model-independent.  This is consistent with earlier weak-coupling
results \cite{finitetBFKL,KirschLipat} and is illustrated in
\reffig{Vruncplg}.  
\FIGURE[ht]{
%\begin{center}
\includegraphics[width = 4.3in]{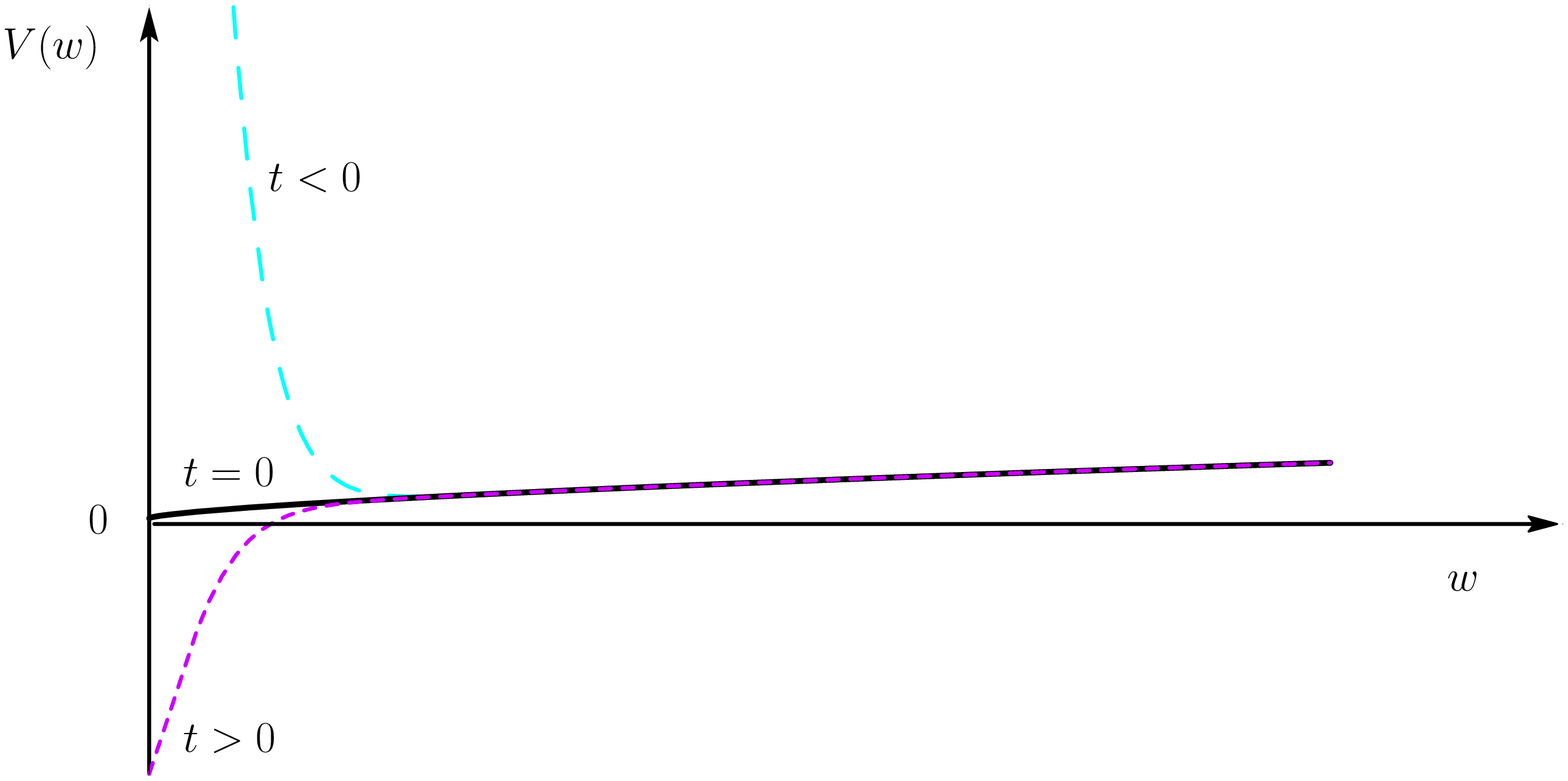}
%\end{center}
%
\caption{The form of the potential  $V(w)$ in a model
with a running coupling, where confinement is implemented with a hard wall.
Curves are shown for positive,
zero, and negative values of $t$.
Compare to Fig.~7.
}
\label{fig:Vruncplg}
}%\end{figure}

The potential, for $t<0$, grows to infinity at both small and large
$w$, which implies that the spectrum of $\Delta_2$ consists of
discrete bound states.  More precisely, this is true only within the
supergravity approximation, where $|E|\ll \sqrt\lambda$, and
$|j-2|\sim 1/\sqrt\lambda$.  At large $w$, where the coupling constant
becomes small, the supergravity approximation breaks down and
perturbative field theory becomes valid.  In this regime one can match
on to existing results for BFKL with a running coupling
\cite{Fadin:1998py, Camici:1997ij, Salam:1998tj,
Ciafaloni:1999au,Ross:1998xw,Levin:1999nm, Brodsky:1998kn}.  It 
is well-known \cite{finitetBFKL}
that in this case the discrete poles~\footnote{For the non-vacuum $q\bar q$ system,  the corresponding poles end at $j=0$\cite{McGuigan:1992bi}.} end at $j=1$, where a
cut begins and extends to $j\to -\infty$.  

For the bound states lying within the supergravity approximation, the
slowly-growing potential at large $w$ ensures that the spacing of
their energy eigenvalues decreases at higher eigenvalues.  The lowest
eigenvalue, not surprisingly, lies near
\bel{j0oft} j_1(t) \sim 2 - {2\over
\sqrt{\lambda\left(\mu\right)|_{\mu=\sqrt{|t|}}}} - \ { \cal O}(\sqrt{B})
= 2 - {\sqrt{2B
\ln \Big[z_0^2 |t|\Big]}} - \ {\cal O}(\sqrt{B}) \  .
\ee 
This is simply the conformal result for the beginning of the cut but
with $\lambda$ replaced with the running coupling $\lambda(\mu)$,
evaluated at $\mu\sim \sqrt{|t|}$, with corrections from zero-point
fluctuations around the minimum of the effective
potential.\footnote{To see this it is sufficient to do a variational
calculation or harmonic-oscillator-approximation for the ground state
in the potential \eref{newVeff}.  One must use the fact that $C\sim
B^{1/3}$ is small, that $h(w)$ is slowly varying, that $Cw^{2/3}\sim
\sqrt{Bu}\sim 2/\sqrt{\lambda(u)}$, and that the answer must be
consistent with our previous result for the conformal regime, $j_0 =
2-2/\sqrt{\lambda}$, in the limit $B\to 0$, $t\to \infty$ with
$\lambda(\sqrt{|t|})$ fixed.  
Higher states are better
described using the WKB approximation, in a form which is quite
similar to the weak coupling calculations of \cite{KirschLipat}.}

Strictly speaking, none
of these calculations can be done entirely within the
large $\lambda$ regime, since $\lambda(\mu)\to 0$ in the ultraviolet.
However, because $B\sim 1/N$, this occurs at very large $u$, in
particular $u\gg N$. The leading BFKL poles, associated with eigenstates of
$\Delta_2$, have exponentially damped eigenfunctions at large $u$,
and are not sensitive to this region.

For sufficiently
large $-t$, the region of small $u$ makes an exponentially suppressed
contribution, but the above calculations will have to be modified at
small $u$ as
$|t|$ decreases.  There are two possible effects that should be
accounted for: at small $u$ our adiabatic approximation may break
down, and also
ultra-strong-coupling effects and confinement become
important.  In fact these two conditions are related, and both occur
around $u\sim 1$, or more precisely at $u\sim 1/BN$.  Therefore
either $-t$ is large enough that both effects can be neglected,
or $-t$ is small enough that infrared effects such as confinement
must be 
accounted for.
In short we expect the results just
obtained for large negative $t$ will be reliable for $-t\gg
\Lambda^2$, requiring modification only at the scale where confinement
effects set in.\footnote{
This can be seen in particular examples. For example, in
the orientifold model, the subleading terms become important
where the dilaton reaches the value of order 1; this is also where
large deviations in the metric are expected. In a confining version of the
orientifold model, along the lines of the ${\cal N}=1^*$ model
\cite{nonestar}, the confinement regime must set in at or above this
scale.}

It is straightforward to repeat this exercise for the case of a
positive beta function.  The result is quite different, because the
effective potential is bounded as $u\to\infty$.  The spectrum again
consists of a cut, which begins at $j_0 = 2 - 2/\sqrt{\lambda_{max}}$
where $\lambda_{max}$ is the largest value obtained by the coupling.
In the case of the duality cascade \cite{ks}, the 't Hooft coupling
formally runs to infinity and $j_0=2$.

\FIGURE[ht]{
%\begin{center}
\includegraphics[width = 0.82\textwidth]{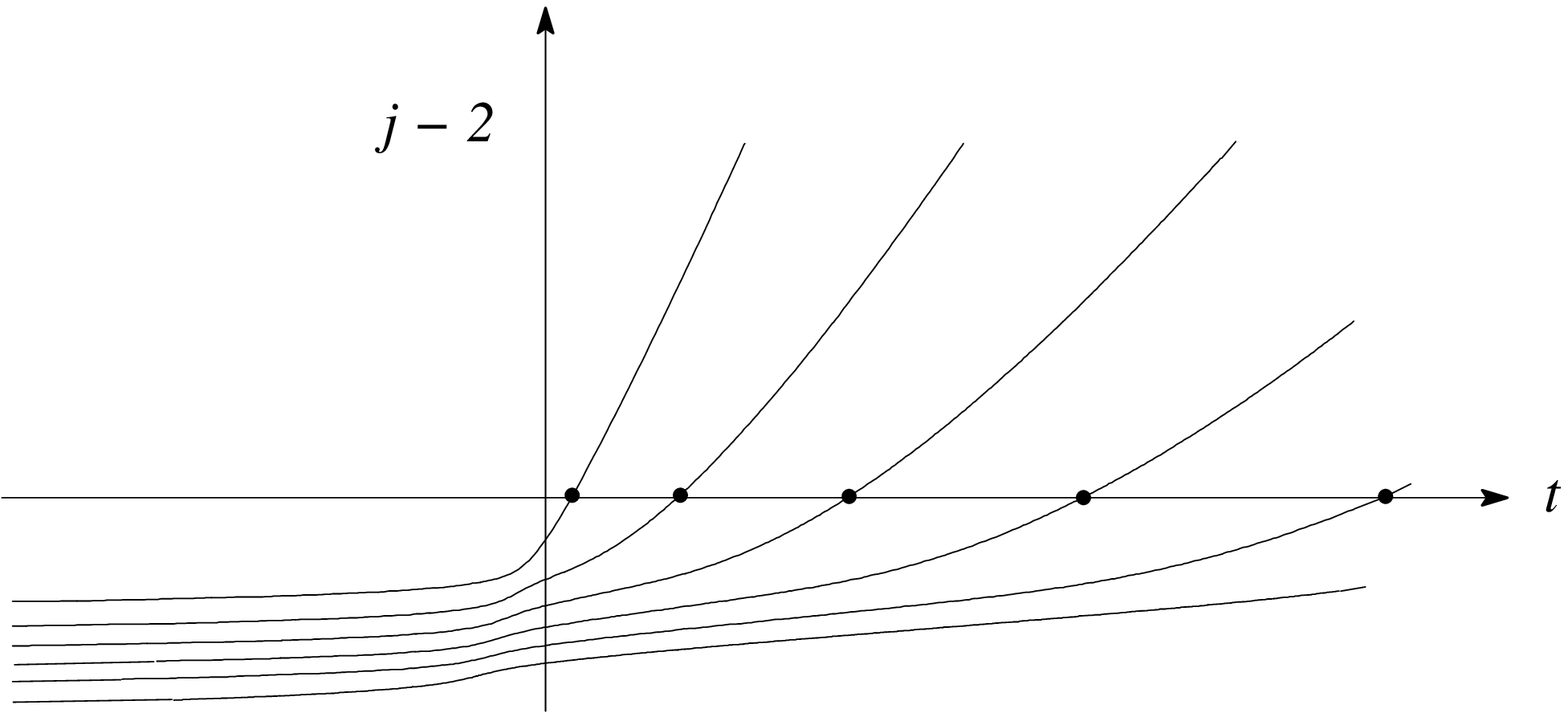}
%\end{center}
%
\caption{The analytic behavior of  Regge trajectories
with a running coupling. The figure
represents the spectrum of a hard-wall model 
with potential \eref{newVeff}; for definiteness we have set
$h(w)=1$ and put the wall at a point where the exponential
term is of order one.  
For $t>0$ the bound-state
poles are only logarithmically changed from Fig.~8,
but for $t<0$ the cut in Fig.~8 
disintegrates into
poles which are the continuation of the Regge poles at positive $t$.
As $t$ decreases, the poles slowly descend; weak
coupling considerations indicate they move toward $j=j_{min}\geq
1$, where a cut begins.}
\label{fig:runcplg}
}%\end{figure}

\subsection{The analytic structure at all $t$ (running coupling)}

Now let us turn to the properties of the kernel at values of $t$ where
confinement is important. As before, we focus on the general
properties of confining theories, whose essential feature is the
ending of the dual spacetime at some $u\sim 1$, or equivalently, some
$w\sim N^{1/4}$.  In Sec.~\ref{sec:UVconf} we considered the hard-wall model
with a conformal ultraviolet, and inferred general lessons from it.
Rather than pursue a similar strategy here, we will now modify the
lessons of Sec.~\ref{subsec:scathad} as required for the case of a
running coupling.

Many features of ultraviolet-conformal infrared-confining theories
continue to apply here.  Again the quantum mechanics problem in the
large positive $t$ region is characterized by a set of bound states
with negative energy eigenvalues (that is, with positive values of
$j-2$) which are well-separated and form the Regge trajectories of the
supergravity regime.  Again the $|t|\sim\Lambda^2$ regime is
model-dependent in its details.  The major new feature is that the
continuum of states with positive eigenvalues (the cut at $j\leq j_0$)
is no longer present, for any $t$, because of the growing potential at
large $w$.  The operator has a discrete spectrum at any $t$,
consisting of an infinite number of closely spaced states with a
positive eigenvalue $(j<2)$, and for $t>0$ a finite number of
well-spaced negative-eigenvalue $(j>2)$ states.  As $t$ is increased,
the eigenvalue $E_n(t)$ of any given state will move continuously from
positive to negative; correspondingly its spin $j_n(t)$ will move
smoothly from below 2 to above 2.  This is shown
in \reffig{runcplg}.
Thus, in contrast to the conformal case, where the spin $j$ of each
Regge trajectory, as $t$ is decreased, moves down to $j=j_0$ and
disappears below the BFKL-type cut, here each trajectory moves down to
become one of the BFKL-type poles.  In particular, the leading Regge
trajectory (often called the soft Pomeron) smoothly becomes the
leading pole (the hard Pomeron) as $t$ moves from positive to
negative.  This is consistent with the suggestion of
\cite{levintan}.  
It is interesting to compare the figure with the
results in \cite{Hancock:1992xh}; see also \cite{FR}.

As in the UV-conformal case, this basic form of \reffig{runcplg} is
model-independent, while some details, such as the precise nature of
the transition near $|t|\sim \Lambda^2$, are not.  Specifically, the
value $j_1(t)$ of the leading pole at and near $t=0$ is sensitive to
the details of confinement, as was also true in the ultraviolet
conformal case.   

It seems likely that the analytic structure shown in \reffig{runcplg}
is preserved into theories with a parametrically larger beta function,
including large-$N$ QCD, for which the small-$\lambda$ regime is much
closer to the confinement regime and the large-$\lambda$ regime is
very small.  QCD data \cite{rho} and lattice results on the hadron
spectrum \cite{Morningstar:1999rf,Meyer:2004jc,Meyer:2004gx}, and BFKL
calculations at $-t\gg \Lambda^2$
\cite{finitetBFKL,KirschLipat,Hancock:1992xh}, suggest that our results
match the analytic structure of QCD at $|t|\gg \Lambda^2$, for both
signs of $t$.  The logarithmically-violated conformal invariance of
the theory continues to put constraints on the form of the kernel at
large negative $t$, while the Regge trajectories at positive $t$ are
not expected to be much affected by the beta function.  A smooth
transition between the two behaviors, as in \reffig{runcplg}, is not
required theoretically, but seems plausible.

An interesting feature is that the leading Regge trajectory in
\reffig{runcplg} has $dj/dt>0$ everywhere, and so the $t=0$ behavior
of the amplitude must have faster growth with $s$ than the $t<0$
behavior.  The data are ambiguous as to
whether this applies in real-world QCD, as we now discuss.

\section{Outlook}

We have obtained the $j$-plane singularity structure of the Pomeron,
as a function of $t$, at large 't Hooft coupling.  One may ask if our
strong-coupling results are consistent with those which have already
been obtained at weak coupling.  In particular, since the location
$j_0$ of the leading singularity has been obtained to second- or
third-order in certain weakly-coupled theories, using BFKL
computational methods, one may ask if the weak- and strong-coupling
results can be suitably compared.  A theory in which this comparison
is well-posed is \nfour\ Yang-Mills theory.  This theory has a
constant and fully adjustable coupling $\alpha$, and the quantity
$\lambda = R^4/(\alpha')^2 = 4 \pi \alpha N$, where $\alpha$ is the
constant Yang-Mills coupling.  Can our result $j_0=2-2/\sqrt{\lambda}$
be interpolated with the weak-coupling result? The answer is shown in
Fig.~\ref{fig:N4j0}.  The leading-order BFKL computation of $j_0$,
\Eref{BFKLj}, grows from 1 toward 2 as $\alpha N$ increases, but
dramatically overshoots our result well before one would trust the
strong-coupling calculation.  However, the next-to-leading-order
correction to the coefficient $j_0$ is substantial and negative 
\cite{Kotikov:2002ab} --- though
much smaller than the correction in QCD itself (on which we will have
more to say below.)  The two-loop formula for the BFKL exponent $j_0$
\be
j_0 = 1 +{ 4 \ln 2}\ {
 \alpha N\over \pi}\left(1 - 7.58\ {\alpha N\over 4\pi}\right)
\ee
does not overshoot the strong-coupling result, and indeed matches on
to it quite reasonably at $\alpha N\sim 1$.  Thus, at least in \nfour\
Yang-Mills, the results appear compatible.

\FIGURE[ht]{
%\begin{center}
\includegraphics[width = 3.7in]{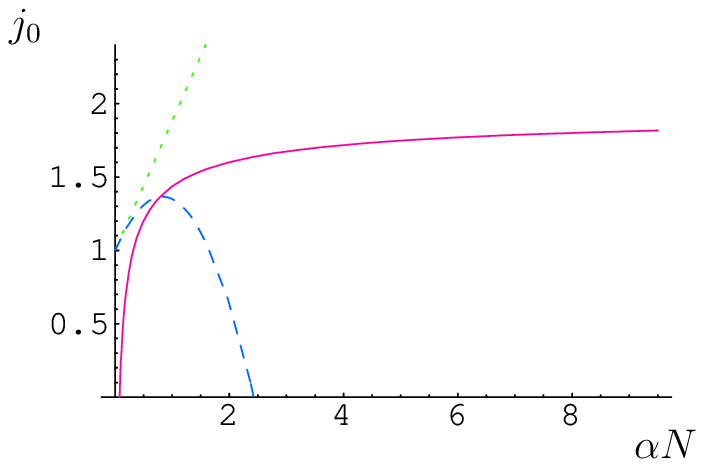}
%\end{center}
%
\caption{In \nfour\ Yang-Mills theory,
the weak- and strong-coupling calculations of the
position $j_0$ of the leading singularity for $t\leq 0$, 
as a function of $\alpha N$.
Shown are the leading-order BFKL calculation (dotted), the
next-to-leading-order calculation (dashed), and the strong-coupling
calculation of this paper (solid).  Note the latter two
can be reasonably interpolated.
}
\label{fig:N4j0}
}%\end{figure}

Let us now try to relate our dual string picture to the Pomeron physics 
of QCD itself.  The latter is still a subject of great confusion; we cannot 
fully resolve this here, but the simple and unified picture that we 
have found gives a useful framework for organizing the discussion.

One of the most striking aspects of high-energy hadronic scattering is
the rise in the total cross section, $\sigma_T(s)$, at the highest
available energies to date.  In Regge language, this requires a
leading $j$-plane singularity with vacuum quantum numbers, i.e. the 
Pomeron, and an intercept above $j=1$.  For instance, in the well-known 
work of Donnachie-Landshoff \cite{DL}, hadron cross-sections are fitted to a 
single-Pomeron exchange model
that gives $j_0=1.08$ for the Pomeron at $t=0$.  On the other hand, 
there is also evidence, for example from the small-$x$ behavior of 
parton distribution functions, for a much larger intercept, perhaps as 
large as 1.5 \cite{Andersen:2003xj}. 

It is common to ascribe these
behaviors to two distinct components of the Pomeron.
The exchanged object relevant for processes dominated by infrared physics
is called the ``soft'' Pomeron, while in processes in which scales
above the confinement scale are dominant, it is the ``hard'' Pomeron
which is relevant.  This 
distinction has a simple meaning in our picture, where the Pomeron 
depends on the fifth coordinate $r$: this degree of freedom corresponds 
to the overall size of the hadron wavefunction, $\delta x \propto 
R^2/r$, so the soft Pomeron has a size set by the confinement scale 
while the hard Pomeron is much smaller.

The two-component Pomeron
still presents a significant puzzle, however.
Consider the large-$N$ limit of QCD, where we can isolate the
contribution from single Pomeron exchange.  The leading Pomeron is a
pole, due to the running coupling, and we have the sharp question:
what is its intercept --- is it near 1.08, or much larger?  The
present theoretical understanding is not sufficient to answer this.
If we start at large negative $t$, where the potential barrier forces
the Pomeron to be small, then the perturbative BFKL analysis applies
and gives an exponent
\begin{equation}
j_0 = 1 +  {4\ln 2\over \pi}\alpha(t) N 
\left(1 + {\cal O}\left[\alpha(t)N\over \pi\right]\right)
\ .
\end{equation}
As we reduce $|t|$, the effective coupling increases and so does the
exponent, until at some point infrared effects take over and the
growth stops, leading to a finite intercept at $t=0$.\footnote{In our
strong-coupling potential model with running coupling, the exponent is
monotonically increasing with $t$.  This is likely to be true at
all $\lambda$, since even if $|t|$ is small we can take a small
Pomeron wavefunction as a variational approximation, giving a lower
bound on the exponent.}  If we make even the seemingly conservative
assumption that the BFKL result holds down to $\alpha = 0.25$ (for
$N=3$ QCD), we obtain the large exponent $j_0 \sim 1.6$.  
However, it is known that the two-loop correction to the exponent is very large
and negative, 
\be
j_0 = 1 +{ 4 \ln 2}\ { \alpha N\over \pi}\left(1 - \left[25.8+0.2{N_f\over 
N}\right]
\ {\alpha N\over 4\pi}\right) \ ,
\ee
and within the regime in which the calculation is
reliable --- at most $\alpha<0.1$, or more usefully $\alpha N<0.3$ ---
the value of $j_0$ does not exceed 1.10 \cite{Fadin:1998py}.  The
leading- and next-to-leading expressions for $j_0$ as a function of
$\alpha N$ are shown in Fig.~\ref{fig:QCDj0}, along with a horizontal
line at 1.08.  Thus there is no reliable indication that the true
intercept as $t\to 0$ must exceed 1.10, and perhaps the all-orders
BFKL calculation would predict nothing larger than 1.08.  We cannot
resolve this issue, but let us discuss how the various possibilities
could be consistent with the two-component
Pomeron picture.

\FIGURE[ht]{
%\begin{center}
\includegraphics[width = 3.7in]{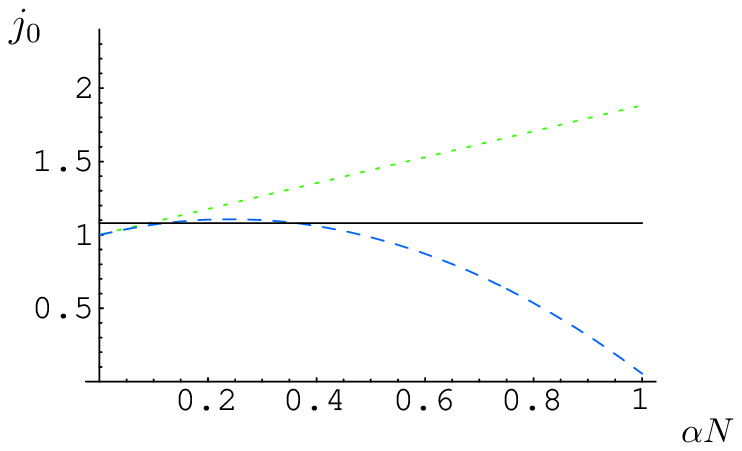}
%\end{center}
%
\caption{In QCD, the leading-order and next-to-leading-order
calculations of the position  $j_0$ of the leading
singularity as a function of $\alpha N$.   Shown are the leading-order 
BFKL calculation (dotted), the next-to-leading-order calculation (dashed), 
with $N_f=3$, and the value 1.08 of the
phenomenological soft-Pomeron intercept as extracted from data
\cite{DL} (solid).
 There is no convincing evidence that $j_0$ ever
exceeds 1.08.  }
\label{fig:QCDj0}
}%\end{figure}

If the true exponent for the hard Pomeron is large compared to 1.08,
how could we see a much smaller exponent in the total cross section?
There are two possibilities here: that the total cross section is
still in a regime dominated by single Pomeron exchange, or that we are
seeing the effects of multiple Pomeron exchange.  If we are seeing
single-Pomeron exchange, then the soft Pomeron must be some sort of
resonance.  Recall that the energy in the potential model appears with
a negative sign in the exponent, so a lower exponent is a higher
energy.  In the potential model, there could be a barrier between
the small-$r$ and large-$r$ region, with the true Pomeron ground state
(the hard Pomeron) concentrated at large $r$, and the soft Pomeron
being an excited state at small $r$.  Despite the fact that
$dj_0/dt>0$ for the leading pole, the effect of the resonance would be
to make it appear that $dj_0/dt<0$ in a region where the resonance
begins to dominate but before the asymptotic Regge behavior is
reached.  We should emphasize that at small $\lambda$ the sharp
locality in $r$ no longer holds, and the reduction to the single
degree of freedom of the Pomeron size is no longer a precise
statement, but we can imagine that there are different regions in the
Pomeron wavefunction that mix only weakly \cite{levintan}.

The other possibility is that the single-Pomeron exponent really is 
large.  This leads to rapid growth of the cross 
section with energy and the apparent value of 1.08 must then be due to 
unitarization effects, with multi-Pomeron exchange pulling the exponent down 
toward the Froissart bound of 1-plus-logarithms.  We will briefly 
discuss unitarization in the string picture below.  For QCD, if 
multi-Pomeron exchange dominates, there is the puzzling question of why 
factorization works so well, in particular that data indicates that 
$\sigma_{a,b}\simeq \gamma_a
\gamma_b$ to the level of $10-20\%$.

If instead the true exponent for the hard Pomeron is no larger, and
perhaps even smaller, than that of the soft Pomeron, why do we see a
larger exponent in some processes?  It is possible that this is a
transient effect due to diffusion.  If we have a process where some
external states are hard and some soft, as in deep inelastic
scattering, then the initial overlap of the wavefunctions is small,
but as we go to larger $s$ the Pomeron diffusion kernel leads to an
increasing overlap, and thereby gives an amplitude that increases
faster than the exponent in the kernel.  An example of this effect
is discussed in \cite{jpmsDIS}.

Regardless of the situation in QCD at current energies, it must be true 
asymptotically that multi-Pomeron exchange is important.  The BFKL 
calculation, in the extreme UV where it must be valid, gives a 
variational lower bound on the Pomeron intercept that is strictly 
greater than 1.  Thus single Pomeron exchange will necessarily violate 
the Froissart bound.
If we consider very large but fixed $N$, each Pomeron exchange costs a 
factor of $1/N^2$ but increases without bound at high energy, so at 
some point multiple Pomeron exchange will dominate.

It is interesting to consider this regime on the dual string side.   
Giddings studied hadronic total cross sections via gauge-string 
duality, and argued that the dominant process was black hole 
production, and moreover that this would saturate the Froissart 
bound~\cite{Giddings:2002cd} (for some recent followup see 
refs.~\cite{Kang:2005bj,Nastase:2005bk}).  Our work in this paper 
applies in a different region of parameter space; that is, we take $N$ 
large compared to all other quantities so that we are strictly limited 
to one-Pomeron exchange, while black holes are produced as $s \to 
\infty$ at fixed $N$.  For completeness we provide here a brief 
discussion of the latter limit.

Looking first at the scattering process in ten dimensions, the 
amplitude in the supergravity approximation is
$$
{\cal T} \sim G\tilde s^2/ \tilde t\ ,\quad G \sim g_{\rm string}^2 
\alpha' \sim R^8/N^2\ ;
$$
note that we are ignoring dimensionless constants.  To get a 
dimensionless measure of the size of this amplitude we rescale to 
canonical normalization and go to impact parameter space,
$$
{\cal T}' \sim R^8 \tilde s/b^6 N^2\  .
$$
where $b$ is the impact parameter of the collision.  
At any fixed $b$ and $N$, this is large when
$
\tilde s \gg b^6 N^2/  R^8 .
$
The minimum effective impact parameter is $\sqrt{\alpha'}$, so 
perturbation theory first breaks down when
$\tilde s \sim N^2 \alpha'^3/R^8$.  This condition is reached first at 
$r=r_0$, corresponding to $s \sim \Lambda^2 N^2 \lambda^{-3/2}$.

When perturbation theory breaks down we can go further by the eikonal 
summation, which in impact-parameter space takes the form
$$
{\cal T}_{\rm eik} = -i (e^{i {\cal T}' } - 1)\ .
$$
There is a simple interpretation: one-graviton exchange breaks down 
because the center of mass energy is large, and the resummed amplitude 
represents the interaction of the particles through their coherent 
gravitational fields.

The eikonal approximation breaks down when the momentum transfer 
$\partial_b {\cal T}' $  is of order $\sqrt s$, or
$$
\tilde s^{1/2} R^8 / b^7 N^2 \sim \tilde s^{1/2} G / b^7 \gg 1 \ .
$$
In ten dimensions this is the same as the condition that the impact 
parameter is less than the Schwarzschild radius, so the system forms a 
black hole.  Thus at given $b$ there are three parametric regimes with 
increasing energy: one-graviton exchange, eikonalized graviton 
exchange, and black hole formation.  Note also that nonlinearities of 
the gravitational field are also small below the black hole scale, so 
only $s$-channel ladder plus crossed-ladder graphs are important (the 
eikonal approximation).

We have neglected Regge shrinkage, which does not qualitatively affect 
the discussion, as well as the effect of AdS curvature, which slightly 
reduces the exponent and also gives the graviton a mass;
to obtain a Froissart bound we must take the latter into account.  
Ref.~\cite{Giddings:2002cd} argued that black hole production would 
saturate the Froissart bound.  However, we should note that the eikonal 
approximation also saturates this bound \cite{Chang:1971je, Fried:1990ei},
so the amplitude will take the Froissart form even 
before black hole production.   Note however that the transition from 
eikonal to black hole behavior is a genuine phase transition, from 
states with order $N^0$ degrees of freedom to states with $N^2$ degrees 
of freedom (a gluon plasma; see~\cite{Aharony:2005cx} for a recent
discussion.)

It is interesting to follow the discussion to smaller $\lambda$.  For 
exponent $j_0$, the condition for breakdown of the eikonal 
approximation is $s^{j_0 - 3/2} f(b)/ N^2 \gg 1 $.  Once $j_0$ drops 
below 1.5, the energy dependence cannot overcome the $N^2$, and so 
there should be no analog of the black hole phase; rather the eikonal 
approximation is valid to all energies.  In large-$N$ QCD it seems 
likely, according to the previous discussion, that the effective 
exponent is always less than 1.5; then there is no production of a 
gluon plasma.  Ref.~\cite{Aharony:2005cx} came to a similar conclusion 
by other reasoning.

Let us now comment on a couple of important issues that we have left
unresolved.  The matching at negative $t$ of the weak-coupling and
strong-coupling conformal kernels, discussed at the end of Sec.~2, is
subtle, because the two kernels are functions of somewhat different
variables.  The leading-order weak-coupling kernel is a function of
the momenta of two gluons, while the strong
coupling kernel is a function of collective coordinates of a string
built from an indefinite number of gluons.  To make a precise
match would require a more complete understanding of how partons
emerge from the string description of the theory, which is a question
far outside the supergravity approximation.  However, it should be
possible to clarify the relation of the results without a full
understanding of the partonic limit. Note there is a formal
similarity between the weak-coupling BFKL amplitude (\ref{weakkernel2})
and the general string result~(\ref{formal}); this should be explored
further.

Also unaddressed are the formal issues surrounding the adiabatic
approximation and the slow running of the coupling.  In a conformal
theory, both at weak coupling and at strong coupling, one finds a
stable $t$-independent cut, and there is a clear understanding
of how this follows from conformal invariance.  Meanwhile, a running
coupling breaks the cut into a set of running poles, rather than a
$t$-dependent cut.  The fact that this occurs must follow from a
formal argument involving weakly-broken conformal invariance, but we
are not aware of the existence of any direct proof.  It should also be
possible to study the spacing of the poles and properties of their
residues using methods that are valid at any $\lambda$.

To conclude, we have in this paper concerned ourselves with a unified
treatment of large-$N$ QCD-like theories at high energy, by
concentrating on the single Pomeron kernel over the entire range of
$t$. We have limited our discussion to the $2$-to-$2$ scattering of
spinless particles, but the formalism can be generalized to treat
processes involving particles with low spin, multi-particle
production, and others. Our kernel can be directly used to address
quarkonium-quarkonium and $\gamma^*\gamma^*$ scattering, with external
states involving wave functions which are strongly peaked at large
$u$, to extend earlier discussions of deep inelastic scattering
\cite{jpmsDIS}, and connect these regimes to the large-angle
scattering physics of \cite{hardscat}. Exclusive production processes
involving a moderate number of final particles or jets can also be
studied \cite{Brower:1974yv}.  We can also generalize our analysis to
apply to inclusive particle production
\cite{Detar:1971gn,Detar:1971dj}.  One particular promising area of
study is the inclusive diffractive production of jets, which is
expected to have a significant cross section at LHC. Since such events
will likely involve a wide range of momentum transfer squared, our
unified framework, capable of describing simultaneously both infrared
and ultraviolet features, offers a unique vantage point.  In
particular, the idea of a ``Pomeron structure function", which has
been a controversial notion from a weak-coupling analysis
\cite{Ingelman:1984ns,Goulianos:1995wy,factorization}, can be
addressed from a fresh perspective. Other specific examples of
experimental importance include the study of diffractive production of
vector mesons at large $t$, which has been thoroughly analyzed from a
weak coupling approach \cite{Bartels:1996fs}.  We hope to return to
these issues in future publications.

\

\

\

\

\

\

\

\

Acknowledgements: We are pleased to acknowledge useful conversations
with O. Andreev, A. H. Mueller, R. Janik, E. Levin, L. N. Lipatov, and
M. Strickman.  The work of R.B. was supported by the
Department of Energy under Contract. No.  DE-FG02-91ER40676,
that of C.-I.T. was supported in part by the
Department of Energy under Contract No. DE-FG02-91ER40688, Task-A,
that of J.P. by NSF grants PHY99-07949 and PHY04-56556, and that of
M.J.S. by U.S. Department of Energy grants DE-FG02-96ER40956 and
DOE-FG02-95ER40893 and by an Alfred P. Sloan Foundation award.  We are
grateful to the Kavli Institute for Theoretical Physics for its
support of the 2004 QCD and String Theory Program, where this work was
initiated.

\newpage

\bibliographystyle{utphys}
\bibliography{bfkl_2007}

\end{document}